\documentclass[prb,pre,aps,twocolumn,amsmath,amssymb,floatfix,
superscriptaddress]{revtex4-1}
\usepackage[dvips]{graphics}
\usepackage{graphicx}
\usepackage{color}
\usepackage{blindtext}
\usepackage{amsmath}
\definecolor{dred}{rgb}{0.75,0,0}
\usepackage{graphicx}
\usepackage{dcolumn}
\usepackage{bm}
\usepackage{xcolor}
\usepackage{braket}
\usepackage{physics}
\usepackage{appendix}
\usepackage{hyperref}
\hypersetup{
 colorlinks=true,
citecolor=magenta,filecolor=blue,
 linkcolor=magenta,
 }

\usepackage{graphicx}
\usepackage{dcolumn}
\usepackage{bm}
\usepackage{xcolor}

\definecolor{codegreen}{rgb}{0,0.6,0}
\definecolor{codegray}{rgb}{0.5,0.5,0.5}
\definecolor{codepurple}{rgb}{0.58,0,0.82}
\definecolor{backcolour}{rgb}{0.95,0.95,0.92}

\begin{document}

\preprint{APS/123-QED}

\title{\textcolor{blue}{Topological properties of a class of generalized Su-Schrieffer-Heeger networks:  chains and meshes}} 

\author{Sougata Biswas}
\affiliation{Department of Physics, Presidency University, 86/1 College Street, Kolkata, West Bengal - 700 073, India}
\affiliation{sougata.rs@presiuniv.ac.in}
\author{Arunava Chakrabarti}
\affiliation{Department of Physics, Presidency University, 86/1 College Street, Kolkata, West Bengal - 700 073, India}
\affiliation{arunava.physics@presiuniv.ac.in}
\date{\today}

\begin{abstract}
We analyze the topological properties of a family of generalized Su-Schrieffer-Heeger (SSH) chains and mesh geometries. In both the geometries the usual staggering in the distribution of the {\it two} overlap integrals is {\it delayed} (in space) by the inclusion of a third (additional) hopping term.  A tight-binding Hamiltonian is used to unravel the topological phases, characterized by a topological invariant. While in the linear chains, the topological invariant (the Zak phase) always appears to be quantized, in the quasi-one dimensional strip geometries and the generalized SSH mesh patterns the quantization of the Zak phase is sensitive to the strength of the {\it additional} interaction (the `extra' hopping integral). We study its influence thoroughly and explore the edge states and their robustness against disorder in the cross-linked generalized SSH mesh geometries. The systems considered here can be taken to model (though crudely) two-dimensional polymers where the cross-linking brings in non-trivial modification of the energy bands and transport properties. In addition to the topological features studied, we provide a prescription to unravel any {\it flat}, non-dispersive energy bands in the mesh geometries, along with the structure and distribution of the compact localized eigenstates. Our results are analytically exact.
\end{abstract}

\maketitle

\section{Introduction}
\label{intro}
The Su-Schrieffer-Heeger (SSH) model~\cite{su,heeger,asboth} stands out to be the paradigmatic representative of a one-dimensional (1D) version of a topological insulator~\cite{thouless}. The model, in the shape of a one-dimensional lattice and described conveniently in a tight binding formalism, is characterized by a staggered distribution of two different values of the overlap integral (popularly called the `hopping integrals', and designated by $v$ and $w$ in the present work), alternating periodically, exposing a two-sublattice structure of the parent lattice. A control over the ratio of these two hopping amplitudes, namely the `intra-cell' and the `inter-cell' ones that connect atomic sites within a unit cell, and between two neighbouring unit cells respectively, can lead to both topologically trivial and non-trivial phases. The nontrivial topological phase is marked by the existence of a nonzero topological invariant, which is the so-called `winding number' ($\nu$), that is closely related to the integral of Berry curvature over the Brillouin zone (BZ). The integral is popularly known as the Zak phase~\cite{zak}. The bulk-boundary correspondence~\cite{asboth} explains the emergence of symmetry-protected zero-energy states which are found localized at the edge(s) of the system. The transition between two topologically different phases is accompanied by a vanishing of the band gap at the phase transition point - a feature that comes out naturally through the SSH model.

The realization of topological effects has been achieved in recent times through exciting experiments involving photonic systems~\cite{hening,malzard,weimann,yang,alex}, or using lasers~\cite{bandres,harder}. The interface states that are induced by topology are observed in dielectric resonator chains~\cite{poli}, for example. The Bloch band topology is experimentally understood using Aharonov-Bohm interferometry~\cite{duca}, and reports of recent experimental determination of the Zak phase are available in literature~\cite{atala,longhi} strengthening the subject. 

\begin{figure}[ht]
\centering
\includegraphics[width=\columnwidth]{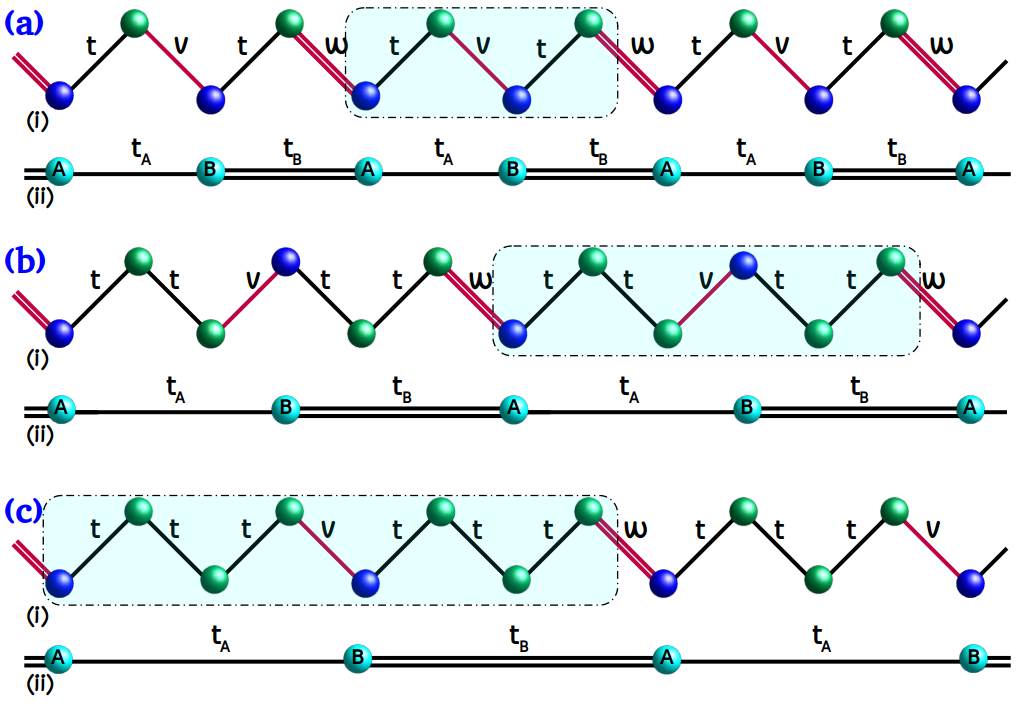}
\caption{(Color online) (a)(i), (b)(i), and (c)(i) show three different types of generalized SSH chains (GSSHC), referred to in the text as GSSHC-I, GSSHC-II and GSSHC-III respectively. Each GSSHC is built by arranging three different nearest neighbor hoppings, viz,  $t$, $v$, and $w$ in the orders shown. After decimating the green-colored sites, the GSSHC variants are transformed (renormalized) into chains with two different sublattices $A$ and $B$, as shown in  (a)(ii), (b)(ii) and (c)(ii) respectively. The unit cells are marked by the shaded box in each case.}  
\label{fig1}
\end{figure}

\begin{figure}[ht]
\centering
(a)\includegraphics[width=.9\columnwidth]{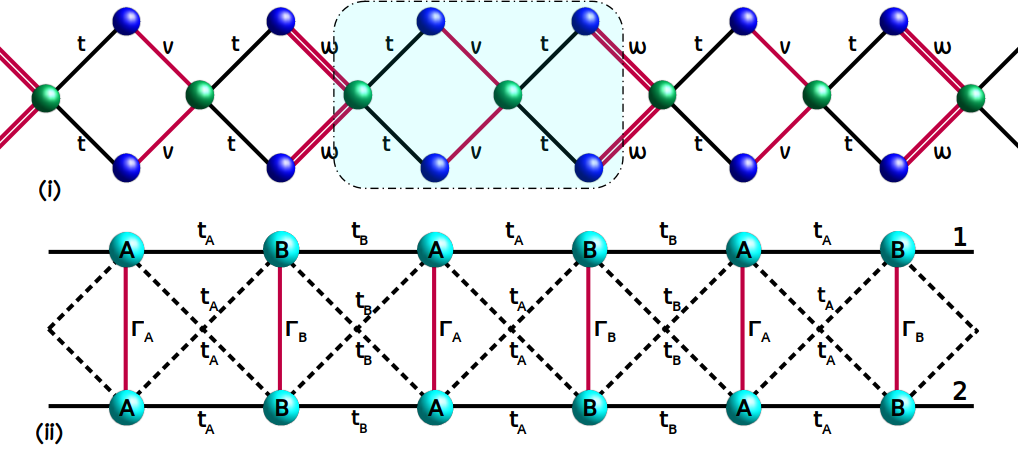}
(b)\includegraphics[width=.9\columnwidth]{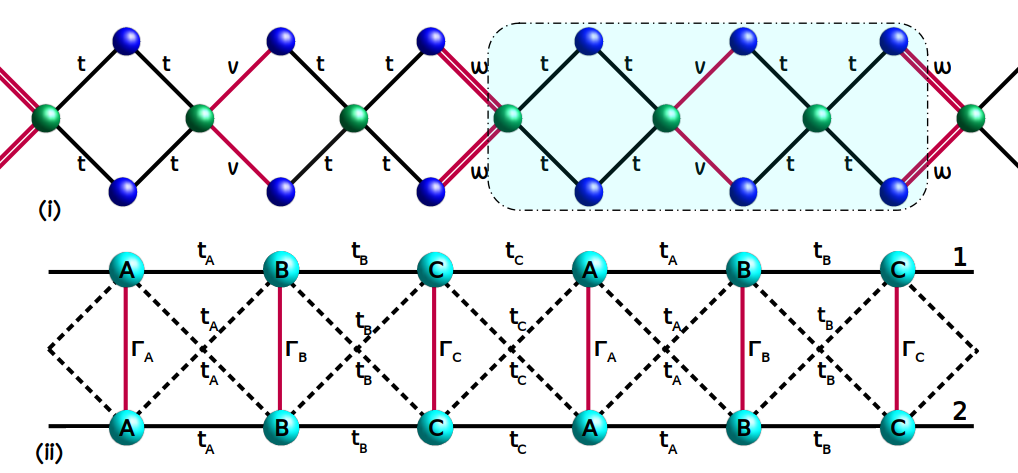}
(c)\includegraphics[width=.9\columnwidth]{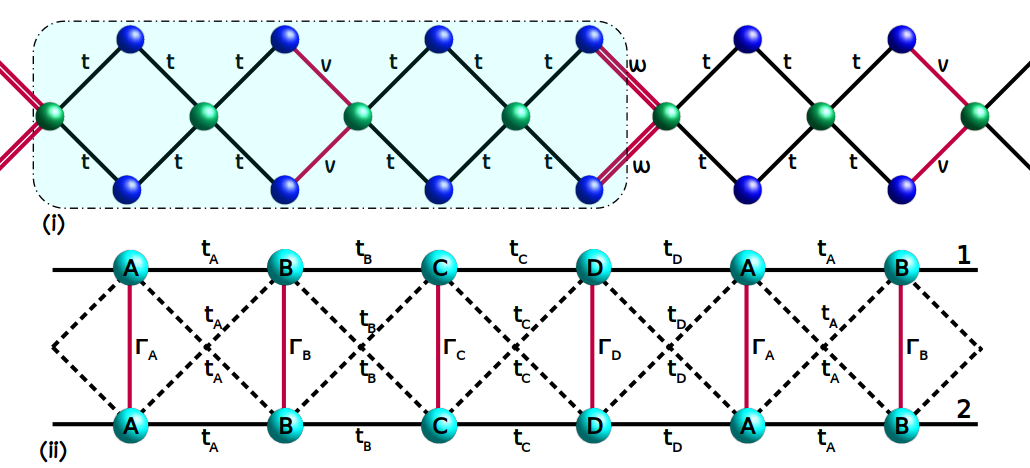}

\caption{(Color online) (a)(i), (b)(i), and (c)(i) show three variants of two cross-linked generalized SSH chains (GSSHC) of types I, II and III as shown in  Fig.~\ref{fig1} (a(i), b(i), c(i)). The unit cells are marked by a shaded box in each case. After decimating out the green-colored sites (having coordination number equal to four), these cross-linked GSSHC are mapped into two-strand ladder networks with two ($A$, $B$), three ($A$, $B$, $C$), and four ($A$, $B$, $C$, $D$) sublattice structures, as shown in (a)(ii), (b)(ii) and (c)(ii) respectively. The numerical values of on-site potentials and hopping integrals assigned to the models are described in the text. }  
\label{fig2}
\end{figure}

Coming back to the SSH model, which initially played an important role in understanding the properties of conjugated polymers~\cite{lu,baeriswyl}, we find that the simplicity of the model has motivated researchers to look deeper into the effect of variations of dimensionality or of the pattern of periodicity on the topological properties of the model. The variations include, for example, coupled SSH chains~\cite{li}, chains with non-local coupling leading to non-monotonous edge states~\cite{mirosh}, generalized SSH model~\cite{xu}, an SSH `trimer' lattice~\cite{anastasia,alvarez}, or a four-bond SSH model~\cite{bid}, to name a few. Multistarnd Creutz ladder network~\cite{amrita}, topological insulators with non-centered inversion symmetry~\cite{ricardo1}, topological properties of two bosons in flat band systems~\cite{ricardo2}, or the interesting case of a $2^n$ root topological insulator~\cite{ricardo3} have presented a few more (and not all, of course) intriguing cases with unanticipated insight into the physics of such systems. This observation makes the extensions of the basic SSH model worth exploring further.

In this communication we propose to inspect the topological properties of a class of generalized SSH chains (GSSHC), followed by an in-depth study of the band and topological aspects of a mesh designed out of such GSSHC's that crisscross each other forming a quasi-one-dimensional network. The network is infinitely extended along the $x$-direction and is finite, but arbitrarily wide along the $y$-direction. We call such a geometry a {\it generalized SSH mesh} (GSSHM). The constituent chains in the GSSHM are geometrically entangled, mimicking the situation in a polymeric system. This aspect is  addressed in a very recent work on cross-linked SSH chains~\cite{sivan}, where multiple SSH chains cross each other at a single site. 

In our work, both for the one-dimensional GSSHC family and the GSSHM geometries, we include an {\it additional} overlap integral, marked by $t$ in Fig.~\ref{fig1}. The inclusion of such an extra hopping element {\it delays}, in space, the staggering (alternating) pattern of the two other hopping integrals, marked $v$ and $w$, that are hallmarks of a traditional SSH chain. The effect of this third interaction, on the topological properties of the systems studied is not obvious, and we explore this area in some detail. We find that, even with a delayed staggering effect caused by the presence of one or multiple `extra' hopping integral ($t$) the bands in one-dimensional GSSH models show quantized topological invariant. The story of the quasi-one-dimensional GSSHM systems is not so trivial. There we need a threshold value of $t$ beyond which the Zak phase becomes quantized, in contrast to their one-dimensional counterparts.

\begin{figure}[ht]
\centering
(a)\includegraphics[width=.9\columnwidth]{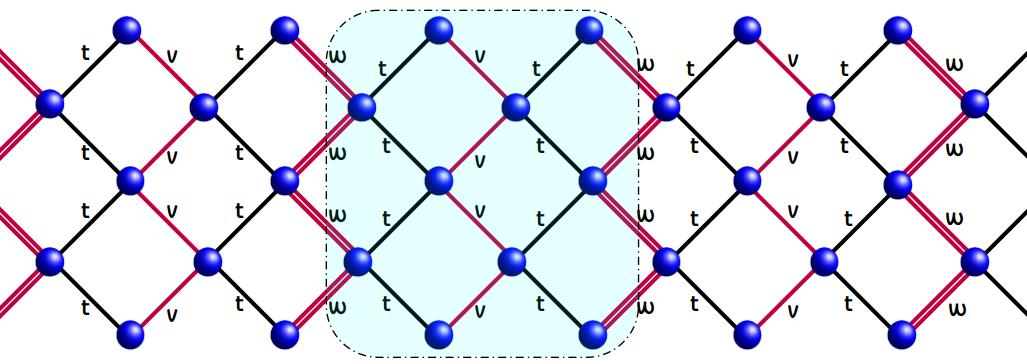}
(b)\includegraphics[width=.9\columnwidth]{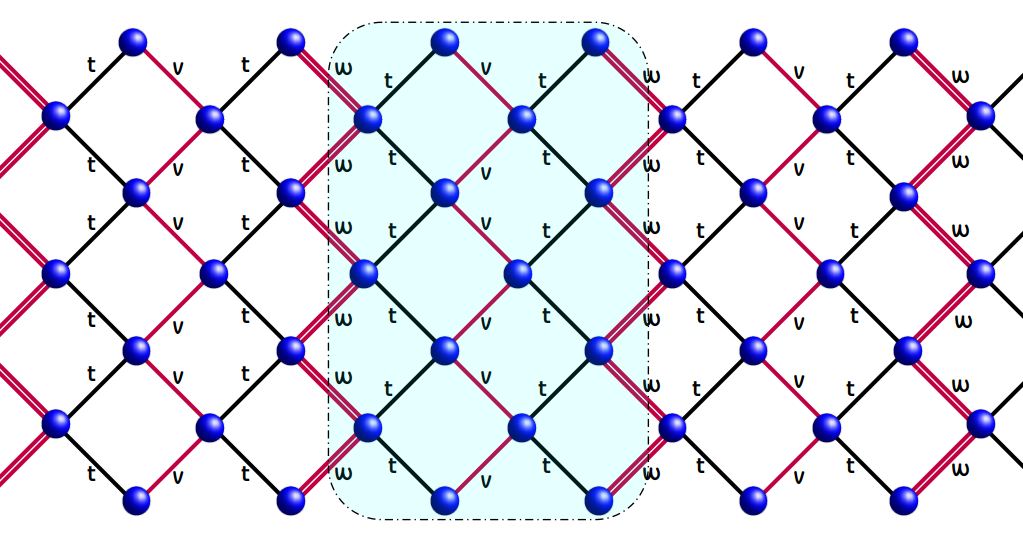} 
\caption{(Color online) Cross-linking of (a) four and (b) six GSSHC-I (as shown in  Fig.~\ref{fig1} a(i)). The unit cells are marked by a cyan-shaded box.}  
\label{fig3}
\end{figure}

To end the introduction it is worth mentioning that, the weaving of multiple cross-linked generalized SSH chains is likely to enrich the spectral properties of the pure SSH lattices and one can look forward to the tailoring of the eigenvalue-spectrum and the eigenstates. The model proposed in the present communication goes beyond the ones reported primarily within one-dimensional frameworks, and the effects of entanglement of the generalized SSH chains in higher-dimensional systems were not rigorously studied, to the best of our knowledge. We intend to take a step forward in bridging the gap in understanding the effects of an increased dimensionality mingled with a generalization of the SSH arrays - a variation that presents a naive but plausible model for realistic polymeric systems.

\vspace{0.2cm}
\section{The energy bands}
\subsection{The Hamiltonian and the Difference Equations}
The tight binding Hamiltonian in standard form can be written as, 
\begin{equation}
    H = \sum_j \epsilon_j c_j^\dag c_j + \sum_{<j,k>} t_{jk} c_j^\dag c_k + h.c.
    \label{ham}
\end{equation}
The operator $c_j^\dag$ ($c_j$) creates (destroys) a particle (electron, say) at the $j$-th atomic site. The `on-site' potential $\epsilon_j$ is chosen as $\epsilon$  for all sites, and we will set $\epsilon=0$ throughout. $t_{jk}$ is the nearest hopping integral, set as $t$, $v$ and $w$. In the cases of the one-dimensional GSSHC family, $t$, $v$ and $w$ are periodically repeated in any chosen order. We show our choices in  Fig.~\ref{fig1} (a), (b) and (c). 

To extract the energy bands and to obtain the eigenvalues at which the energy gaps close at the Brillouin zone boundaries, we will diagonalize the Hamiltonian in $k$-space, and parallelly we shall make use of a real space decimation method that will be based on a set of {\it difference equations}, which represents a discretized version of the Schr\"{o}dinger equation on a lattice. The difference equations read, 
\begin{equation} 
(E - \epsilon_j) \psi_j = \sum_k t_{jk} \psi_k
\label{diff}
\end{equation}
where, the amplitude of the wavefunction at the $j$-th atomic site is $\psi_{j}$, and $k$ runs over the nearest neighbors of the $j$-th site. 

\subsection{ Energy bands and the gap-closure energies of the GSSH chains: a real space decimation scheme}
\subsubsection{Type-I}
The decimation method yields the gap-closing (gap-opening) energies {\it exactly}, and in an anlytical form. We explain the scheme by choosing the first member of the GSSHC family, as shown in Fig.~\ref{fig1}(a). We name this lattice GSSHC-I.
Using the set of Eqs.~\eqref{diff} we evaluate $\psi_j$ corresponding to every green-colored vertex in Fig.~\ref{fig1}(a) in terms of the amplitudes of the wave function on its nearest neighboring sites. These amplitudes are then {\it eliminated} (decimated) in terms of the remaining (undecimated) sites, and the set of Eqs.~\eqref{diff} are re-written in terms of the amplitudes on the un-decimated sites.
The decimation of the subset of $\psi_j$'s corresponding to the green sites maps the GSSHC-I lattice onto a traditional-looking SSH lattice with just two `bonds' alternating periodically. This is depicted in Fig.~\ref{fig1}(a)($ii$). The mapping is exact and no underlying physics is lost. 

In Fig.~\ref{fig1}(a)($ii$) one easily identifies two sublattices ($A$ and $B$). The process of decimation renders the on-site potentials and the effective nearest neighbor hopping integrals on the {\it new} lattice energy dependent. These new, {\it renormalized} on-site potentials are named $\epsilon_A$, $\epsilon_B$ respectively, and the new nearest neighbor hopping integrals are termed $t_A$, $t_B$, which alternate periodically. They are given by, 
\begin{eqnarray}
\epsilon_A & = & \epsilon + \frac{w^2+t^2}{E-\epsilon} \nonumber \\
\epsilon_B & = & \epsilon + \frac{v^2+t^2}{E-\epsilon} \nonumber \\
t_A & = &  \frac{ v t }{E-\epsilon} \nonumber \\
t_B & = &  \frac{ w t }{E-\epsilon} 
\label{rsrg-1a}
\end{eqnarray}

A set of difference equations, written for this renormalized two-sublattice chain, given in Fig.~\ref{fig1}(a)($ii$), is easily written as, 
\begin{eqnarray}
    (E - \epsilon_A) \psi_{j \subset A} & = & t_B \psi_{j-1 \subset B} + t_A \psi_{j+1 \subset B} \nonumber \\
    (E - \epsilon_B) \psi_{j \subset B} & = & t_A \psi_{j-1 \subset A} + t_B \psi_{j+1 \subset A}
    \label{diff2}
\end{eqnarray}

The above set of equations represents an effective SSH chain with on-site, energy-dependent potentials $\epsilon_A$ and $\epsilon_B$, and periodically alternating `hopping amplitudes' $t_A$ and $t_B$. Keeping in mind the analysis of a traditional SSH chain it is easy to appreciate that the energy spectrum of this system will have band gaps closed at the Brillouin zone boundaries when $E-\epsilon_{A(B)}=0$, and $t_A=t_B$. This leads to,
\begin{eqnarray}
    E & = & \epsilon \pm \sqrt{v^2 + t^2} \nonumber \\
    v & = & w
    \label{bandclosing-1}
\end{eqnarray}


\begin{figure*}[ht]
\centering
(a)\includegraphics[width=0.6\columnwidth]{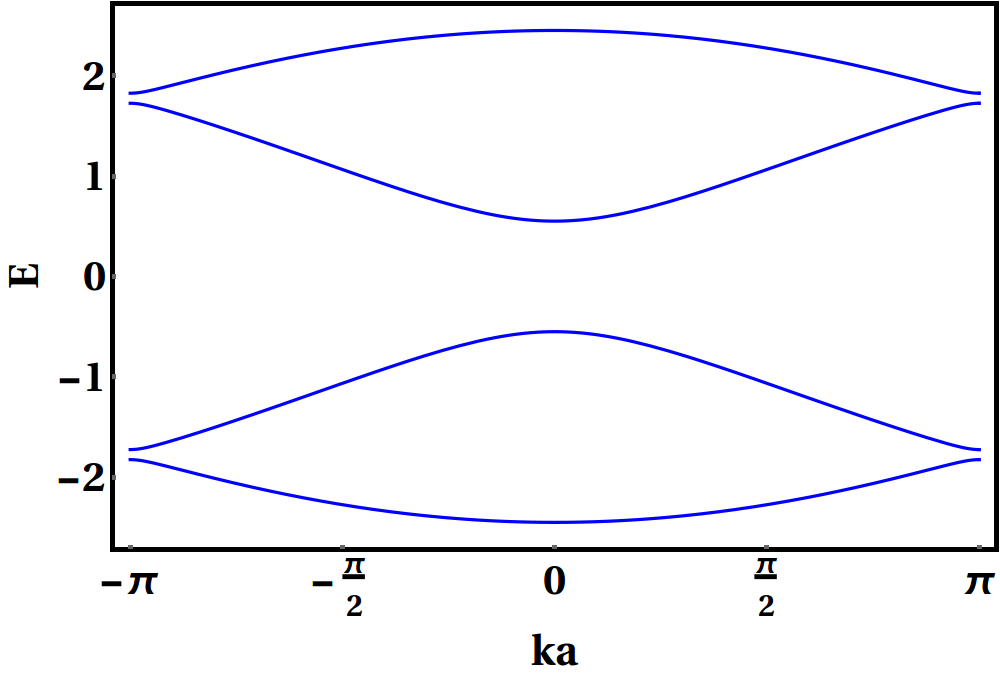}
(b)\includegraphics[width=0.6\columnwidth]{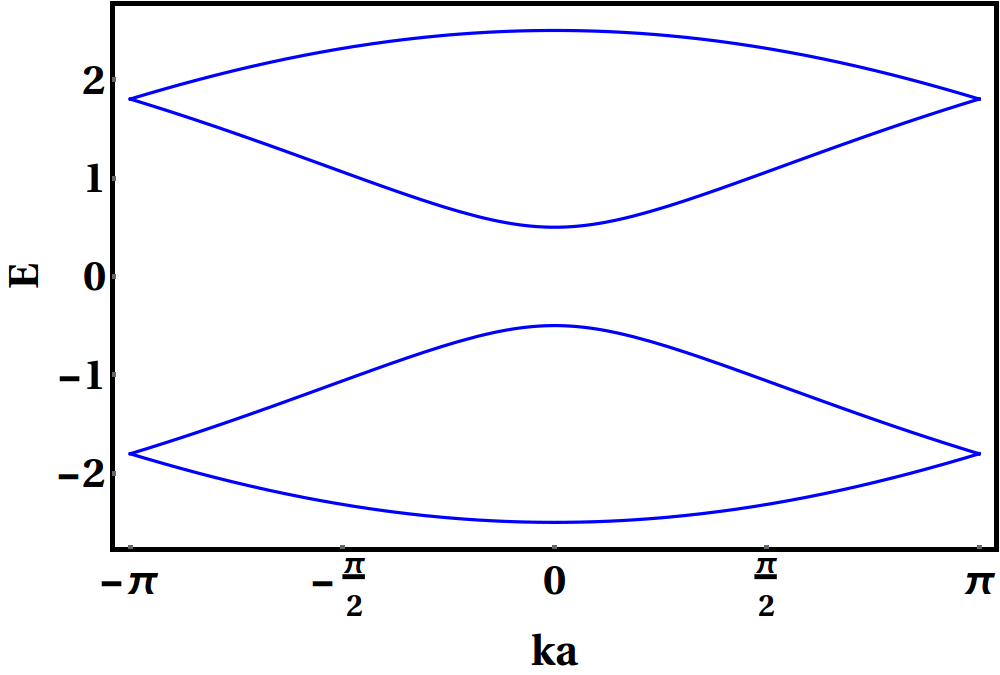}
(c)\includegraphics[width=0.6\columnwidth]{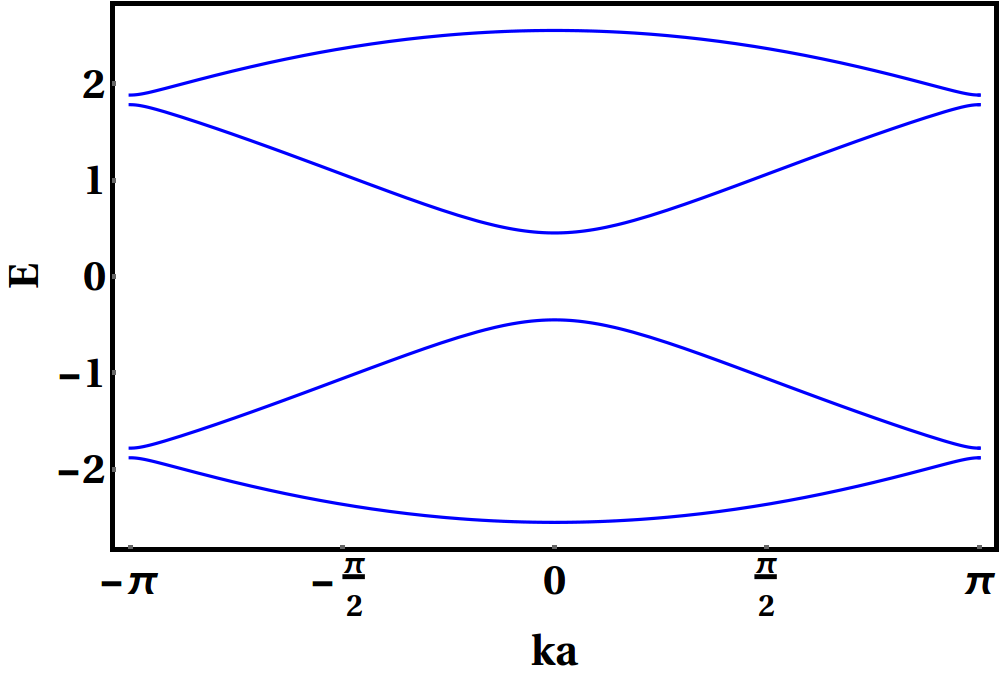}
(d)\includegraphics[width=0.6\columnwidth]{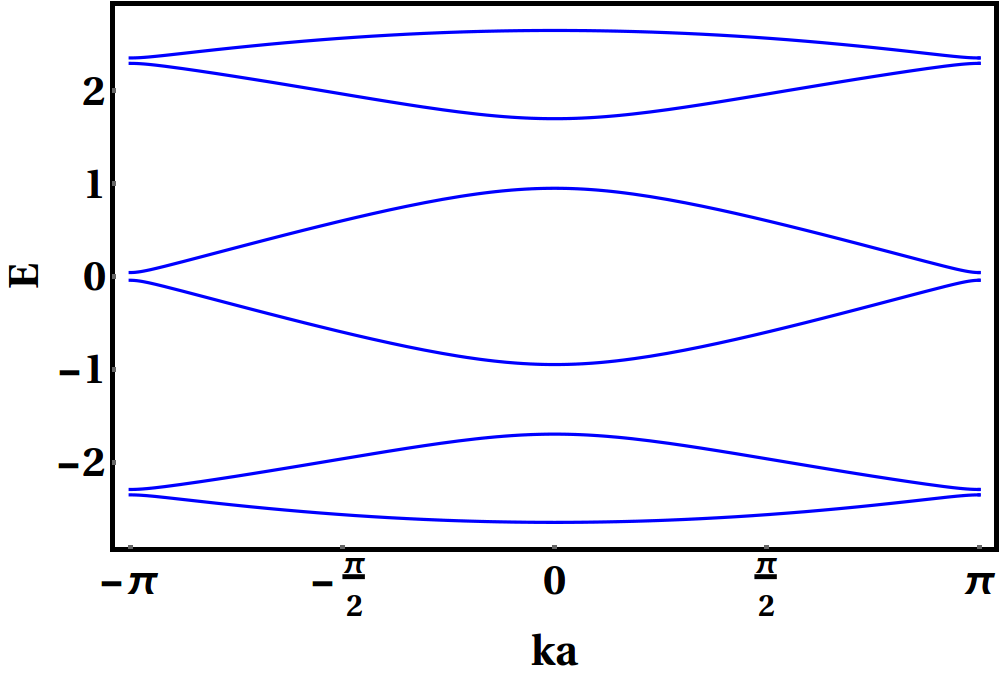}
(e)\includegraphics[width=0.6\columnwidth]{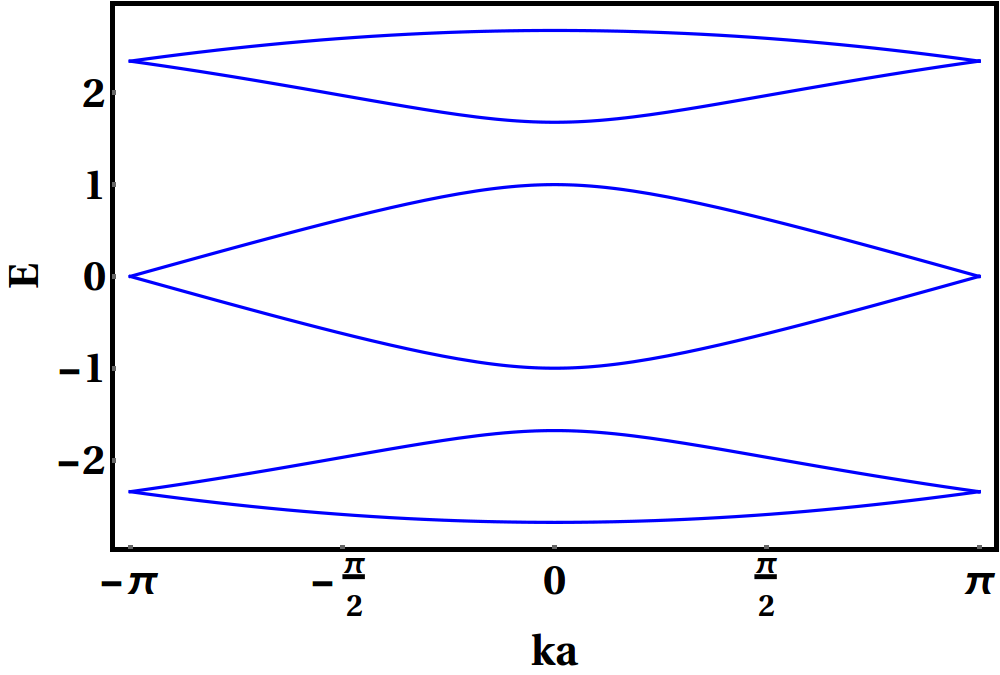}
(f)\includegraphics[width=0.6\columnwidth]{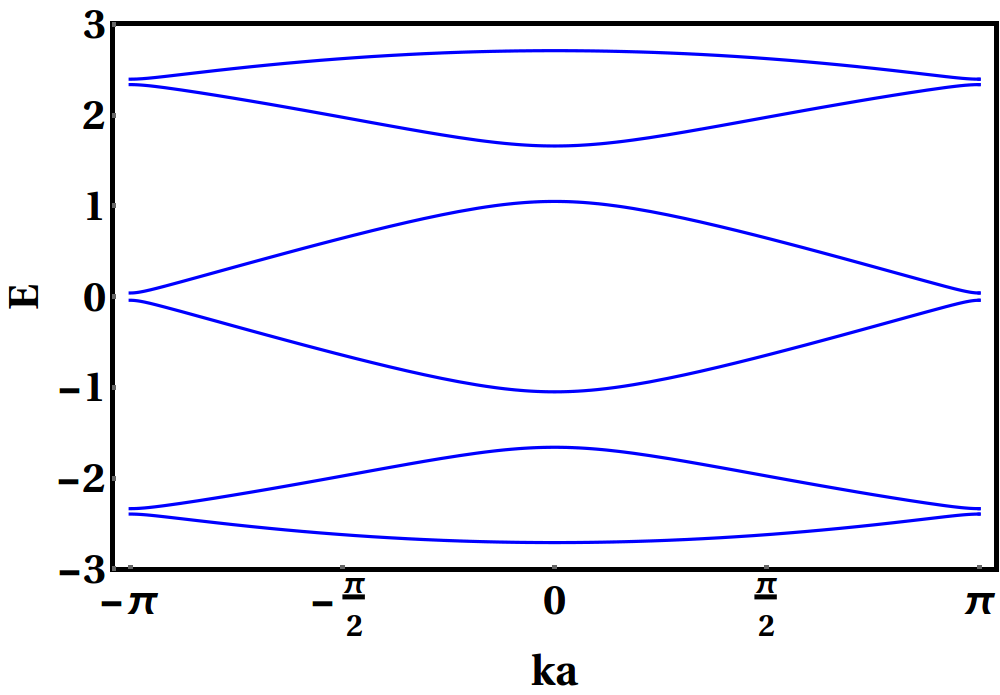}
(g)\includegraphics[width=0.6\columnwidth]{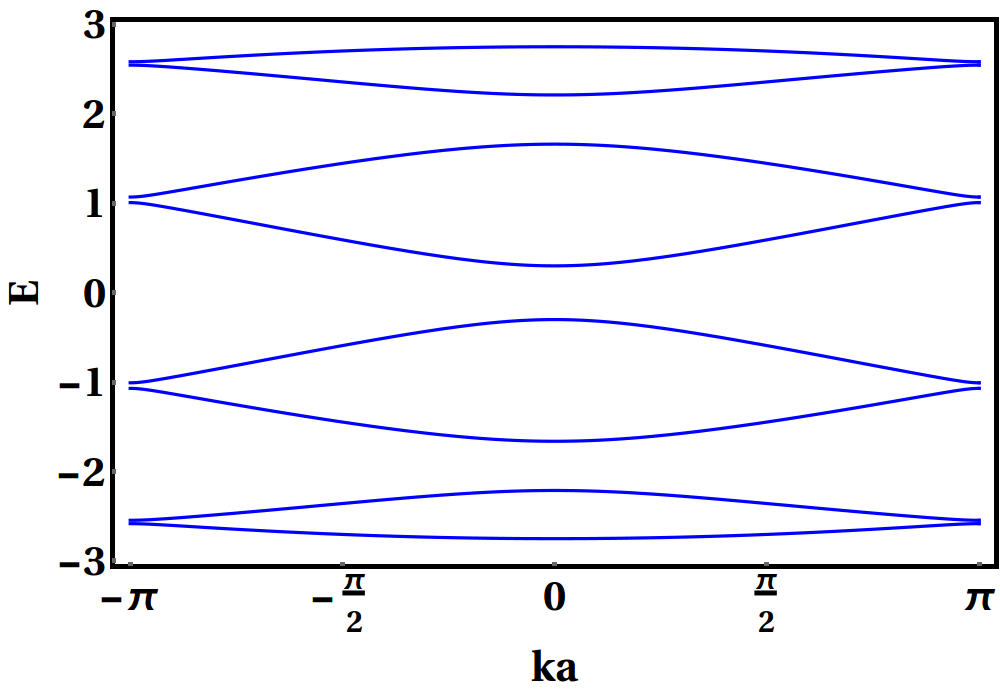}
(h)\includegraphics[width=0.6\columnwidth]{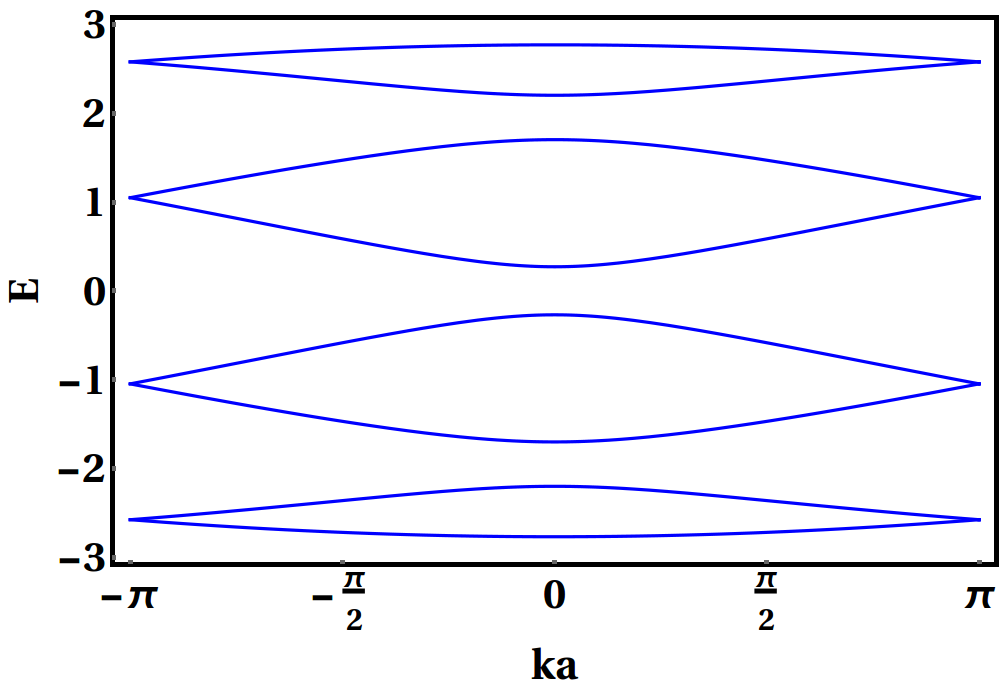}
(i)\includegraphics[width=0.6\columnwidth]{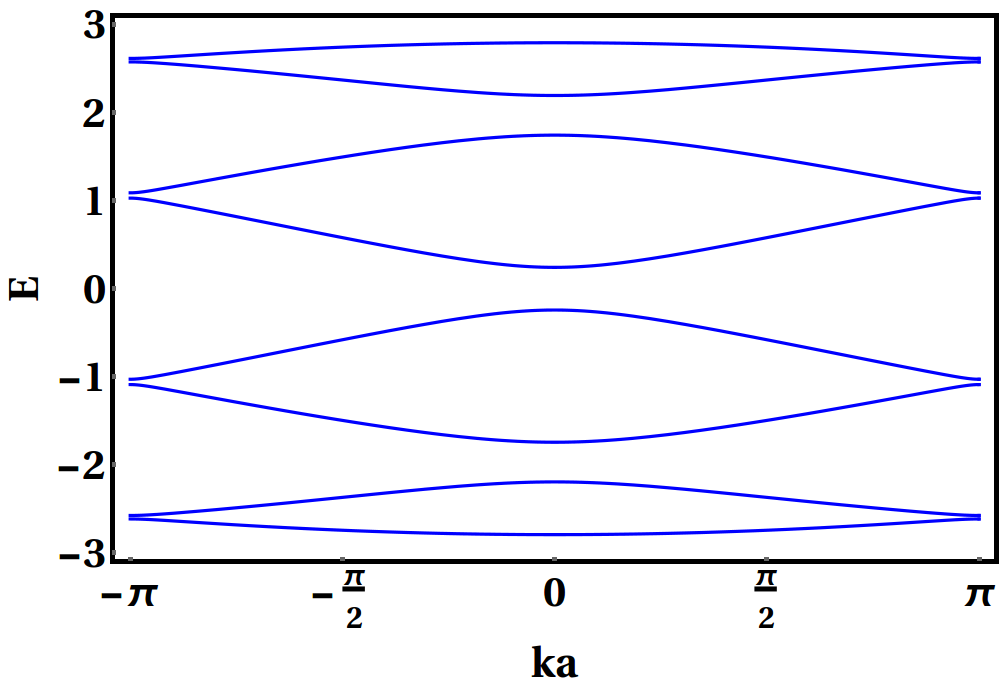}
\caption{(Color online) Energy bands (a,b,c) for GSSHC-I ( Fig.~\ref{fig1}(a)), (d,e,f) for GSSHC-II (Fig.~\ref{fig1}(b)) and (g,h,i) for GSSHC-III (Fig.~\ref{fig1}(c)) respectively.  We have chosen (a,d,g) $t = 1.5$, $\epsilon = 0$, $v = 1$, $w = 0.9$, (b,e,h) $t = 1.5$, $\epsilon = 0$, $v = 1$, $w = 1$ and (c,f,i) $t = 1.5$, $\epsilon = 0$, $v = 1$, $w = 1.1 $. For both the cases $v>w$ and $v<w$ all these variants have a quantized Zak phase for all bands, as will be discussed later.}  
\label{e-k1}
\end{figure*}
  

To work out the dispersion relation in general, for any combination of $v$ and $w$, we first simplify Eqs.~\eqref{diff2} using Eqs.~\eqref{rsrg-1a}, to
\begin{eqnarray}
[(E - \epsilon)^2 - \beta] \psi_{j \subset A} & = & wt~ \psi_{j-1 \subset B} + vt ~\psi_{j+1 \subset B} \nonumber\\\
[(E - \epsilon)^2 - \alpha] \psi_{j \subset B} & = & vt ~\psi_{j-1 \subset A} + wt ~\psi_{j+1 \subset A}
\label{diff3}
\end{eqnarray}
where, $\alpha = (v^2+t^2)$ and $\beta = (w^2+t^2)$.
A further decimation of the alternate sites on this effective chain maps it to a simple 1-d perfectly periodic chain with a new onsite potential $\tilde{\epsilon}$ and a uniform hopping integral $\tilde{t}$, given by,
\begin{eqnarray}
\tilde{\epsilon}  & = & \epsilon_{A} + \frac{{t_A}^2 + {t_B}^2}{E - \epsilon_B} \nonumber \\
\tilde{t}  & = & \frac{t_A t_B}{E - \epsilon_B}
\label{rsrg2-1a}
\end{eqnarray}
It is now simple to work out the dispersion relations for the effective 1-d  chains. The dispersion relation is written as, 
\begin{eqnarray}
 E & = & \tilde{\epsilon} + 2 \tilde{t} \cos~ka'
\label{dispersion-1a}
\end{eqnarray}

Here, `$a'$' represents the effective lattice spacing of the renormalized 1-d periodic chain and $k$ is the wave vector. 
Using Eqs.~\eqref{rsrg-1a}, and Eqs.~\eqref{rsrg2-1a}, the dispersion relation  Eq.~\eqref{dispersion-1a} of  the effective 1-d period chain is expanded to obtain
\begin{equation}
(E - \epsilon)^4 - (E-\epsilon)^2(\beta+\alpha) + \alpha\beta - \gamma t^2 - 2 v w t^2 \cos~ka'  =  0 
   \label{dispersion2-1a}
\end{equation}
where, $\alpha = (v^2+t^2)$, $\beta = (w^2+t^2)$ and $\gamma = (v^2+w^2)$.
All the dispersive bands of such a GSSHC-I type lattice are easily obtained from the solutions of the Eqs.~\eqref{dispersion2-1a}.

We display the dispersive energy bands corresponding to the GSSHC-I chain in Fig.~\ref{e-k1} (a), (b) and (c) for the three cases, $ v > w$, $v=w$, and $v < w$ respectively. The bands are obtained by a direct diagonalization of the matrix written in $k$-space, corresponding to the unit cell. The on-site potential is set at $\epsilon=0$, and the hopping integral is chosen as $t=1.5$. For $v=w=1$, and $t=1.5$ the gaps at the BZ boundaries are found to close exactly at $E = \pm\sqrt{v^2 + t^2} = \pm\sqrt{w^2 + t^2}$, as we expect from our earlier arguments. All the dispersive bands, obtained from an exact diagonalization, matches exactly the results obtained from the decimation scheme, including the magnitudes of the gap-closing energies.

\subsubsection{Type-II}

A similar decimation scheme implemented on the second variety of generalized SSH chain, namely, the GSSHC-II (as shown in  Fig.~\ref{fig1} b(i)), converts it into a conventional looking SSH model (as shown in  Fig.~\ref{fig1} b(ii)) with an energy-dependent onsite potential $\epsilon_A$, $ \epsilon_B $ and a staggered distribution of hopping integrals $t_A$ and $t_B$. These are given by,
\begin{align}
\epsilon_A  =  \epsilon + \frac{w^2+t^2}{\Delta} &&
\epsilon_B  =  \epsilon + \frac{v^2+t^2}{\Delta} \nonumber \\
t_A  =   \frac{ v t^2}{\Delta(E-\epsilon)} &&
t_B  =   \frac{ w t^2 }{\Delta(E-\epsilon)} 
\label{rsrg-1b}    
\end{align}
where, $\Delta = E - \epsilon - \frac{t^2}{(E-\epsilon)}$.


\begin{figure*}[ht]
\centering
(a)\includegraphics[width=0.6\columnwidth]{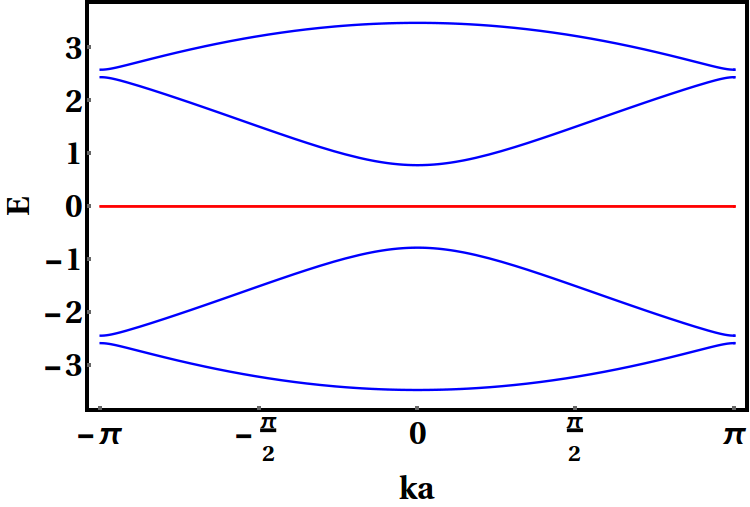}
(b)\includegraphics[width=0.6\columnwidth]{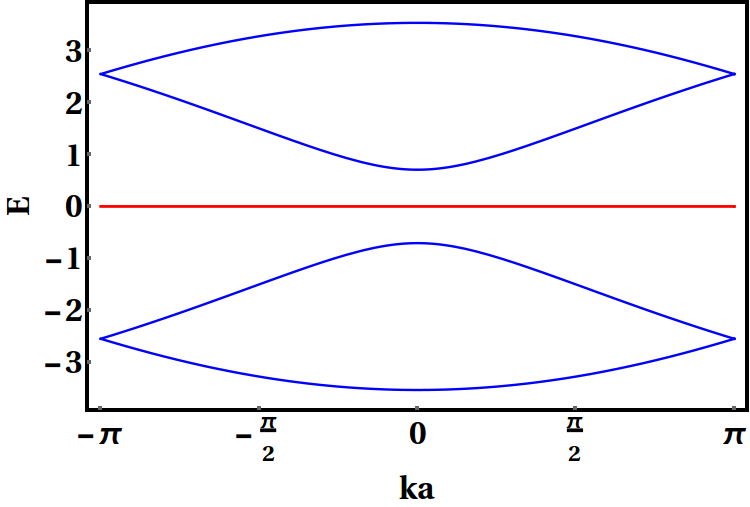}
(c)\includegraphics[width=0.6\columnwidth]{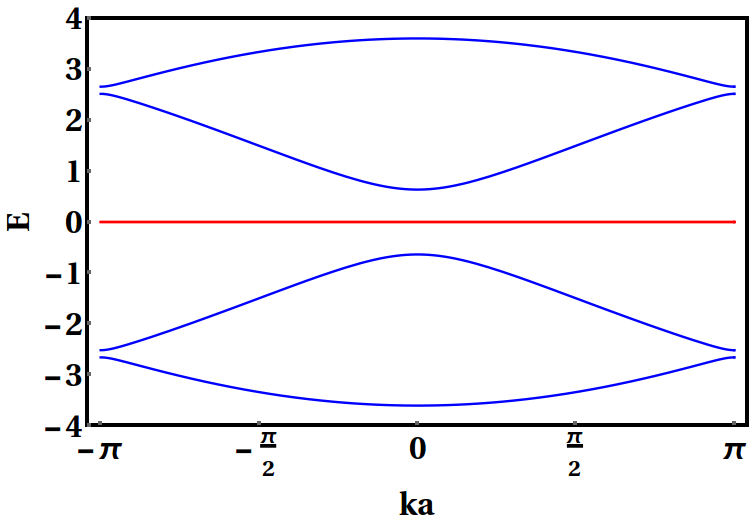}\\
(d)\includegraphics[width=0.6\columnwidth]{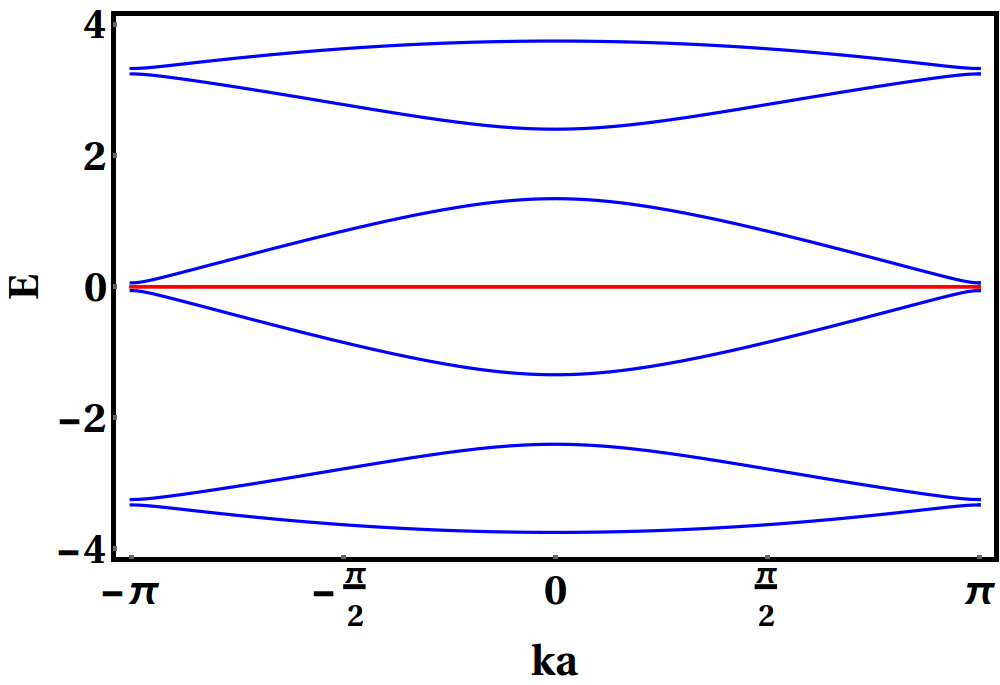}
(e)\includegraphics[width=0.6\columnwidth]{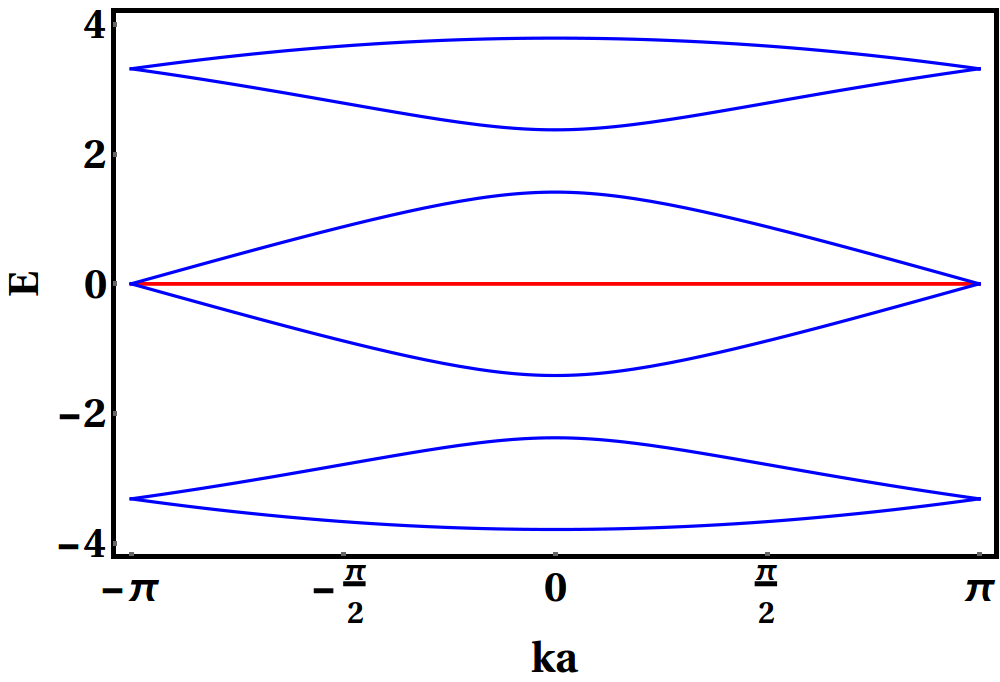}
(f)\includegraphics[width=0.6\columnwidth]{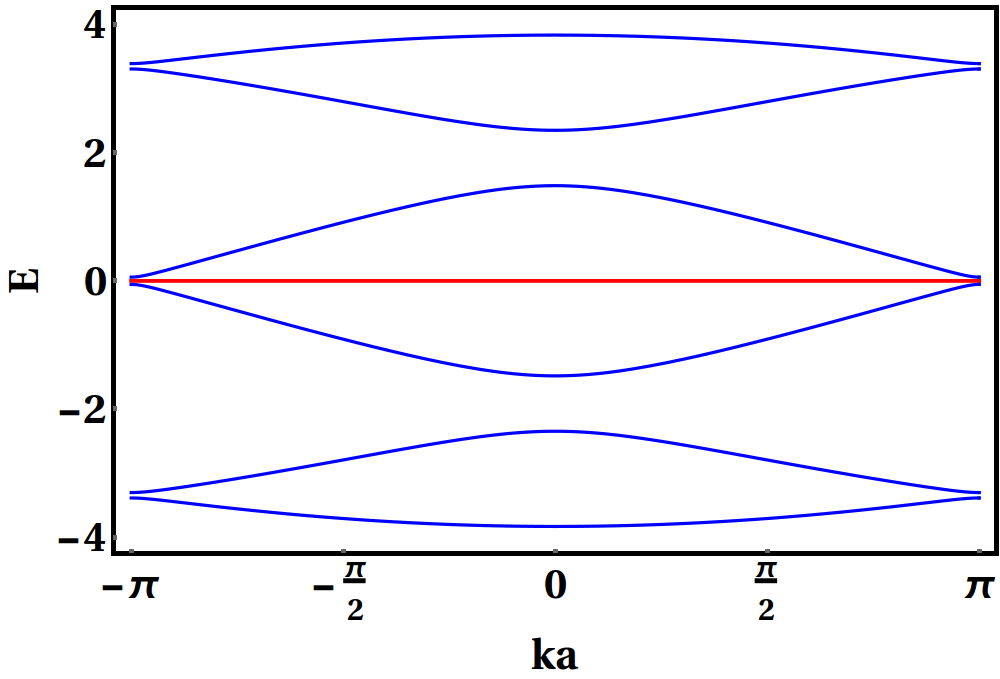}\\
(g)\includegraphics[width=0.6\columnwidth]{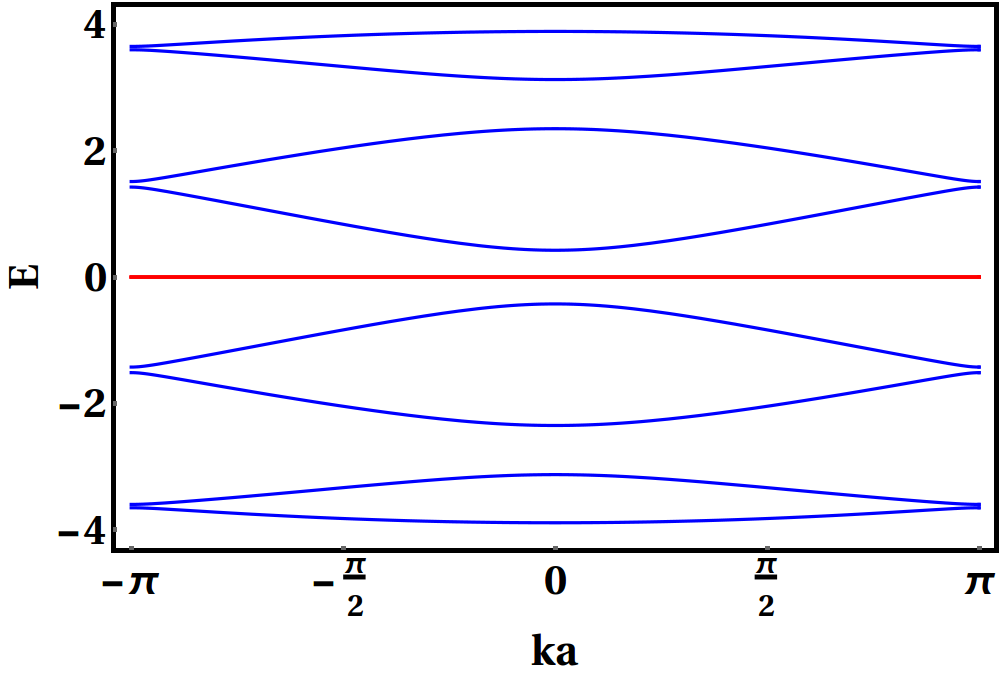}
(h)\includegraphics[width=0.6\columnwidth]{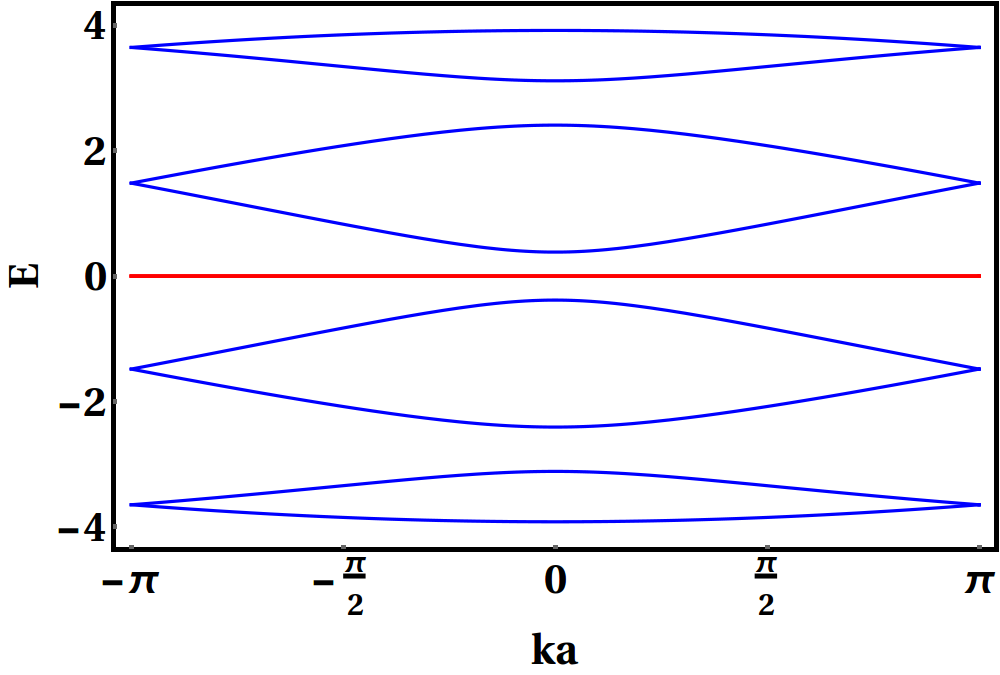}
(i)\includegraphics[width=0.6\columnwidth]{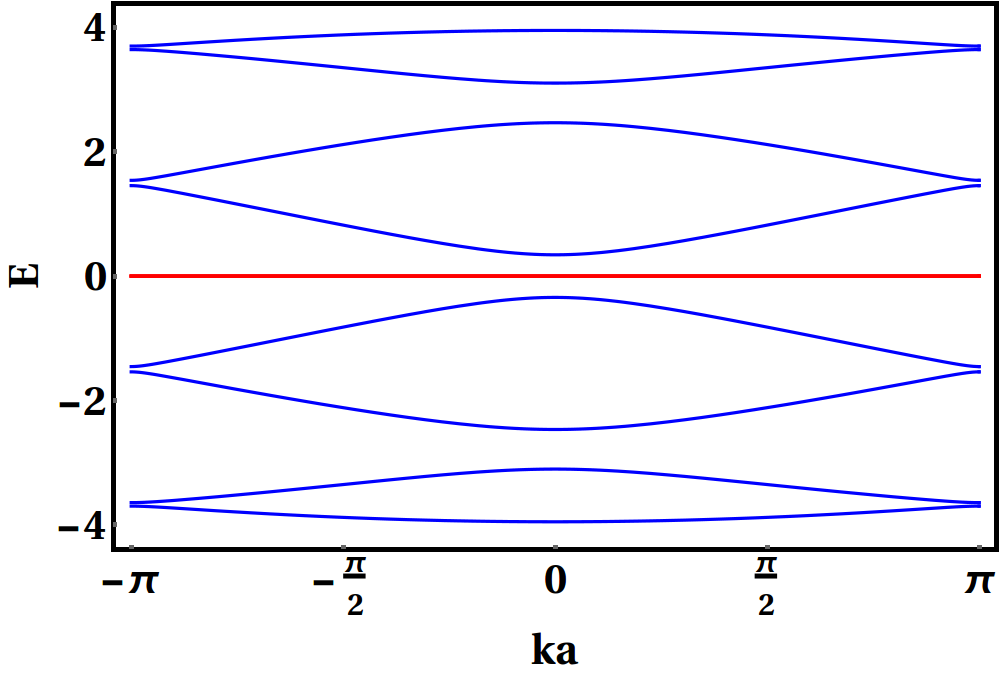}
\caption{(Color online) Energy-wave vector (E vs. ka) dispersion relation (a,b,c) for two cross-linked GSSHC-I (Fig.~\ref{fig2}(a)),  (d,e,f) for two cross-linked GSSHC-II (Fig.~\ref{fig2}(b)) and (g,h,i) for two cross-linked GSSHC-III (Fig.~\ref{fig2}(c)) respectively.   We have chosen in (a,d,g) $t = 1.5$, $\epsilon = 0$, $v = 1$, $w = 0.9$, in (b,e,f) $t = 1.5$, $\epsilon = 0$, $v = 1$, $w = 1$ and in (c,f,i) $t = 1.5$, $\epsilon = 0$, $v = 1$, $w = 1.1 $. The Flat bands are marked by red color which are two-fold degenerate for (a,b,c), three-fold degenerate for (d,e,f), and four-fold degenerate for (g,h,i). For both cases $v>w$ and $v<w$ all these networks have a quantized zak phase for all dispersive bands.}  
\label{e-k2}
\end{figure*}

From  Eqs.~\eqref{rsrg-1b} one can easily confirm, following the procedure laid down in the previous subsection, that the band gap vanishes at the BZ boundaries when $v=w$. The gap-closing energy eigenvalues are obtained as the solutions of the equation $(E-\epsilon) [1 - \frac{v^2+t^2}{(E-\epsilon)^2-t^2}] = 0$, which leads to $E=0$, and $E= \epsilon \pm \sqrt{v^2+2t^2}$ (with $v=w$).


The dispersion relation, following the procedure already discussed, turns out to be,
\begin{widetext}
\begin{equation}
(E - \epsilon)^6 - (E-\epsilon)^4[2t^2+\alpha+\beta] + (E-\epsilon)^2[t^4+t^2\alpha+t^2\beta+\alpha\beta] - \gamma t^4 - 2 v w t^4 \cos~ka' =  0 
   \label{dispersion2-1b}
  \end{equation}
\end{widetext}
where, $\alpha = (v^2+t^2)$, $\beta = (w^2+t^2)$ and $\gamma = (v^2+w^2)$ and provides with all the energy bands in this case.

The dispersive energy bands corresponding to the GSSHC-II, obtained by directly diagonalizing the corresponding Hamiltonian, and exactly matching the bands obtained from Eq.~\eqref{dispersion2-1b} are depicted in Fig.~\ref{e-k1} (d), (e) and (f) for the three cases, $ v > w$, $v=w$, and $v < w$ respectively. The on-site potential and the hopping integrals are chosen to have the same values as the Type-I case, and gap-closing energy matches the results obtained by a direct diagonalization of the Hamiltonian in $k$-space. The results obtained from the decimation scheme match exactly in this case also.
\subsubsection{Type - III}
Using the same process as discussed above, the third variant of a GSSHC, namely, GSSHC-III (Fig.~\ref{fig1} c(i)) is treated. Fig.~\ref{fig1} c(ii)) shows the traditional SSH structure with the on-site potentials and the nearest neighbor hopping integrals, all functions of energy $E$,  now given by,
\begin{align}
\epsilon_A  =  \epsilon + \frac{w^2+t^2}{\delta} &&
\epsilon_B  =  \epsilon + \frac{v^2+t^2}{\delta} \nonumber \\
t_A  =   \frac{ v t^3}{\delta\Delta(E-\epsilon)} &&
t_B  =   \frac{ w t^3}{\delta\Delta(E-\epsilon)} 
\label{rsrg-1c}
\end{align}
where, $\Delta = E-\epsilon-\frac{t^2}{(E-\epsilon)}$ and $\delta = E-\epsilon-\frac{t^2}{\Delta}$. The gap-closing energies are obtained as the zeros of the equation $E-\epsilon_{A(B)} = 0$, which, on using Eqs.~\eqref{rsrg-1c}, becomes
\begin{equation} 
(E-\epsilon)^4 - (E-\epsilon)^2(3t^2+v^2) + t^2(v^2+t^2) = 0
\end{equation}
and, by seting $t_A=t_B$ simultaneously. The last condition is actually equivalent to making $v$ and $w$ equal. The solutions of these equations are,  
    \begin{equation}
        E  =  \epsilon\pm\frac{\sqrt{3t^2+v^2\pm\sqrt{5t^4+2t^2v^2+v^4}}}{\sqrt{2}} 
                \label{sol-1c}
    \end{equation}
From these solutions as shown in Eqs.~\eqref{sol-1c} with any values of parameters, one is able to calculate all energies at which gaps are closed at the BZ boundaries.


The dispersion relation can be worked out as before, through an implementation of the decimation scheme that makes this third variant completely equivalent to a simple SSH chain, and reads,
\begin{widetext}
\begin{equation}
(E - \epsilon)^8 -(E-\epsilon)^6[4t^2+\alpha+\beta]+(E-\epsilon)^4[4t^4+3t^2(\alpha+\beta)+\alpha\beta]-(E-\epsilon)^2[2t^4(\alpha+\beta)+2t^2\alpha\beta]+t^4\alpha\beta- \gamma t^6 -2vwt^6\cos~ka =  0 
   \label{dispersion2-1c}
 \end{equation}
\end{widetext}
where $\alpha = v^2+t^2$, $\beta = w^2+t^2$ and $\gamma = v^2+w^2$. 
All the eight dispersive bands, obtained by an exact diagonalization are found to be identical to the solutions of the Eqs.~\eqref{dispersion2-1c}.\\

\begin{figure}[ht]
\centering
(a)\includegraphics[width=0.43\columnwidth]{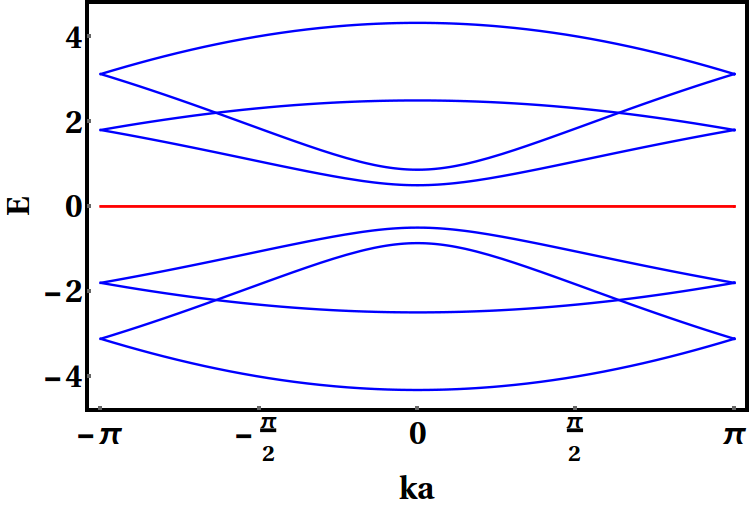}
(b)\includegraphics[width=0.43\columnwidth]{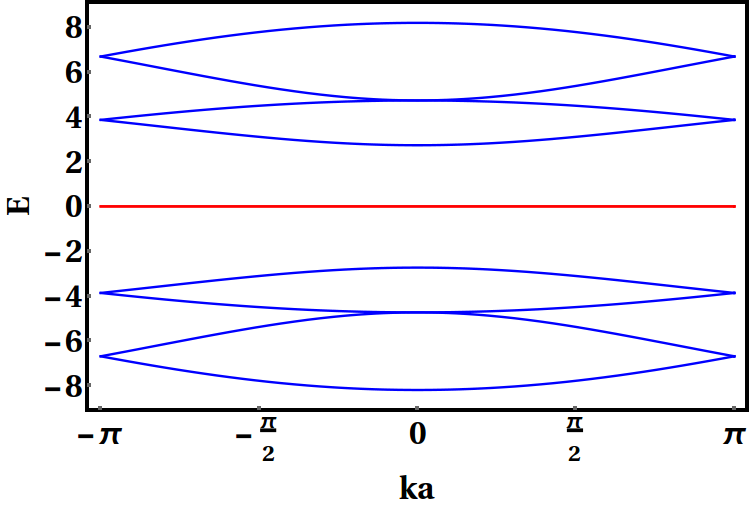}
(c)\includegraphics[width=0.43\columnwidth]{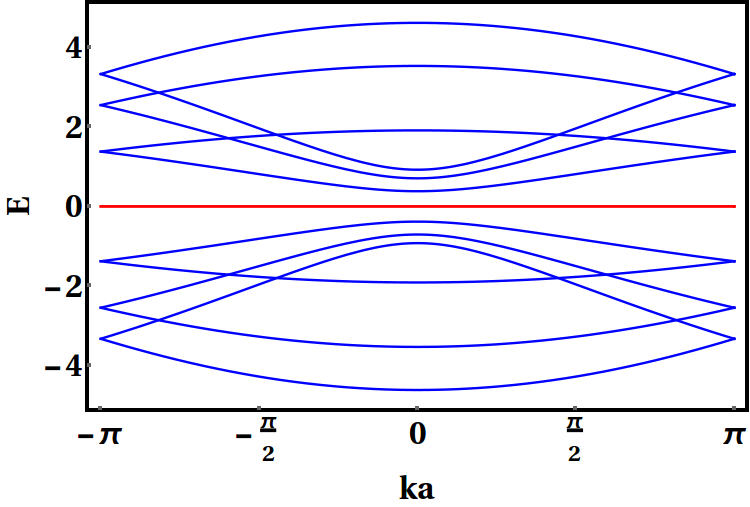}
(d)\includegraphics[width=0.43\columnwidth]{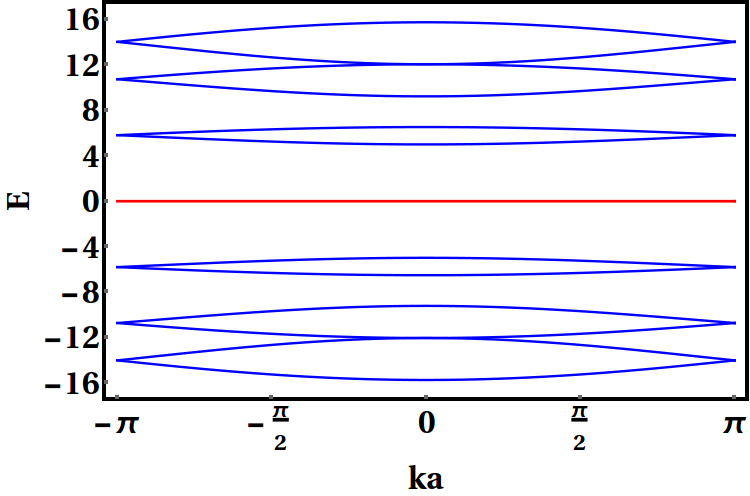}

\caption{(Color online) Energy-wave vector (E vs. ka) dispersion relation (a,b) four cross-linked GSSHC-I ( Fig.~\ref{fig3}(a)) and  (c,d) six cross-linked GSSHC-I( Fig.~\ref{fig3}(b)) respectively.  We have chosen (a,c) $t = 1.5$, $\epsilon = 0$, $v = 1$, $w = 1$, 
 (b) $t = 3.73205 $, $\epsilon = 0$, $v = 1$, $w = 1$ and (d) $t = 7.52395 $, $\epsilon = 0$, $v = 1$, $w = 1$. The magnified version of (b) and (d) are shown in Appendix~\ref{appendixC}  (Fig.~\ref{mag-e-k3}). }  
\label{e-k3}
\end{figure}


\section{A generalized SSH mesh: the flat and the dispersive bands}
\subsection{Decimation and the commuting matrices}

In this section we explicitly discuss the decimation method using an example of just two geometrically entangled GSSHC-I systems (as shown in Fig.~\ref{fig2} a(i)).  The extension to wider mesh geometries is in the same line, but obviously with more intricate matrix structures. These details are laid out in the appendix~\ref{appendixA}.

The two cross-linked chain system in the GSSHC-I family is first converted into a two-strand ladder network (as shown in Fig.~\ref{fig2} a(ii)) by decimating out the vertices having a coordination number of four (green sites). Now we have a two-strand ladder with a two-sublattice structure ($A$ and $B$). The onsite potentials are  $\epsilon_A$ and $ \epsilon_B$. The decimation generates intra-strand nearest neighbor hopping integrals $t_A$ and $ t_B$ and the inter-strand hopping integrals $\Gamma_A$, $\Gamma_B$, shown by the red vertical lines in Fig.~\ref{fig2} a(i). There will also be a second-neighbor (diagonal) hopping (black dotted lines) of the same magnitudes as $t_A$ and $ t_B$, and the equality of the diagonal hopping integrals to the nearest neighbor intra-strand hopping are going to unravel the flat, non-dispersive energy bands, as we shall see shortly.

The conversion of the cross-linked GSSHC-I systems into a two-strand ladder generates the following energy-dependent quantities:
\begin{align}
\epsilon_A  =  \epsilon + \frac{t^2 + v^2}{E - \epsilon}, &&
\epsilon_B  =  \epsilon + \frac{t^2 + w^2}{E - \epsilon} \nonumber\\
\Gamma_A  =  \frac{t^2 + v^2}{E - \epsilon}, &&
\Gamma_B  =  \frac{t^2 + w^2}{E - \epsilon}\nonumber\\
t_A  =  \frac{v t}{E - \epsilon}, &&
t_B  =  \frac{w t}{E - \epsilon} 
 \label{rsrg-2a}
\end{align}


The corresponding difference equations for the two-sublattice, two-strand ladder network are written conveniently in matrix form, viz, 

\begin{eqnarray}
    [E.\mathbb{I}_{2\cross2} -\tilde{\epsilon}_{A}] \Psi_{A,n} & = & \tilde{t}_A \Psi_{B,n+1} + \tilde{t}_B \Psi_{B,n-1} \nonumber\\\
    [E.\mathbb{I}_{2\cross2} -\tilde{\epsilon}_{B}] \Psi_{B,n} & = & \tilde{t}_B \Psi_{A,n+1} + \tilde{t}_A \Psi_{A,n-1} 
  \label{eq1}
\end{eqnarray}

\begin{figure*}[ht]
\centering
(a)\includegraphics[width=0.6\columnwidth]{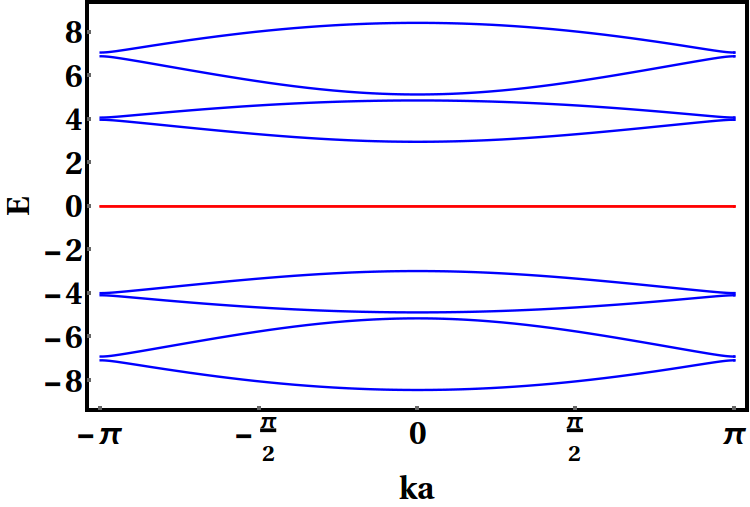}
(b)\includegraphics[width=0.6\columnwidth]{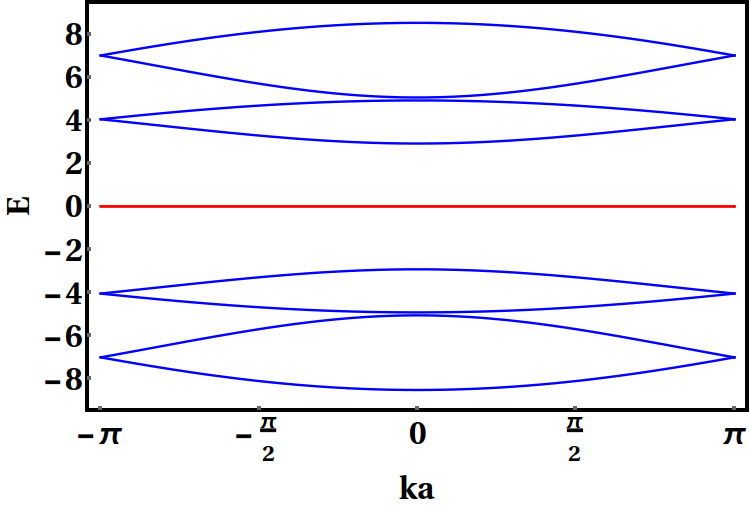}
(c)\includegraphics[width=0.6\columnwidth]{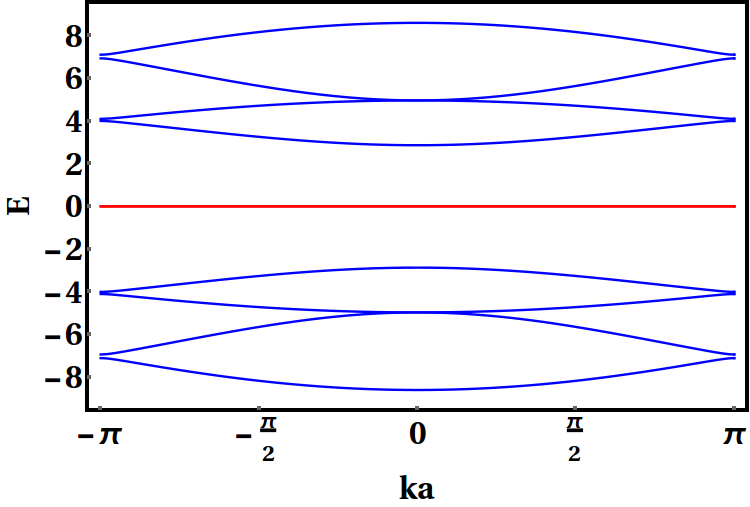}\\
(d)\includegraphics[width=0.6\columnwidth]{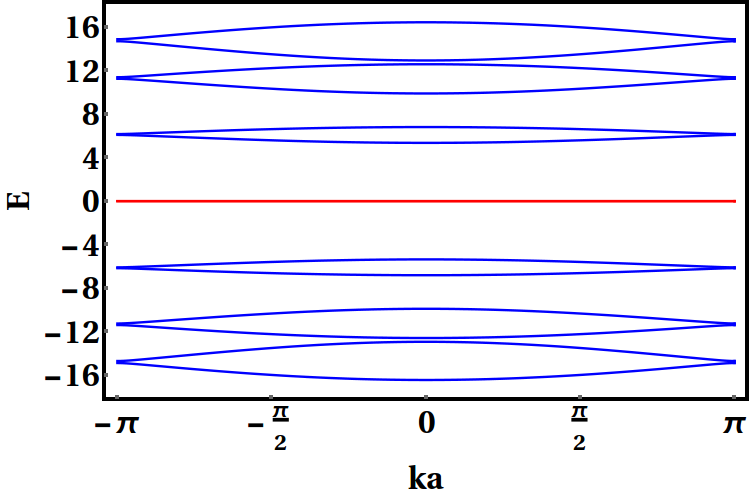}
(e)\includegraphics[width=0.6\columnwidth]{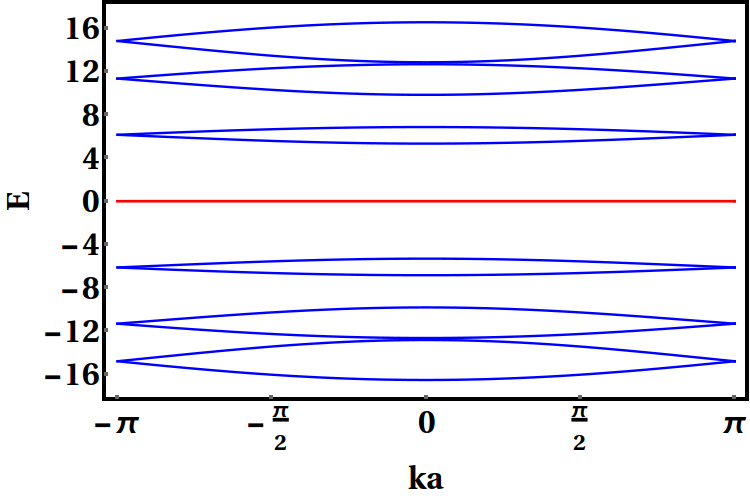}
(f)\includegraphics[width=0.6\columnwidth]{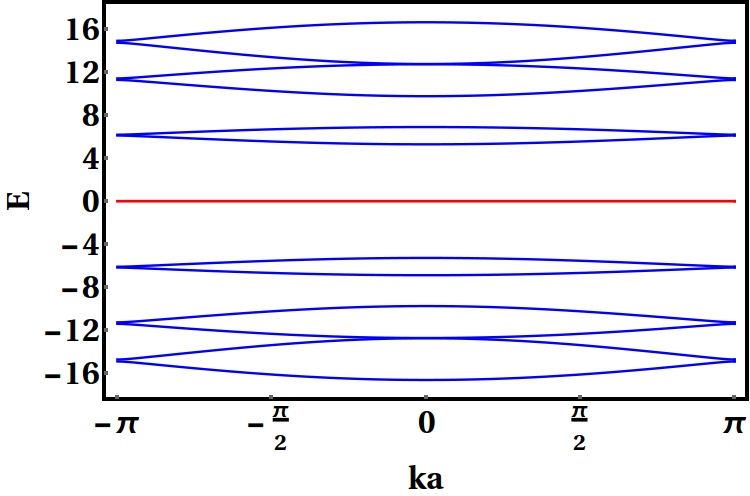}\\

\caption{(Color online) Energy-wave vector (E vs. ka) dispersion relation (a,b,c) for cross-linking of four,  (d,e,f) for cross-linking of six GSSHC-I respectively.  We have chosen (a) $t = 3.92$, $\epsilon = 0$, $v = 1$, $w = 0.9$, (b) $t = 3.92$, $\epsilon = 0$, $v = 1$, $w = 1$,  (c) $t = 3.92$, $\epsilon = 0$, $v = 1$, $w = 1.1$, (d) $t = 7.95$, $\epsilon = 0$, $v = 1$, $w = 0.9$, (e) $t = 7.95$, $\epsilon = 0$, $v = 1$, $w = 1$ and  (f) $t = 7.95$, $\epsilon = 0$, $v = 1$, $w = 1.1$,. The Flat bands (doubly degenerate) are marked by red color. For both cases $v>w$ and $v<w$ all these networks have a quantized zak phase for all dispersive bands. The magnified versions of all these diagrams are available in Appendix~\ref{appendixC}  Fig.~\ref{zoom}.}  
\label{e-k4}
\end{figure*}

where,
\begin{equation}
\Psi_{A(B),n} = \left[ \begin{array}{cccccccccccccccc}
\psi_{A(B),n,1}\\
 \psi_{A(B),n,2}
\end{array}
\right ] 
\end{equation}
 and,
 \begin{equation}
\tilde{\epsilon}_{A(B)} = \left[ \begin{array}{cccccccccccccccc}
\epsilon + \Gamma_{A(B)} & \Gamma_{A(B)}\\
 \Gamma_{A(B)} & \epsilon + \Gamma_{A(B)} 
\end{array}
\right ] 
\end{equation}

\begin{eqnarray}
 \tilde{t}_{A(B)} & = & t_{A(B)} \mathbf{\Lambda}_{2\times2} 
\end{eqnarray}

with $\mathbf{\Lambda}_{2 \times 2}^{ij}= 1 $ for all $(i,j)$.

It is easily verified that, all combinations of the `potential matrix' and the `hopping matrix' commute {\it independent of  energy} $E$. That is, $ [\tilde{\epsilon}_{(A,B)} , \tilde{t}_{(A,B)}]  =  0$ irrespective of energy $E$. These matrices can thus be diagonalized simultaneously using the same matrix, $\mathcal{M}$ say, and the difference equations written in the matrix form above are easily written down in a new basis defined by $\Phi_{A(B),n}=\mathcal{M}^{-1}$ $\Psi_{A(B),n}$. The equations for the $A$ and the $B$ sub-lattices are now completely decoupled in the new basis~\cite{sil}, and read 

\begin{eqnarray}
(E - \epsilon)  \phi_{A,n,1} & = & 0 \nonumber \\
\left (E - \epsilon - 2 \Gamma_A \right ) \phi_{A,n,2} & = & 2 t_B \phi_{B,n-1,2}+ 2 t_A \phi_{B,n+1,2}\nonumber\\
(E - \epsilon)  \phi_{B,n,1} & = & 0  \nonumber \\
\left (E - \epsilon - 2 \Gamma_B \right ) \phi_{B,n,2} & = & 2 t_A \phi_{A,n-1,2}+ 2 t_B \phi_{A,n+1,2}\nonumber\\
\label{sub-AB}
\end{eqnarray}
From the first and the third of the set of Eqs.~\eqref{sub-AB} one can identify a two-fold degenerate, sharply localized eigenstate at $E=\epsilon$. This energy corresponds to a flat, non-dispersive band in the $E-k$ spectrum, and is clearly seen in Fig.~\ref{e-k2} (a), (b) and (c) for two cross-linked GSSHC-I, obtained through a direct diagonalization of the Hamiltonian. The energy bands are closed at the  Brillouin zone boundary for $ v = w$ with the gap closing energy $ E = \epsilon\pm\sqrt{2(t^2+v^2)} = \epsilon\pm\sqrt{2(t^2+w^2)}$, as can be worked out easily from Eqs.~\eqref{sub-AB} on setting $t_A=t_B$ (which gives $v=w$), and $E-\epsilon-2\Gamma_{A(B)}=0$. 

Solutions of the second and the fourth equations in Eq.~\eqref{sub-AB} yield the dispersion relation which, on simplification, reads,



\begin{equation}
(E - \epsilon)^4 - 2 (E-\epsilon)^2(\alpha + \beta) + 4 \alpha \beta - 4 \gamma t^2 - 8 v w t^2 \cos~ka'  =  0 
   \label{dispersion2-2a}
\end{equation}
where $\alpha = v^2+t^2$, $\beta = w^2+t^2$ and $\gamma = v^2+w^2$. 
All the dispersive bands of such a network are easily identified by the solutions of the Eqs.~\eqref{dispersion2-2a}.

A similar real space renormalization group (RSRG) decimation scheme is applied to each type of two cross-linked GSSHC-II and III (as shown in Fig.~\ref{fig2}(b),(c)) and corresponding dispersion relations, gap closing energy, the degeneracy of flat bands are calculated analytically. All details are in Appendix~\ref{appendixA}.  

\begin{figure}[ht]
\centering
(a)\includegraphics[width=0.43\columnwidth]{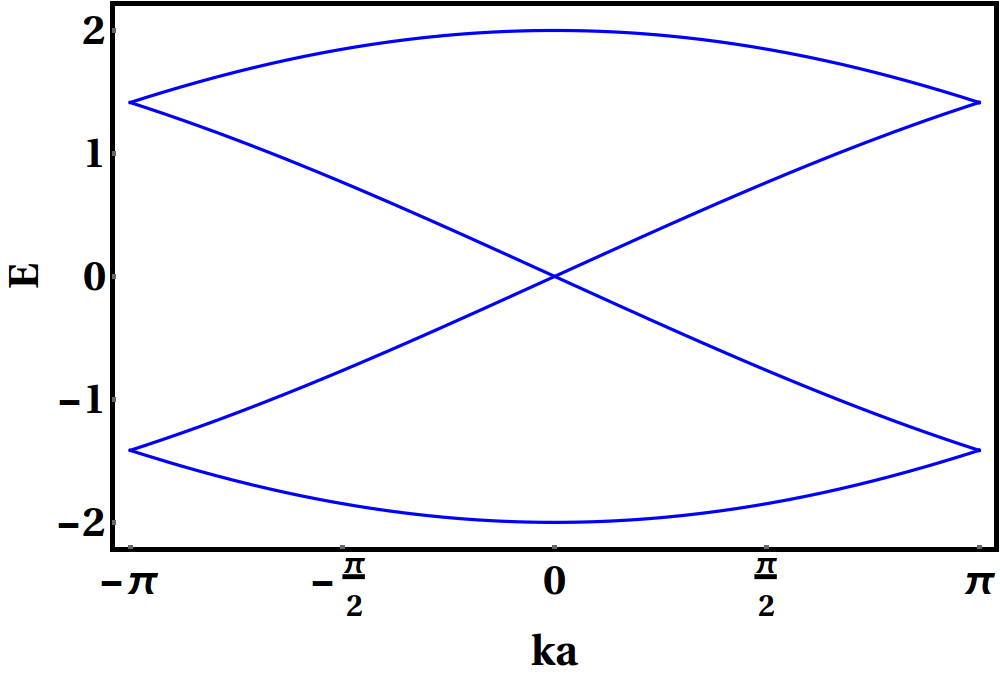}
(b)\includegraphics[width=0.43\columnwidth]{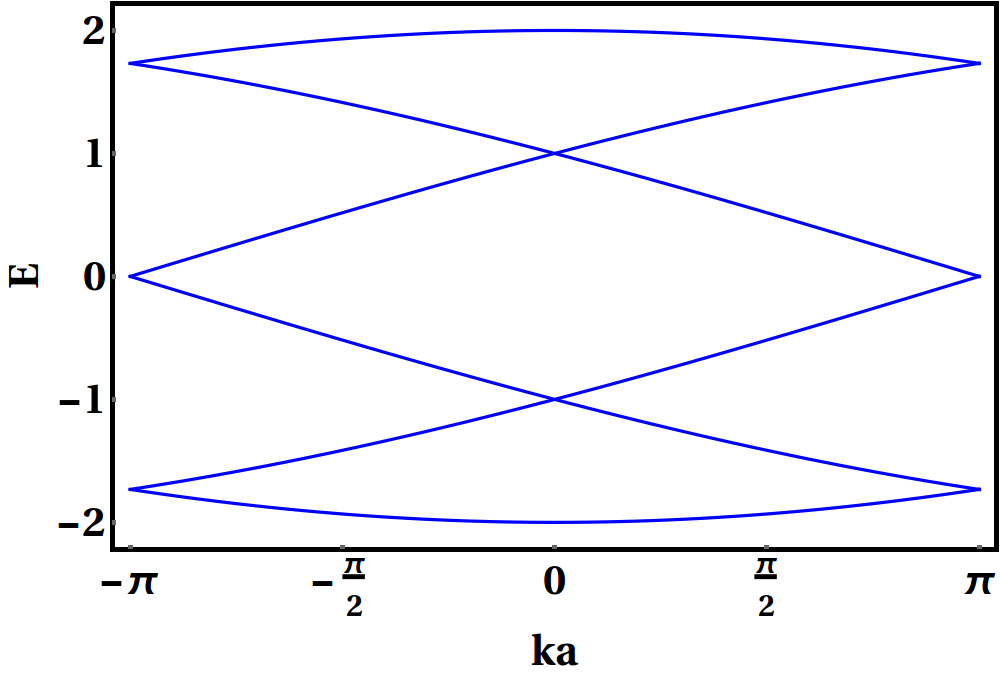}
(c)\includegraphics[width=0.43\columnwidth]{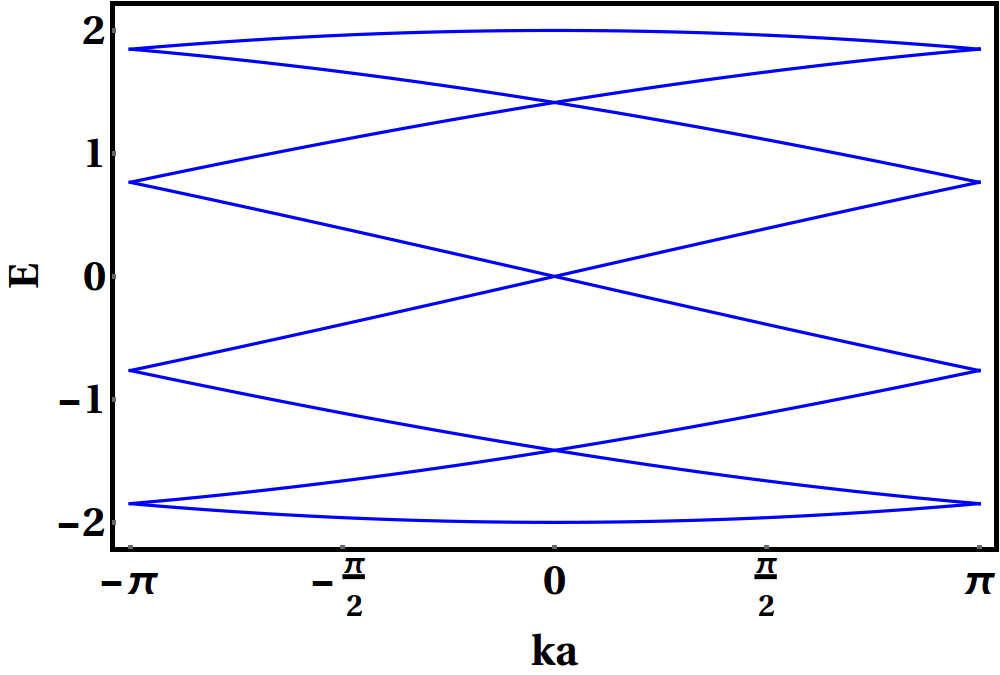}
(d)\includegraphics[width=0.43\columnwidth]{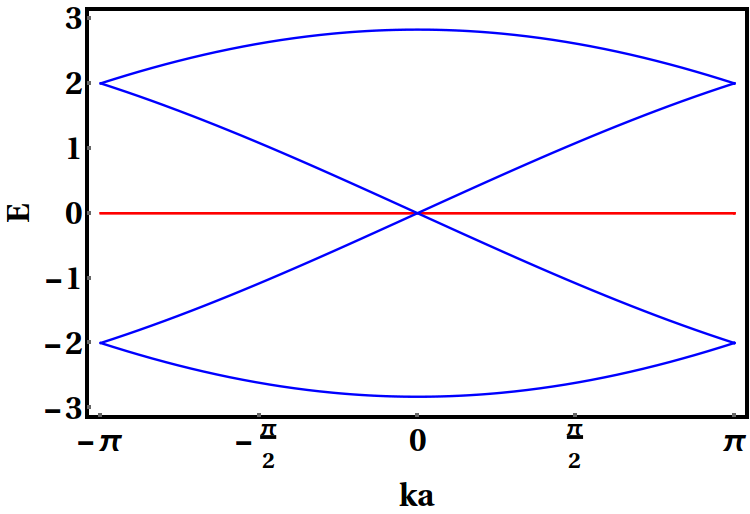}
(e)\includegraphics[width=0.43\columnwidth]{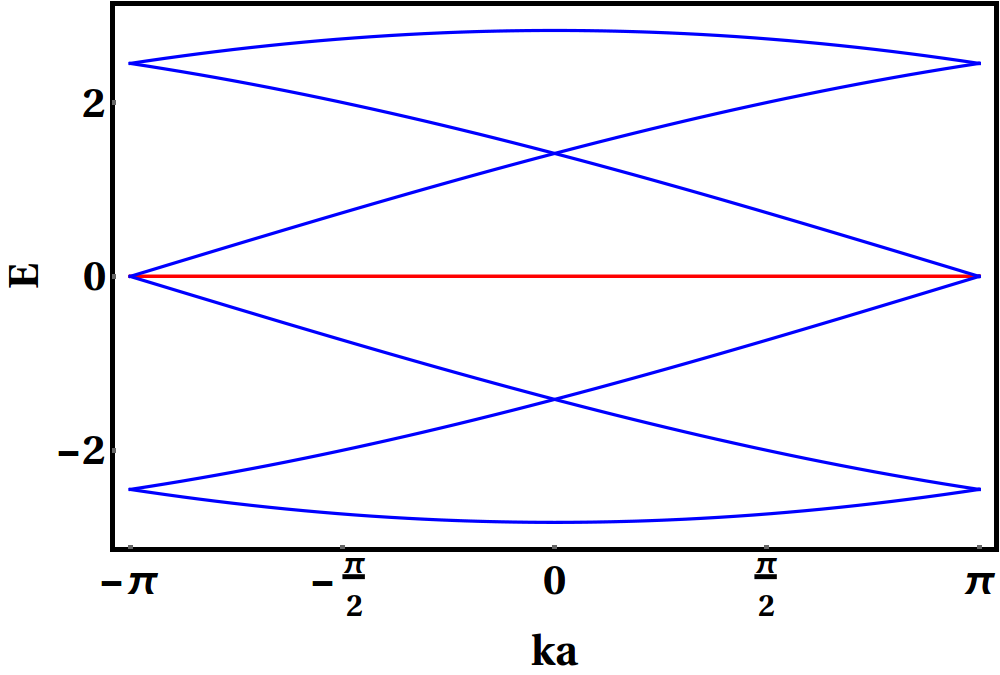}
(f)\includegraphics[width=0.43\columnwidth]{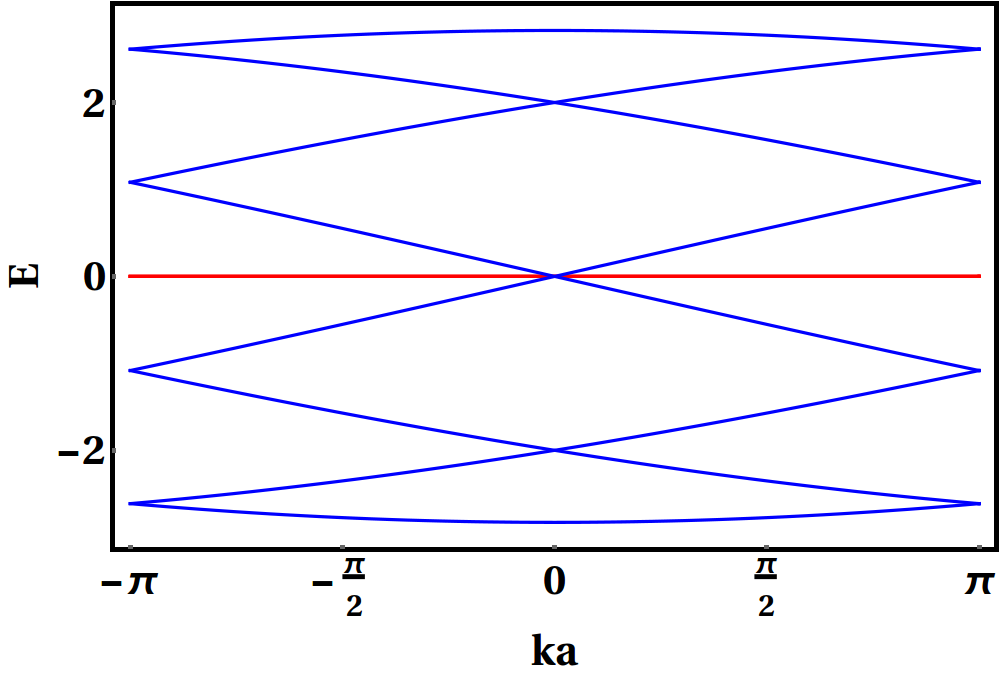}
(g)\includegraphics[width=0.43\columnwidth]{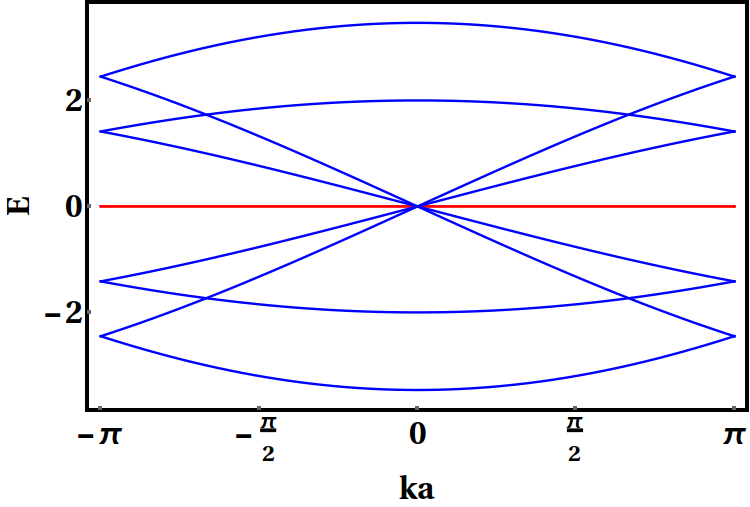}
(h)\includegraphics[width=0.43\columnwidth]{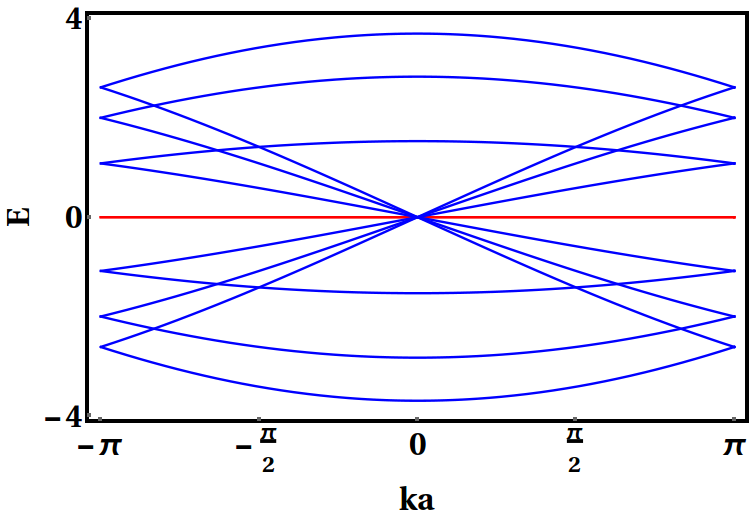}
\caption{(Color online) Energy-wave vector (E vs. ka) dispersion relation (a, b, c) GSSHC-I, GSSHC-II, GSSHC-III, and (d, e, f) for cross-linking of two  GSSHC-I, GSSHC-II, GSSHC-III, respectively. (g, h) for cross-linking of four and six  GSSHC-I. The parameters are chosen as $t = v = w = 1$, $ \epsilon = 0 $.  }  
\label{dirac}
\end{figure}
\subsection{Results from a direct diagonalization of the Hamiltonian}

We have gone for an exact diagonalization of the Hamiltonian corresponding to every example cited in this work, and the dispersive and the flat bands thus obtained, are compared with the results obtained out of the decimation procedure. The match is exact.

The kernels of the Hamiltonian ($ \hat{\mathcal{H}}_{1}(k)$, $ \hat{\mathcal{H}}_{2}(k)$, $ \hat{\mathcal{H}}_{3}(k)$) for the unit cell of the  GSSHC-I, GSSHC-II, GSSHC-III (as shown in  Fig.~\ref{fig1}) are given by in Eqs.~\eqref{ham-1a}, Eqs.~\eqref{ham-1b}, and Eqs.~\eqref{ham-1c} respectively.
Diagonalization of the above three Hamiltonian gives all the eigenvalues as a function of $k$ and corresponding dispersion relations (E .vs. ka) are plotted in Fig.~\ref{e-k1}((a)-(i)).  
Similarly, the kernels of the Hamiltonian ($ \hat{\mathcal{H}}_{4}(k)$, $ \hat{\mathcal{H}}_{5}(k)$, $ \hat{\mathcal{H}}_{6}(k)$) for the unit cell of two cross-linked GSSHC-I, GSSHC-II, GSSHC-III (as shown in  Fig.~\ref{fig2}) are shown in Eqs.~\eqref{ham-2a}, Eqs.~\eqref{ham-2b}, and Eqs.~\eqref{ham-2c} respectively.
\begin{figure}[ht]
\centering
(a)\includegraphics[width=.43\columnwidth]{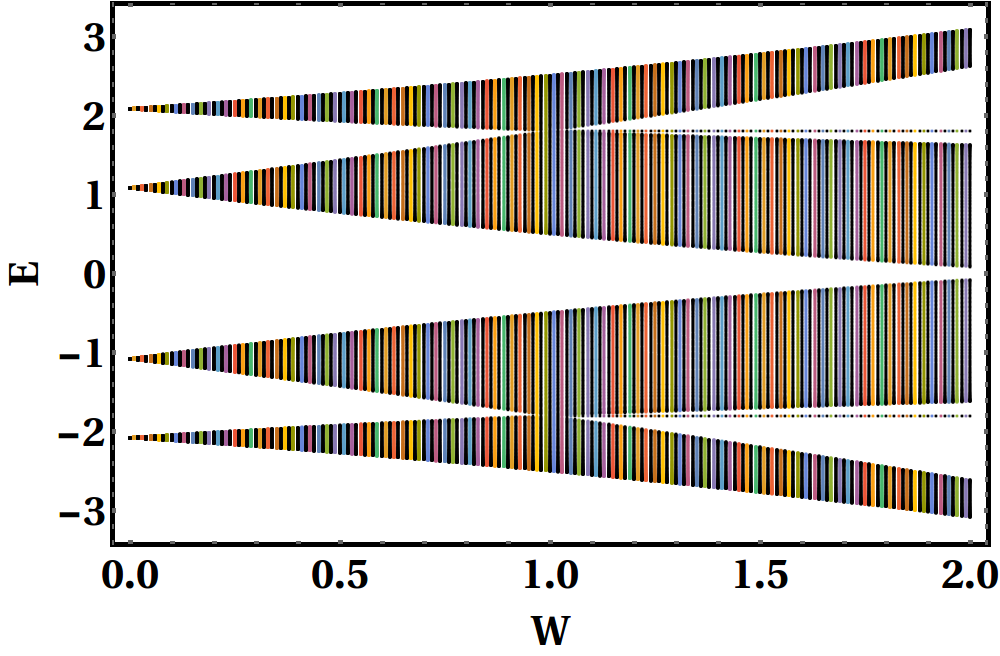}
(b)\includegraphics[width=.43\columnwidth]{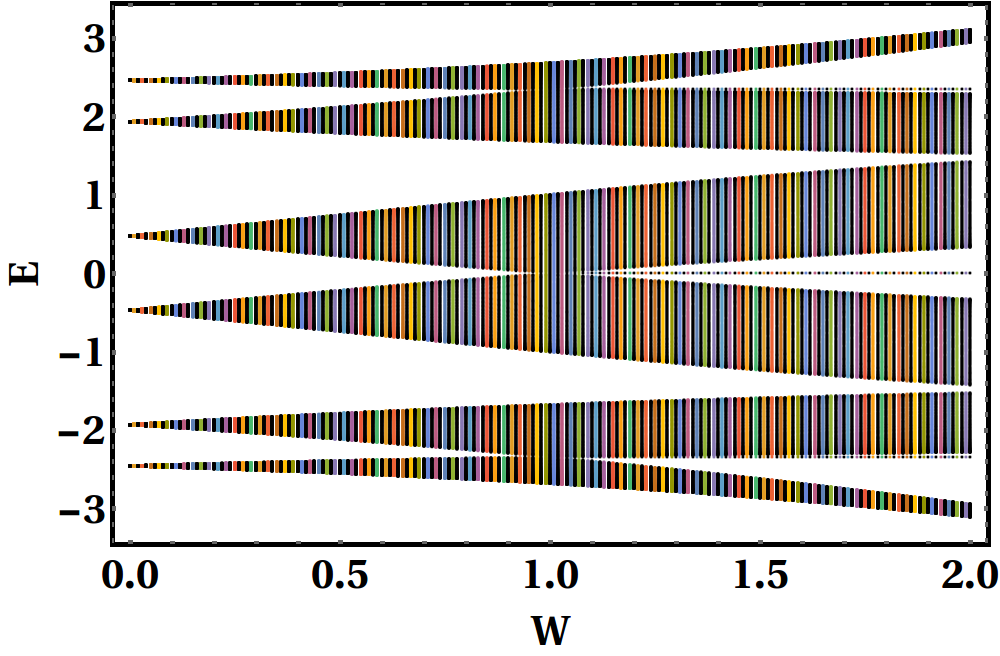}
(c)\includegraphics[width=.43\columnwidth]{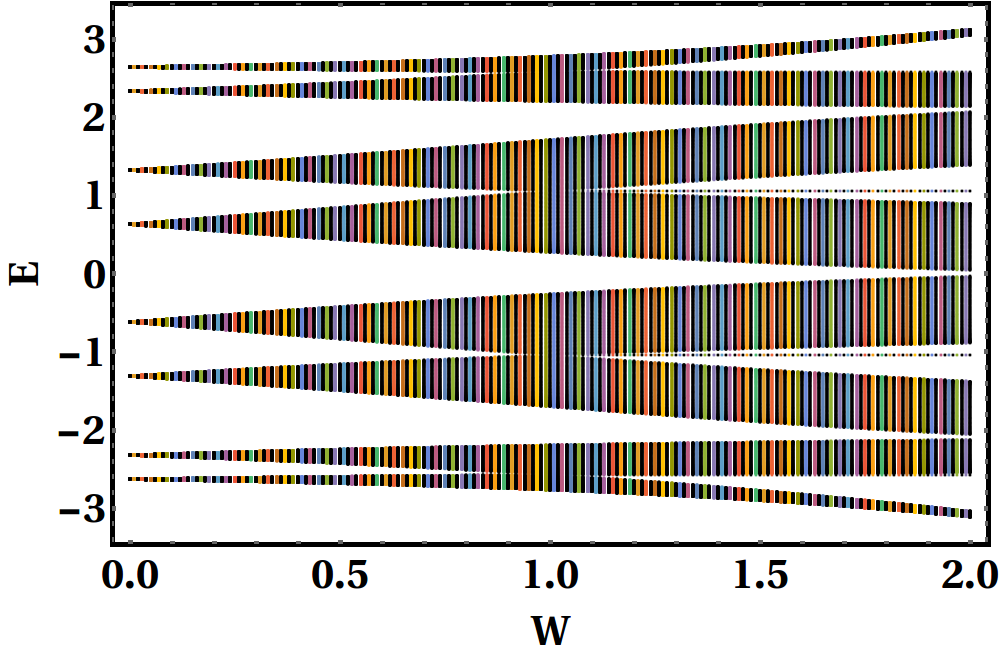}
(d)\includegraphics[width=.43\columnwidth]{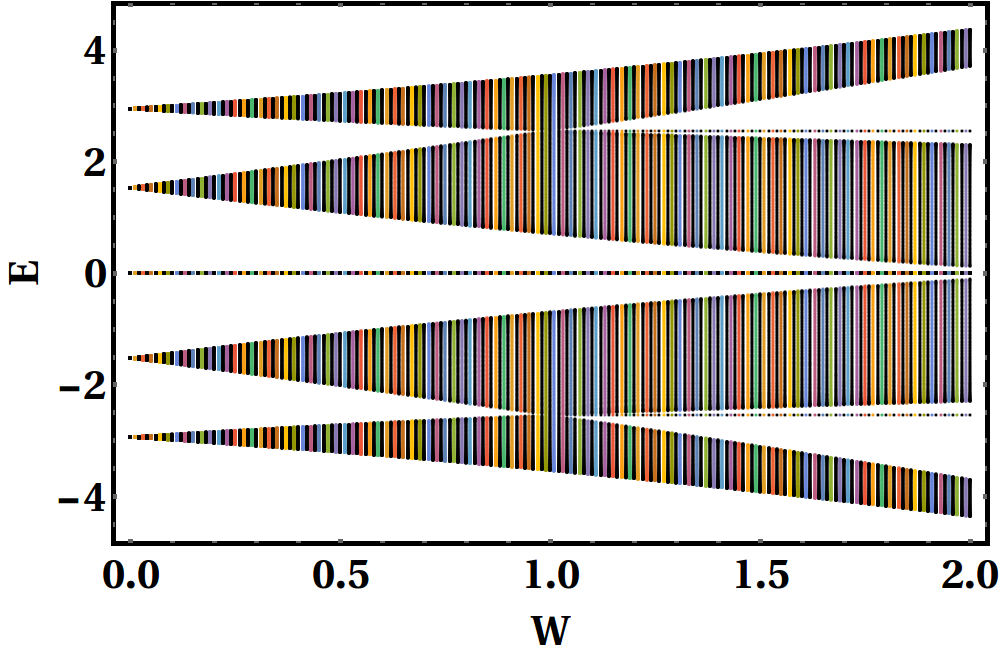}
(e)\includegraphics[width=.43\columnwidth]{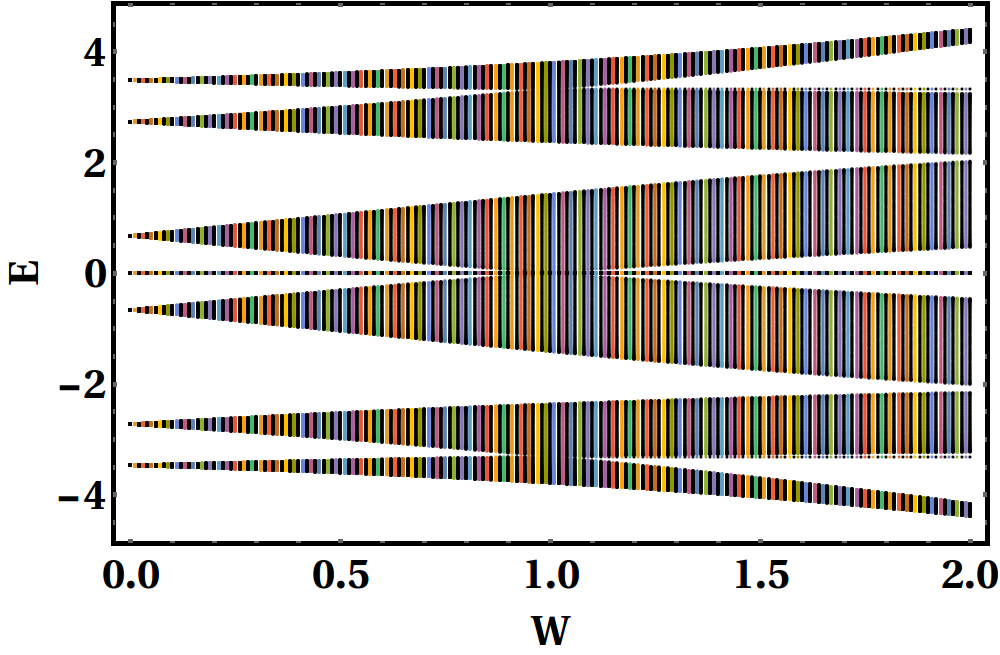}
(f)\includegraphics[width=.43\columnwidth]{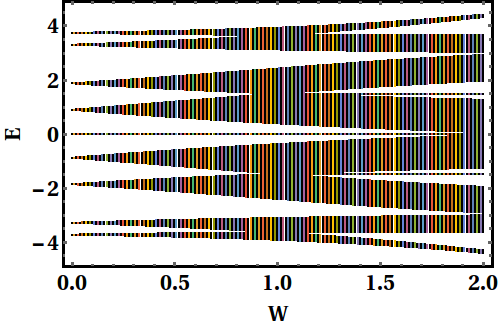}

\caption{(Color online) Energy spectra for (a)  GSSHC-I (Fig.~\ref{fig1}(a)), (b)  GSSHC-II ( Fig.~\ref{fig1}(b)), (c)  GSSHC-III ( Fig.~\ref{fig1}(c)), (d) cross-linking of two  GSSHC-I (Fig.~\ref{fig2}(a)), (e)  GSSHC-II (Fig.~\ref{fig2}(b)) and (f)  GSSHC-III ( Fig.~\ref{fig2}(c) ) with the coupling $w$ varies. We have used open boundary conditions for $N_x=200$, where $N_x$ denotes the number of unit cells taken along the $x$-direction. The values of the parameters are chosen as $t = 1.5$, $\epsilon = 0$, and $v = 1$ for all networks. The edges states are clearly seen, separated from the bulk bands.}  
\label{evsw1}
\end{figure}

\begin{figure}[ht]
\centering
(a)\includegraphics[width=.44\columnwidth]{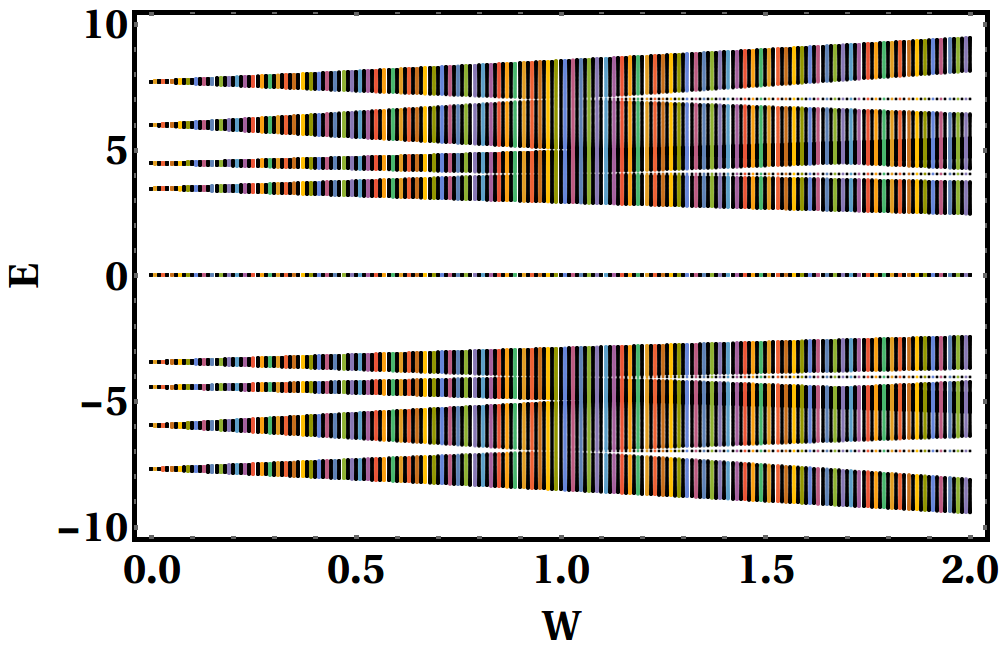}
(b)\includegraphics[width=.44\columnwidth]{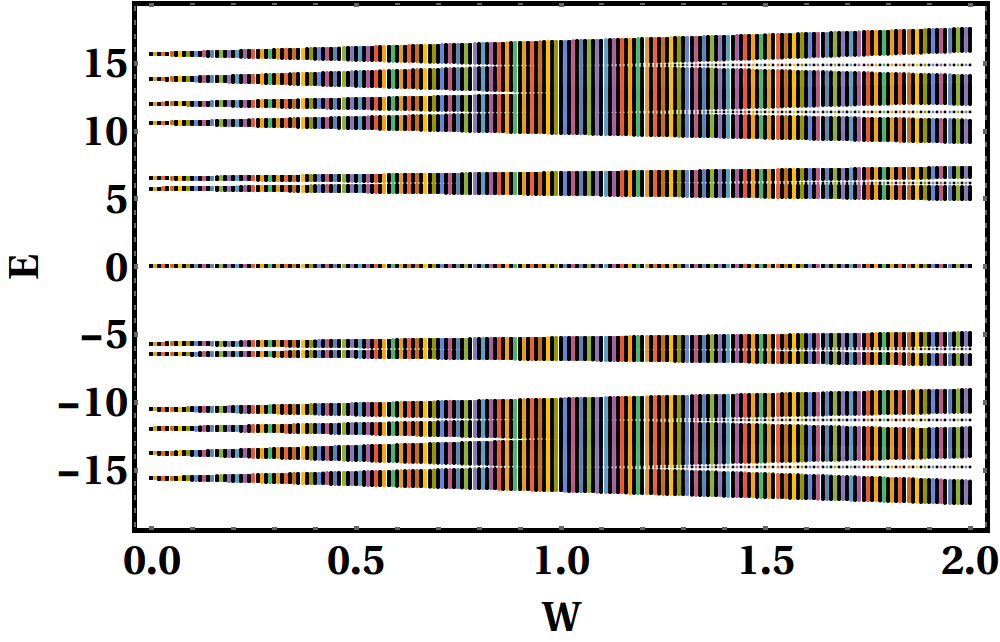}
\caption{(Color online) Energy spectra for (a)  cross-linking four  GSSHC-I (Fig.~\ref{fig3}(a)) and  (b) cross-linking six  GSSHC-I ( Fig.~\ref{fig3}(b)) with the coupling $w$ varies. We have used open boundary conditions for $N_x=200$, where $N_x$ denotes the number of unit cells taken along the $x$-direction. The values of the parameters are chosen as (a) $t = 3.92$, $\epsilon = 0$, $v = 1$, (b)  $t = 7.95$, $\epsilon = 0$, $ v = 1 $. The edge states are apparent in this case also.}  
\label{evsw2}
\end{figure}
Diagonalizing these three latter Hamiltonians we obtain the dispersive and the flat bands. The results are depicted in Fig.~\ref{e-k2}((a)-(i)). In all these cases the flat bands appear at an energy $ E = \epsilon $, which are degenerate with two, three, and four-fold degeneracy for two  cross-linked  GSSHC-I,  GSSHC-II, and GSSHC-III systems respectively. The energy eigenvalues corresponding to the flat bands as well as their degeneracies are extracted analytically using the RSRG scheme, which is discussed using Eqs.~\eqref{sub-AB}, Eqs.~\eqref{sub-ABC}, and Eqs.~\eqref{sub-ABCD}.

We can now have a critical look at the role of the relative magnitudes of the hopping integrals in determining the topological invariants. 

It is apparent that every Hamiltonian has three hopping integrals, viz. $t$, $v$, and $w$. Among these $v$ and $w$ are responsible for gap-closing ( for $v=w$ ) or opening ( for $v \neq w $ ) at the Brillouin zone boundary. For all these models extensive numerical search reveals that a Dirac cone appears at $ k = 0 $ as shown in Fig.~\ref{dirac}(a)-(f), depending on a specific relationship between $t$, $v$, and $w$. After evaluating the dispersion relations it is possible to work out the exact mathematical relationship between the hopping integrals $t$, $v$ and $w$ for which the Dirac cone appears at $k=0$. The condition for the appearance of the Dirac cone turns out to be,  
\begin{eqnarray}
t & = & \pm \sqrt{v} \sqrt{w}
\label{tvw}
\end{eqnarray}
To observe a topological phase transition, we need a gapped system. This requires that the strength of the hopping integral $t$ be set to any value except $ \pm \sqrt{v} \sqrt{w}$ as shown in Eqs.~\eqref{tvw}. Once this is implemented, all the bands get separated from each other and each of them has a quantized value of the Zak phase. The system exhibits a topological phase transition. This is true for even higher order mesh structures, such as the four and six  GSSHC-I cross-linked mesh families, as shown in Fig.~\ref{fig3}.
The energy-bands of these, as obtained by diagonalizing the corresponding Hamiltonians ($\hat{\mathcal{H}}_{7}(k)$, $ \hat{\mathcal{H}}_{8}(k)$, given by Eqs.~\eqref{ham-3a} and Eqs.~\eqref{ham-3b})  are shown in Fig.~\ref{e-k3}. For both the cases, the Dirac-cone at $k = 0 $ is seen in Fig.~\ref{dirac}(g)-(h), if the relation in Eq.~\eqref{tvw} is maintained by the parameters $t$, $v$, and $w$. \\

It is important to appreciate that, the external parameter $t$ plays a very important role in separating the energy bands. There is a particular value of $t$, a `threshold' ($t_{th}$), above which all dispersive bands separate out from each other. precisely at $t=t_{th}$ pairs of dispersive bands just touch each other (as shown in Fig.~\ref{e-k3}((b) and (d))) and just above it the band-overlap is totally absent. The numerical value of $t_{th}$ of course, is sensitive to the number of chains forming a mesh, that is, the width of the mesh. 
We have anlytically calculated this threshold value $t_{th}$ in terms of two other variables $v$ and $w$. The $t_{th}$ for the cross-linking of four  GSSHC-I lattices is, 
\begin{equation}
   t_{th}  =  \pm\frac{\sqrt{3 v^2 + 8 v w + 3 w ^2 + 2 \sqrt{3}\sqrt{(v+w)^4}}}{\sqrt{2}}
    \label{t1}
\end{equation}
Similarly, for cross-linking of six  GSSHC-I, it becomes,

\begin{eqnarray}
   t_{th} & = & \pm\sqrt{a + 2\sqrt{2} \sqrt{b} + \frac{1}{2} \sqrt{c + 8 a^2 - 2d +\eta}}\nonumber\\
  \eta & = & \frac{64 a^3 - 32 a d - 32 e}{16\sqrt{2}\sqrt{b}}
    \label{t2}
\end{eqnarray}
where,
$a =  4 v^2 + 9  v w + 4 w^2, b  =  v^4 + 4 v^3 w + 6  v^2 w^2 + 4  v w^3 + w ^4, c  =  4 v^4 - 10  v^2  w^2 + 4 w^4, d  =  14 v^4 + 80  v^3  w + 135 v^2  w^2 + 80  v w^3 + 14 w^4,$  and  $e  =  4 v^6 + 10 v^5w + 10v w^5 - 8v^4  w^2 - 8v^2 w^4 - 29 v^3 w^3 +4w^6$ .\\

The threshold values of $t$ due to cross-linking of four and six GSSHC-I for different values of $w$ for a fixed $v$ are calculated using  Eqs.~\eqref{t1} and Eqs.~\eqref{t2}, and are shown in Table~\ref{threshold}.

\begin{table}[ht]
\centering
\caption{Threshold values of $t$ for various $w$}
\vspace{0.5cm}
\begin{tabular}{|p{2cm}|p{2cm}|p{2cm}|}
 \hline
 \multicolumn{3}{|c|}{\textbf{ Four cross-linked GSSHC-I }}\\
 \hline
  $v$ & $w$  & $t_{th}$\\
 \hline
 1.0   &  0.9   & $\pm3.5451$\\
 \hline
 1.0   &   1.0  & $\pm3.73205$ \\
 \hline
 1.0   &   1.1  & $\pm3.91833$\\
 \hline
 \multicolumn{3}{|c|}{ \textbf{Six cross-linked GSSHC-I} }\\
 \hline
  $v$ & $w$  & $t_{th}$\\
 \hline
 1.0   &  0.9   & $\pm7.14757$ \\
 \hline
 1.0   &   1.0  & $\pm7.52395$ \\
 \hline
 1.0   &   1.1  & $\pm7.89998$\\
 \hline
\end{tabular}
\label{threshold}
\end{table}



\begin{figure}[ht]
\centering
(a)\includegraphics[width=0.43\columnwidth]{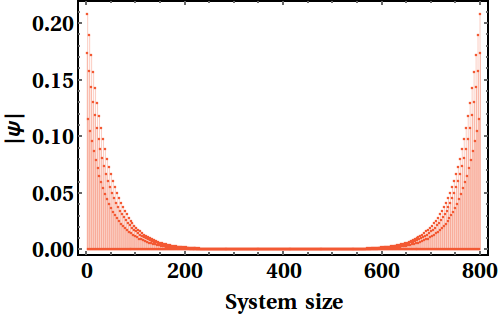}
(b)\includegraphics[width=0.43\columnwidth]{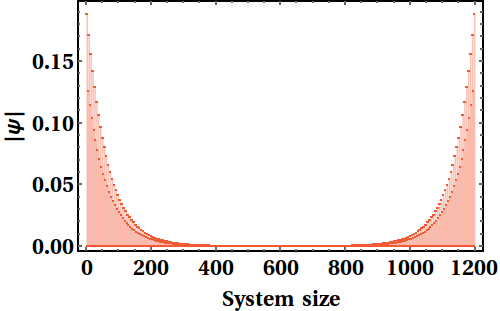}
(c)\includegraphics[width=0.43\columnwidth]{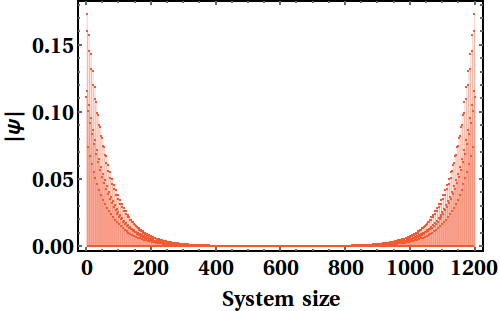}
(d)\includegraphics[width=0.43\columnwidth]{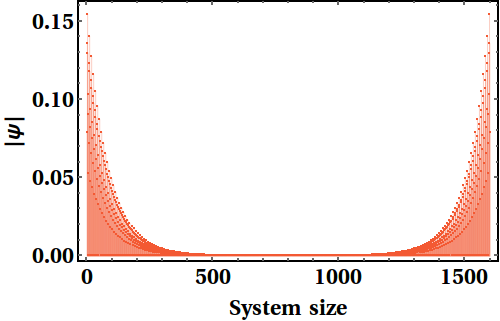}
(e)\includegraphics[width=0.43\columnwidth]{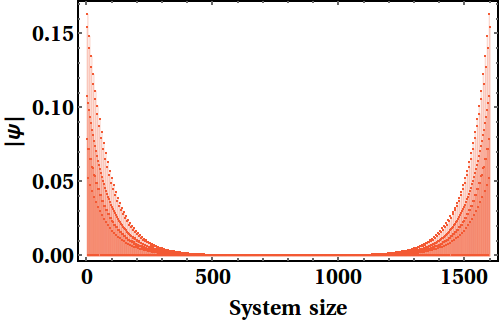}
(f)\includegraphics[width=0.43\columnwidth]{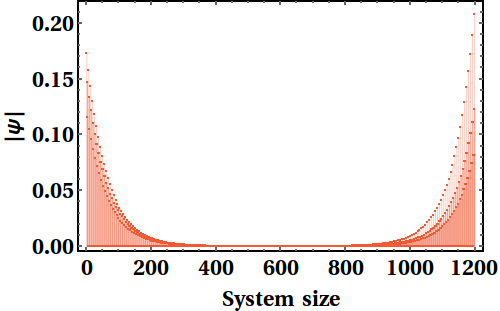}
(g)\includegraphics[width=0.43\columnwidth]{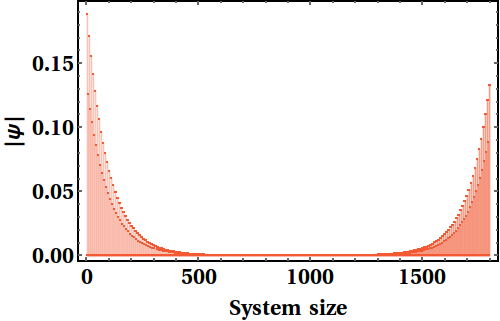}
(h)\includegraphics[width=0.43\columnwidth]{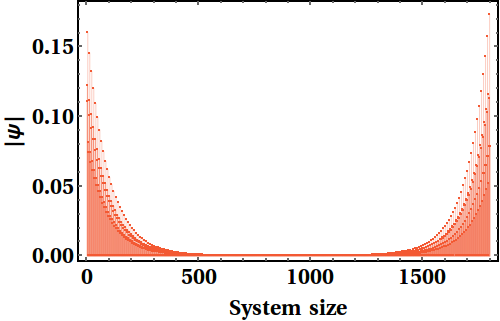}
(i)\includegraphics[width=0.43\columnwidth]{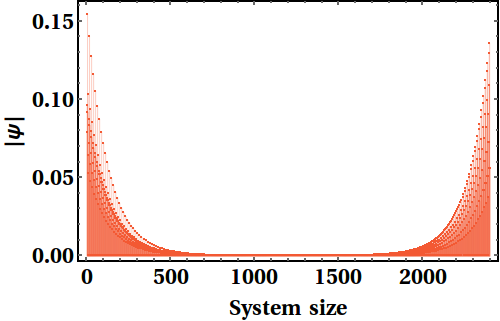}
(j)\includegraphics[width=0.43\columnwidth]{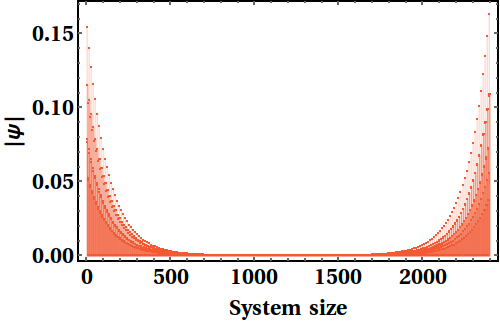}

\caption{(Color online)  Wave function distributions of the edge states (a) for  GSSHC-I (Fig.~\ref{fig1}(a)) with energies $ E = \pm1.80278$,  (b,c) for GSSHC-II ( Fig.~\ref{fig1}(b)) with energies (b) $ E = 0$, (c) $ E =\pm2.34521 $, (d,e) for  GSSHC-III (Fig.~\ref{fig1}(c)) with energies (d) $ E = \pm2.57885$, (e) $ E = \pm1.04859$, (f) for cross-linking of two  GSSHC-I (Fig.~\ref{fig2}(a)) with energies (f) $E = \pm2.54951$, (g,h) for cross-linking of two  GSSHC-II (Fig.~\ref{fig2}(b)) with energies (g) $E = 0$, (h) $E = \pm3.31662$,  and (i,j) for cross-linking of two  GSSHC-III (Fig.~\ref{fig2}(c)) with energies (i) $E = \pm3.64704$, (j) $E = \pm1.48294$, respectively. All of these energies are doubly degenerate. We have used open boundary conditions for $N_x=200$, where $N_x$ denotes the number of unit cells taken along the $x$-direction. The values of the parameters are chosen as $t = 1.5$, $\epsilon = 0$, $v = 1$, $ w = 1.1$ for all networks..}  
\label{edge1}
\end{figure}


\begin{figure}[ht]
\centering
(a)\includegraphics[width=0.43\columnwidth]{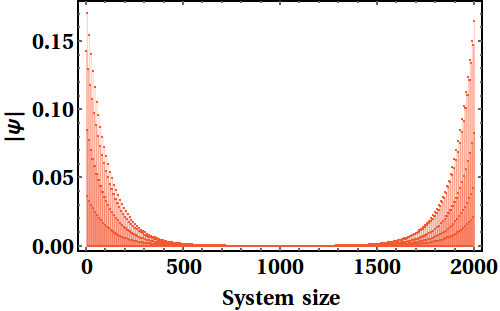}
(b)\includegraphics[width=0.43\columnwidth]{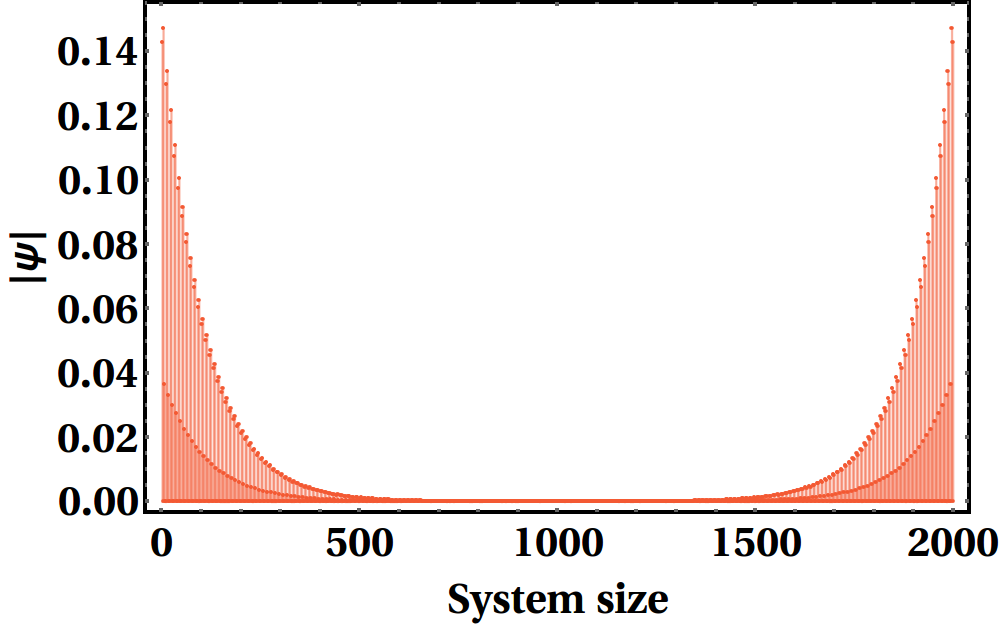}
(c)\includegraphics[width=0.43\columnwidth]{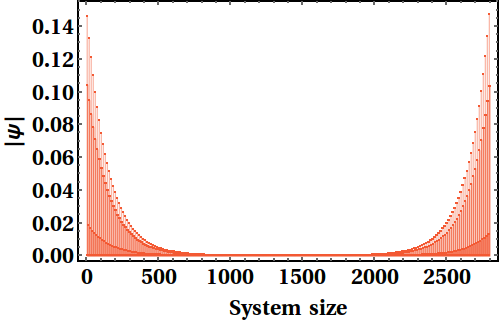}
(d)\includegraphics[width=0.43\columnwidth]{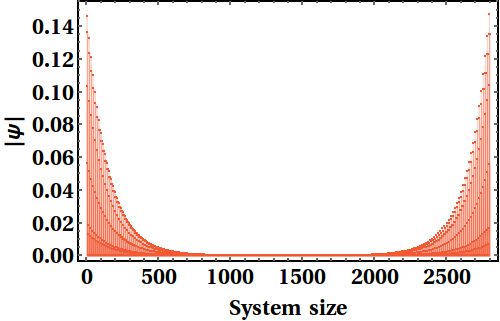}

\caption{(Color online)  Wave function distributions of the edge states (a,b) for cross-linking of four  GSSHC-I( Fig.~\ref{fig3}(a)) with energies (a)$ E = \pm7.00708$, (b) $ E = \pm4.04554 $,  and (c,d) for cross-linking of six  GSSHC-I (Fig.~\ref{fig3}(b)) with energies (c) $ E = \pm11.3316$, (d) $ E = \pm14.8054, E = \pm6.13261 $, respectively. All of these energies are doubly degenerate. We have used open boundary conditions for $N_x=200$, where $N_x$ denotes the number of unit cells taken along the $x$-direction. The values of the parameters are chosen as (a,b) $t = 3.92$, $\epsilon = 0$, $v = 1$, $ w = 1.1$, and (c,d) $t = 7.95$, $\epsilon = 0$, $v = 1$, $ w = 1.1$.}  
\label{edge2}
\end{figure}


For all such GSSHC and cross-linked  GSSHC, the opening or closing of a gap at the Brillouin zone boundary is totally controlled by $v$ and $w$. $t$ has no role in that. The only role of $t$ is in separating out the energy bands. Now if we smoothly transform the Hamiltonian just by tuning $w$, keeping the two other parameters $t$ and $v$ fixed,  gaps open up when $v > w$ (insulating phase), as shown in Fig.~\ref{e-k1}( (a),(d),(g)), Fig.~\ref{e-k2}( (a),(d),(g)), Fig.~\ref{e-k4}( (a),(d)), then close when $v = w$ ( metallic phase)  as shown in Fig.~\ref{e-k1}( (b),(e),(h)), Fig.~\ref{e-k2}( (b),(e),(h)), Fig.~\ref{e-k4}( (b),(e)), and re-open when $v < w$ (insulating phase)  as shown in Fig.~\ref{e-k1}( (c),(f),(i)), Fig.~\ref{e-k2}( (c),(f),(i)), Fig.~\ref{e-k4}( (c),(f)) at the Brillouin zone boundary. So, the crossover from one insulating phase to another is only possible by crossing the metallic phase at least once, which is a primary signature of topological phase transition. 

\section{Topological issues}

\subsection{Symmetry Operation}
 First, for all models, the lattice constant ($a$) and onsite potentials ($\epsilon$) are chosen as unity and zero respectively, throughout the chains. The Hamiltonian $\hat{\mathcal{H}}_{i}(k)$ ($i=1,2...8$) exhibits time-reversal symmetry, viz, $\hat{\mathcal{H}}(-k)^\ast = \hat{\mathcal{H}}(k)$.
 All  the GSSHC and cross-linked  GSSHC also obey the chiral symmetry operation, viz,
 \begin{eqnarray}
     {\mathcal{C}_i}^{-1}~\hat{\mathcal{H}_{i}}(k)~{\mathcal{C}_i} & = &- \hat{\mathcal{H}_{i}}(k)
     \label{ch}
 \end{eqnarray}
 with $i=1,2...,8$. The kernels of the Hamiltonians are already discussed, and the corresponding chiral operators ($\mathcal{C}_{i}$) are defined by,

\begin{eqnarray}
    \mathcal{C}_1 & = & diag[1, -1, 1, -1]\nonumber\\
    \mathcal{C}_2 & = & diag[1, -1, 1, -1, 1, -1]\nonumber\\
    \mathcal{C}_3 & = & diag[1, -1, 1, -1, 1, -1, 1, -1]\nonumber\\
    \mathcal{C}_4 & = & diag[1, -1, -1, 1, -1, -1]\nonumber\\
    \mathcal{C}_5 & = & diag[-1, 1, 1, -1, 1, 1, -1, 1, 1]\nonumber\\
    \mathcal{C}_6 & = & diag[1, -1, -1, 1, -1, -1, 1, -1, -1, 1, -1, -1]\nonumber\\
    \mathcal{C}_7 & = & diag[1, 1, -1, -1, -1, 1, 1, -1, -1, -1]\nonumber\\
    \mathcal{C}_8 & = & diag[1, 1, 1, -1, -1, -1, -1, 1, 1, 1, -1, -1, -1, -1]\nonumber\\
\label{chiral}
\end{eqnarray}


\subsection{The Zak phase}

The transition from one gap opening ($ v > w $) to the other ($ v < w $) is only possible by closing the energy gap ($ v = w) $, which indicates a topological phase transition. A topological invariant must be associated with this, called the Zak phase~\cite{zak} that flips its quantized value from {\it zero} to {\it unity } (in the unit of $\pi$) corresponding to the trivial and the non-trivial insulating phases respectively.

The Zak phase for the $n$-th bulk bands is defined as,
\begin{equation}
    Z = -i \oint_{BZ}  \mathcal{A}_{nk}(k) dk
    \label{zak} 
\end{equation}
where $\mathcal{A}_{nk}$ is called the Berry  curvature of the $n$-th Bloch eigenstate, which is again defined as~\cite{asboth},
\begin{equation}
    \mathcal{A}_{n{k}}(k)= \bra{\psi_{n{k}}}\ket{\frac{d\psi_{nk}}{dk}}
\end{equation}

The integral is done along a closed loop in the Brillouin zone.  $\ket{\psi_{nk}}$ is the $n$-th Bloch state.The Zak phase is totally a bulk property of the system. To calculate the Zak phase of the bands, we use the Wilson loop approach~\cite{fukui, wang}, by converting the integration Eqs.~\eqref{zak} into a summation over the Brillouin zone.  It is a gauge invariant formalism. It protects the numerical value of the Zak phase against any arbitrary phase change of Bloch wavefunction. This summation for non-degenerate $s$-th band ~\cite{fukui, wang} is given by Eqs.~\eqref{wilson} 

\begin{equation} 
Z_s = - Im ~ \left [ \log \prod_{k_n} \bra{\psi_{k_n,s}} \ket{\psi_{k_{n+1},s} } \right ]
\label{wilson}
\end{equation}

We calculate the Zak phase of all non-degenerate dispersive bands using  Eqs.~\eqref{wilson} for each GSSHC and two cross-linked GSSHC. The Zak phase always turns out to be quantized (either $0$ or $\pi$) for any value of the external parameter $t$. Still, the $ t = \pm\sqrt{v}\sqrt{w}$ situation must be avoided due to the formation of the Dirac cone at $k = 0$. 

The calculation of the Zak phase of multiple cross-linked GSSHC gives quantized values (either $0$ or $pi$) for the dispersive bands which have no overlap in between them. For bands that do overlap, the Zak phase can not be defined. The separation of such bands is, as stated earlier, engineered by choosing $t$ appropriately. The values of the Zak phase are displayed for both the `gap-opening' situations (insulating phases, viz for $v > w$ and for $v < w$ corresponding to the topologically trivial and non-trivial phases respectively) are given in Table~\ref{z}.

 \section{The edge states}
To see the edge states we construct a finite-sized  GSSHC and cross-linked  GSSHC just by joining two hundred unit cells in the $x$ direction. The construction ends with a unit cell at the end.  The energy spectra of the  GSSHC-I,  GSSHC-II,  GSSHC-III, and two cross-linked GSSHC-I,  GSSHC-II,  GSSHC-III are plotted in Fig.~\ref{evsw1} when the inter-cell hopping $w$ is varied. Fig.~\ref{evsw2} presents a similar plot for cross-linking of four and six GSSHC-I. In all these models a clear existence of edge states, that are totally decoupled from the bulk bands is observed, just when the strength of $w$ overcomes $v$ ($v<w)$. The values of the edge state energies are found to be the same (for all practical purposes) to the `gap-closing' energies. The comparison is shown in Table~\ref{gap-edge}.

\begin{table}[ht]
\centering
\caption{Gap closing and Edge state energy}
\vspace{0.5cm}
\begin{tabular}{|p{3.5cm}|p{3.5cm}|}
 \hline
\textcolor{blue}{Gap-closing Energies}  & \textcolor{blue}{ Edge State Energies} \\
 \hline
 \multicolumn{2}{|c|}{\textbf{GSSHC-I }}\\
 \hline
 $E = \pm 1.80277$  &   $ E = \pm1.80278 $ \\
 \hline
 \multicolumn{2}{|c|}{\textbf{ GSSHC-II }}\\
 \hline
 E = 0, $ \pm 2.34520$ &  0, $  \pm 2.34521$\\
 \hline
 \multicolumn{2}{|c|}{\textbf{ GSSHC-III }}\\
 \hline
 $E = \pm 2.57885$, $ \pm 1.04859$ &  $E = \pm 2.57885$, $ \pm 1.04859 $\\
 \hline
  \multicolumn{2}{|c|}{\textbf{ Two cross-linked GSSHC-I }}\\
  \hline
 $E = \pm 2.54951$  &   $E = \pm2.54951 $ \\
 \hline
 \multicolumn{2}{|c|}{\textbf{ Two cross-linked GSSHC-II }}\\
 \hline
 E = 0, $ \pm 3.31662$ & E = 0, $  \pm3.31662$\\
 \hline
 \multicolumn{2}{|c|}{\textbf{Two cross-linked GSSHC-III }}\\
 \hline
 $E = \pm1.48294$, $  \pm3.64704$ &  $ E = \pm1.48294$, $ \pm3.64704 $\\
 \hline
 \multicolumn{2}{|c|}{\textbf{ Four cross-linked GSSHC-I}}\\
  \hline
 $ E = \pm4.04554$, $ \pm7.00708$  &   $E = \pm4.04554$, $ \pm7.00708$ \\
 \hline
 \multicolumn{2}{|c|}{\textbf{ Six cross-linked GSSHC-I}}\\
  \hline
 $E = \pm6.13261$, $ \pm11.3316$, $  \pm14.8054$  &   $E = \pm6.13261$, $ \pm11.3316$, $\pm14.8054$ \\
 \hline
\end{tabular}
\label{gap-edge}
\end{table}
On the other hand, when $v>w$, no gap-states are observed in the energy spectra at these edge-state energies. The behavior of such edge states is shown in Fig.~\ref{edge1} and Fig.~\ref{edge2}, which indicates an absence of any state in the bulk portion of these GSSHC and cross-linked GSSHC systems. 

\subsection{Stability of Edge-States}
The chiral operators for all models are already defined in  Eqs.~\eqref{chiral} which leads to the existence of the chirally protected edge states. It is also clear (from Fig.~\ref{evsw1} and Fig.~\ref{evsw2} ) that if the inter-cell coupling $w$ is tuned, keeping other parameters fixed, still there is no change in edge-state energies.\\

We have tested the robustness of the edge states of these GSSHC and cross-linked GSSHC against disorder. The disorder is applied in all unit cells of each array at the inter-cell hopping $w$. We introduce the disorder in $w$ by assigning a random component $\delta w$  chosen from a window between  $\delta =0.01$ to   $\delta=1.0$, which is enough to produce a strong disorder. Now the models have a strong disorder in the bulk as well as the boundary. Interestingly, there is no change in edge-state energies. So, we conclude that all GSSHC and cross-linked GSSHC models exhibit robust edge states even against a substantially strong disorder.
The existence of quantized Zak phase for all the Bloch bands, and the robust edge states that are protected by chiral symmetry, imply that the bulk-boundary correspondence is strictly obeyed by all of these GSSHC and cross-linked GSSHC networks.
\section{Concluding Remarks}
In conclusion, we have studied a large family of cross-linked generalized SSH chains, forming a mesh. The width of the mesh may vary from a trivial two-chain limit to an arbitrarily large value. We find that one particular parameter, namely a hopping integral ($t$ here) that punctuates the usual SSH kind of staggering between the two others ($v$ and $w$) plays a crucial role in making the mesh topologically non-trivial. In-depth analytical calculations have been done and the results are supported by an exact numerical diagonalization of the system Hamiltonians in each case. The edge states, protected by chiral symmetry, and their robustness against disorder have been discussed in detail.

\section{Acknowledgments}
SB is thankful to the Government of West Bengal for the SVMCM Scholarship (WBP221657867058). 
\begin{widetext}

\appendix

\section{Study of a generalized SSH mesh: the flat and the dispersive bands}
\label{appendixA}

\subsection{Cross-linking of two generalized SSH chains of the second kind (GSSHC-II)}
Applying exactly the same decimation process on two entangled GSSHC-II (as shown in Fig.~\ref{fig2} b(i)), the cross-linked geometry maps onto a two-strand ladder network, now containing three-sublattices (as shown in Fig.~\ref{fig2} b(ii)) ($A$, $B$, $C$). The onsite potentials on the ladder strip are now $ \epsilon_A$, $ \epsilon_B$, and $ \epsilon_C$. The inter-strand hopping integrals are $\Gamma_A$, $\Gamma_B$, and $\Gamma_C$, and the intra-strand hoppings are $t_A$, $ t_B $, and $t_C$. These are given by, 
\begin{align}
\epsilon_A   =   \epsilon + \frac{2 t^2}{E - \epsilon},  &&
\epsilon_B  =  \epsilon + \frac{t^2 + v^2}{E - \epsilon}, &&
\epsilon_C  =  \epsilon + \frac{t^2 + w^2}{E - \epsilon}, \nonumber\\
\Gamma_A  =  \frac{2t^2}{E - \epsilon}, &&
\Gamma_B  =  \frac{t^2 + v^2}{E - \epsilon}, &&
\Gamma_C  =  \frac{t^2 + w^2}{E - \epsilon}, \nonumber\\
t_A  =  \frac{v t}{E - \epsilon}, &&
t_B  =  \frac{ t^2}{E - \epsilon},  &&
t_C  =  \frac{w t}{E - \epsilon} 
 \label{rsrg-2b}
\end{align}

The difference equations for the two-strand, three-sublattice-ladder are now written conveniently in matrix form as, 

\begin{eqnarray}
    [E.\mathbb{I}_{2\cross2} -\tilde{\epsilon}_{A}] \Psi_{A,n} & = & \tilde{t}_A \Psi_{B,n+1} + \tilde{t}_C \Psi_{C,n-1} \nonumber\\\
    [E.\mathbb{I}_{2\cross2} -\tilde{\epsilon}_{B}] \Psi_{B,n} & = & \tilde{t}_B \Psi_{C,n+1} + \tilde{t}_A \Psi_{A,n-1} \nonumber\\\
    [E.\mathbb{I}_{2\cross2} -\tilde{\epsilon}_{C}] \Psi_{C,n} & = & \tilde{t}_C \Psi_{A,n+1} + \tilde{t}_B \Psi_{B,n-1} 
  \label{eq2}
\end{eqnarray}

where
\begin{equation}
\Psi_{(A,B,C),n} = \left[ \begin{array}{cccccccccccccccc}
\psi_{(A,B,C),n,1}\\
 \psi_{(A,B,C),n,2}
\end{array}
\right ] 
\end{equation}

and,

 \begin{equation}
\tilde{\epsilon}_{(A,B,C)} = \left[ \begin{array}{cccccccccccccccc}
\epsilon + \Gamma_{(A,B,C)} & \Gamma_{(A,B,C)}\\
 \Gamma_{(A,B,C)} & \epsilon + \Gamma_{(A,B,C)} 
\end{array}
\right ] 
\end{equation}

\begin{eqnarray}
 \tilde{t}_{(A,B,C)} & = & t_{(A,B,C)} \mathbf{\Lambda}_{2\times2} \nonumber \\  
  \end{eqnarray}
with $\mathbf{\Lambda}_{2 \times 2}^{ij}= 1 $ for all $(i,j)$.

In this case also the potential and the hopping matrices commute, {\it independent of the energy} $E$ with all possible combinations $[\tilde{\epsilon}_{(A,B,C)}, \tilde{t}_{(A,B,C)}]  =  0$ irrespective of the energy $E$. Exploiting this commutation, the difference equations are written down in a new basis defined by $\Phi_{(A,B,C),n}=\mathcal{M}^{-1}$ $\Psi_{(A,B,C),n}$. The equations for the $A$, $B$, and $C$ sub-lattices are now completely decoupled in the new basis. 

\begin{eqnarray}
(E - \epsilon)  \phi_{A,n,1} & = & 0 \nonumber \\
\left (E - \epsilon - 2 \Gamma_A \right ) \phi_{A,n,2} & = & 2 t_C \phi_{C,n-1,2}+ 2 t_A \phi_{B,n+1,2}\nonumber\\
(E - \epsilon)  \phi_{B,n,1} & = & 0 \nonumber \\
\left (E - \epsilon - 2 \Gamma_B \right ) \phi_{B,n,2} & = & 2 t_A \phi_{A,n-1,2}+ 2 t_B \phi_{C,n+1,2}\nonumber\\
(E - \epsilon)  \phi_{C,n,1} & = & 0 \nonumber \\
\left (E - \epsilon - 2 \Gamma_C \right ) \phi_{C,n,2} & = & 2 t_B \phi_{B,n-1,2}+ 2 t_C \phi_{A,n+1,2}\nonumber\\
\label{sub-ABC}
\end{eqnarray}

From Eqs.~\eqref{sub-ABC} a three-fold degenerate flat band at energy $ E= \epsilon $ is clearly observed. If we decimate out the $B$ sites again the whole geometry is converted to a chain with two renormalized onsite potentials $\epsilon+2\Gamma_C+\frac{4t_B^2}{E-\epsilon-2\Gamma_B}$ and $\epsilon+2\Gamma_A+\frac{4t_A^2}{E-\epsilon-2\Gamma_B}$ and periodically alternating hopping integrals $2t_C$ and   $\frac{4t_At_B}{E-\epsilon-2\Gamma_B}$.

Now, the energies at which the band gap closes at the Brillouin zone boundary, are obtained as the solution of the equations,
\begin{eqnarray}
    E - \left(\epsilon+2\Gamma_C+\frac{4t_B^2}{E-\epsilon-2\Gamma_B}\right) & = & 0\nonumber\\
    E - \left(\epsilon+2\Gamma_A+\frac{4t_A^2}{E-\epsilon-2\Gamma_B}\right) & = & 0\nonumber\\
    2t_C - \frac{4t_At_B}{E-\epsilon-2\Gamma_B} & = & 0
    \label{gap-1-abc}
\end{eqnarray}
After simplification of Eqs.~\eqref{gap-1-abc} using Eqs.~\eqref{rsrg-2b}, we obtain a set of three equations, viz, 
 \begin{eqnarray}
   (E-\epsilon)\left[ (E-\epsilon)^4 - (E-\epsilon)^2(4t^2+2v^2+2w^2) + 4(t^2+v^2)(t^2+w^2) - 4t^4\right]  =  0 \nonumber\\
   (E-\epsilon)\left[ (E-\epsilon)^4 - (E-\epsilon)^2(6t^2+2v^2) + 8t^2(t^2+v^2) - 4t^2v^2\right]  =  0 \nonumber\\
   (E-\epsilon)\left [2wt(E-\epsilon)^2-4wt(t^2+v^2)-4vt^3\right]  =  0
   \label{gap-2-abc}
\end{eqnarray}
From Eqs.~\eqref{gap-2-abc}, we understand that the energies at which the band gap will close at the Brillouin zone boundary must be the common roots of the three equations. $E=\epsilon$ is one solution, and the others have to satisfy the set of the following equations simultaneously.
\begin{eqnarray}
   E & = & \epsilon\pm\sqrt{2t^2+v^2+w^2\pm\sqrt{4t^4+v^4-2v^2w^2+w^4}} \nonumber \\ 
   E & = & \epsilon\pm t\sqrt{2} \nonumber \\ 
   E & = & \epsilon \pm \sqrt{2}\sqrt{2t^2+v^2} \nonumber \\ 
   E & = & \epsilon \pm \sqrt{2}\sqrt{t^2+v^2+\frac{vt^2}{w}}
   \label{closing}
\end{eqnarray}
 It can be checked that, for $v = w$ all the above three equations have two common roots, given by,
\begin{equation}
    E = \epsilon \pm \sqrt{2} \sqrt{2t^2+ v^2}
     \label{sol-2b}
\end{equation}
The band gaps close at the two energy values. 

One can now use the set of Eqs.~\eqref{sub-ABC} to obtain the dispersion relation for the two cross-linked GSSHC-II, which is given by,


\begin{equation}
(E - \epsilon)^6 -  (E-\epsilon)^4\tau +(E-\epsilon)^2 (8t^2\alpha-4v^2t^2 - 2\beta\xi)-(E-\epsilon)[4t^4 E_A+4w^2t^2E_B] - 16 v w t^4 \cos~ka'  =  0 
   \label{dispersion2-2b}
  \end{equation}
where, $\alpha = v^2+t^2$, $\beta = w^2+t^2$, $ \tau = 2\alpha +4t^2$, $ \xi=E_AE_B-4{t_A}^2$, $ E_A = E-\epsilon-2\Gamma_A$ and $ E_B = E-\epsilon-2\Gamma_B$.


\subsection{Cross-linking of two generalized SSH chains of the third kind (GSSHC-III)}
Using the same technique as before, on two entangled GSSHC-III (as shown in Fig.~\ref{fig2} c(i)), the cross-linked geometry comprising two GSSHC-III is mapped into a two-strand ladder network containing four-sublattices (as shown in Fig.~\ref{fig2} c(ii)) ($A$, $B$, $C$, $D$) with onsite potential $ \epsilon_A$, $ \epsilon_B$, $ \epsilon_C$, $\epsilon_D$ and hopping integrals $\Gamma_A$, $\Gamma_B$, $\Gamma_C$, $\Gamma_D$, $t_A$, $ t_B $, $t_C$, $t_D$ respectively. Their explicit forms are, 

\begin{align}
\epsilon_{A(C)}  =  \epsilon + \frac{2 t^2}{E - \epsilon}, &&
\epsilon_B  =  \epsilon + \frac{t^2 + v^2}{E - \epsilon}, &&
\epsilon_D  =  \epsilon + \frac{t^2 + w^2}{E - \epsilon}, \nonumber\\
\Gamma_{A(C)}   =  \frac{2t^2}{E - \epsilon},&& 
\Gamma_B  =  \frac{t^2 + v^2}{E - \epsilon}, &&
\Gamma_D  =  \frac{t^2 + w^2}{E - \epsilon}, \nonumber\\
t_{A(C)} =  \frac{ t^2}{E - \epsilon}, &&
t_B  =  \frac{ v t}{E - \epsilon}, &&
t_D  =  \frac{w t}{E - \epsilon},  
 \label{rsrg-2c}
\end{align}


The difference equations for four-sublattice, two-strand ladder network are now given by,

\begin{eqnarray}
    [E.\mathbb{I}_{2\cross2} -\tilde{\epsilon}_{A}] \Psi_{A,n} & = & \tilde{t}_A \Psi_{B,n+1} + \tilde{t}_D \Psi_{D,n-1} \nonumber\\\
    [E.\mathbb{I}_{2\cross2} -\tilde{\epsilon}_{B}] \Psi_{B,n} & = & \tilde{t}_B \Psi_{C,n+1} + \tilde{t}_A \Psi_{A,n-1} \nonumber\\\
    [E.\mathbb{I}_{2\cross2} -\tilde{\epsilon}_{C}] \Psi_{C,n} & = & \tilde{t}_C \Psi_{D,n+1} + \tilde{t}_B \Psi_{B,n-1}  \nonumber\\\
    [E.\mathbb{I}_{2\cross2} -\tilde{\epsilon}_{D}] \Psi_{D,n} & = & \tilde{t}_D \Psi_{A,n+1} + \tilde{t}_C \Psi_{C,n-1}
  \label{eq2c}
\end{eqnarray}

where
\begin{equation}
\Psi_{(A,B,C,D),n} = \left[ \begin{array}{cccccccccccccccc}
\psi_{(A,B,C,D),n,1}\\
 \psi_{(A,B,C,D),n,2}
\end{array}
\right ] 
\end{equation}

and,

 \begin{equation}
\tilde{\epsilon}_{(A,B,C,D)} = \left[ \begin{array}{cccccccccccccccc}
\epsilon + \Gamma_{(A,B,C,D)} & \Gamma_{(A,B,C,D)}\\
 \Gamma_{(A,B,C,D)} & \epsilon + \Gamma_{(A,B,C,D)} 
\end{array}
\right ] 
\end{equation}

\begin{eqnarray}
 \tilde{t}_{(A,B,C,D)} & = & t_{(A,B,C,D)} \mathbf{\Lambda}_{2\times2} \nonumber \\  
  \end{eqnarray}
with $\mathbf{\Lambda}_{2 \times 2}^{ij}= 1 $ for all $(i,j)$.

As before, all possible combinations of the potential and the hopping matrices  commute, that is, 
$[\tilde{\epsilon}_{(A,B,C,D)} , \tilde{t}_{(A,B,C,D)}]  =  0$ irrespective of the choice of the energy $E$. In a changed basis, as done before,  defined by $\Phi_{(A,B,C,D),n}=\mathcal{M}^{-1}$ $\Psi_{(A,B,C,D),n}$. The equations for $A$, $B$, $C$, and $D$ sub-lattice are now completely decoupled. They are listed below.  

\begin{eqnarray}
(E - \epsilon)  \phi_{A,n,1} & = & 0 \nonumber \\
\left (E - \epsilon - 2 \Gamma_A \right ) \phi_{A,n,2} & = & 2 t_D \phi_{D,n-1,2}+ 2 t_A \phi_{B,n+1,2}\nonumber\\
(E - \epsilon)  \phi_{B,n,1} & = & 0 \nonumber \\
\left (E - \epsilon - 2 \Gamma_B \right ) \phi_{B,n,2} & = & 2 t_A \phi_{A,n-1,2}+ 2 t_B \phi_{C,n+1,2}\nonumber\\
(E - \epsilon)  \phi_{C,n,1} & = & 0 \nonumber \\
\left (E - \epsilon - 2 \Gamma_C \right ) \phi_{C,n,2} & = & 2 t_B \phi_{B,n-1,2}+ 2 t_C \phi_{D,n+1,2}\nonumber\\
(E - \epsilon)  \phi_{D,n,1} & = & 0 \nonumber \\
\left (E - \epsilon - 2 \Gamma_D \right ) \phi_{D,n,2} & = & 2 t_C \phi_{C,n-1,2}+ 2 t_D \phi_{A,n+1,2}\nonumber\\
\label{sub-ABCD}
\end{eqnarray}

From Eqs.~\eqref{sub-ABCD} a four-fold degenerate flat band at energy $ E= \epsilon $ is apparent. After one more decimation, it converted into a staggered SSH chain with new onsite potentials and hopping integrals. They are given by,

\begin{eqnarray}
\tilde{\tilde{\epsilon}}_B & = & \epsilon+2\Gamma_B+\frac{4{t_A^2}}{E-\epsilon-2\Gamma_A}+\frac{4{t_B^2}}{E-\epsilon-2\Gamma_C} \nonumber\\
\tilde{\tilde{\epsilon}}_D & = & \epsilon+2\Gamma_D+\frac{4{t_C^2}}{E-\epsilon-2\Gamma_C}+\frac{4{t_D^2}}{E-\epsilon-2\Gamma_A} \nonumber\\
\tilde{\tilde{t}}_B & = & \frac{4t_Bt_C}{E-\epsilon-2\Gamma_C} \nonumber\\
\tilde{\tilde{t}}_D & = & \frac{4t_At_D}{E-\epsilon-2\Gamma_A}
\label{eq-2c}
\end{eqnarray}

The band gaps close at the Brillouin zone boundaries, as before for $v=w$, and the corresponding energy values are given by, 
\begin{equation}
E  =  \epsilon \pm\sqrt{3t^2+v^2\pm\sqrt{5t^4+2t^2v^2+v^4}} 
\label{sol-2c}
\end{equation}

The dispersion relation can be obtained following the same procedure as used in the two earlier cases, and now reads,
\begin{equation}
(E - \epsilon)^8 -2\Gamma_D(E-\epsilon)^7 +(\Delta_1-\Delta_2)(E-\epsilon)^5-(\Delta_3+\Delta_4)\xi-16t^6(v^2+w^2)-32t^6vw\cos~ka'  =  0 
   \label{dispersion2-2c}
  \end{equation}

where $ \Delta_1 = (4{\Gamma_A}^2-4(E-\epsilon)\Gamma_A)(E-\epsilon-2\Gamma_D)$, $ \Delta_2 = (E-\epsilon-2\Gamma_A)(4{t_C}^2+4{t_D}^2)$, $\Delta_3 = \frac{2(t^2+v^2)}{(E-\epsilon)}$, $ \Delta_4 = \frac{4(t^4+v^2t^2)}{(E-\epsilon)^2(E-\epsilon-2\Gamma_A)}$, $ \xi = (E-\epsilon)^4(E-\epsilon-2\Gamma_A)[(E-\epsilon-2\Gamma_A)(E-\epsilon-2\Gamma_D)-4({t_c}^2+{t_D}^2)]$ and using solutions of Eqs.~\eqref{dispersion2-2c} all dispersive bands are obtained.

\section{The kernels of the Hamiltonian}
The kernels of the Hamiltonian ($ \hat{\mathcal{H}}_{1}(k)$, $ \hat{\mathcal{H}}_{2}(k)$, $ \hat{\mathcal{H}}_{3}(k)$) for the unit cell of the GSSHC-I, GSSHC-II, GSSHC-III(as shown in  Fig.~\ref{fig1}) are given by, 
\vspace{0.5cm}
\begin{equation}
\hat{\mathcal{H}}_{1}(k) = \left[ \begin{array}{cccccccccccccccc}
 \epsilon & t & 0 & w e^{-ika}\\
 t & \epsilon & v & 0\\
 0 & v & \epsilon & t\\
 w e^{ika} & 0 & t & \epsilon\\
\end{array}
\right ] 
\label{ham-1a}
\end{equation}
\vspace{0.5cm}
\begin{equation}
\hat{\mathcal{H}}_{2}(k) = \left[ \begin{array}{cccccccccccccccc}
 \epsilon & t & 0 & 0 & 0 & w e^{-ika}\\
 t & \epsilon & t & 0 & 0 & 0\\
 0 & t & \epsilon & v & 0 & 0\\
 0 & 0 & v & \epsilon & t & 0\\
 0 & 0 & 0 & t & \epsilon & t\\
 w e^{ika} & 0 & 0 & 0 & t & \epsilon\\
\end{array}
\right ] 
\label{ham-1b}
\end{equation}
\vspace{0.5cm}
\begin{equation}
\hat{\mathcal{H}}_{3}(k) = \left[ \begin{array}{cccccccccccccccc}
\epsilon & t & 0 & 0 & 0 & 0 & 0 & w e^{-ika}\\
t & \epsilon & t & 0 & 0 & 0 & 0 & 0\\
0 & t & \epsilon & t & 0 & 0 & 0 & 0\\
0 & 0 & t & \epsilon & v & 0 & 0 & 0\\
0 & 0 & 0 & v & \epsilon & t & 0 & 0\\
0 & 0 & 0 & 0 & t & \epsilon & t & 0\\
0 & 0 & 0 & 0 & 0 & t & \epsilon & t\\
w e^{ika} & 0 & 0 & 0 & 0 & 0 & t & \epsilon\\
\end{array}
\right ] 
\label{ham-1c}
\end{equation}

Similarly, the kernels of the Hamiltonian ($ \hat{\mathcal{H}}_{4}(k)$, $ \hat{\mathcal{H}}_{5}(k)$, $ \hat{\mathcal{H}}_{6}(k)$) for the unit cell of two cross-linked GSSHC-I, GSSHC-II, GSSHC-III (as shown in  Fig.~\ref{fig2}) read, 
\vspace{0.5cm}
\begin{equation}
\hat{\mathcal{H}}_{4}(k) = \left[ \begin{array}{cccccccccccccccc}
\epsilon & t & t & 0 & w e^{-ika} & w e^{-ika}\\
t & \epsilon & 0 & v & 0 & 0\\
t & 0 & \epsilon & v & 0 & 0\\
0 & v & v &  \epsilon & t & t\\
w e^{ika} & 0 & 0 & t & \epsilon & 0\\
w e^{ika} & 0 & 0 & t & 0 & \epsilon  \\
\end{array}
\right ] 
\label{ham-2a}
\end{equation}

\vspace{0.5cm}
\begin{equation}
\hat{\mathcal{H}}_{5}(k) = \left[ \begin{array}{cccccccccccccccc}
\epsilon & t & t & 0 & 0 & 0 & 0 & w e^{-ika} & w e^{ika} \\
t & \epsilon & 0 & t & 0 & 0 & 0 & 0 & 0\\
t & 0 & \epsilon & t & 0 & 0 & 0 & 0 & 0\\
0 & t & t & \epsilon & v & v & 0 & 0 & 0\\
0 & 0 & 0 & v & \epsilon & 0 & t & 0 & 0\\
0 & 0 & 0 & v & 0 & \epsilon & t & 0 & 0\\
0 & 0 & 0 & 0 & t & t & \epsilon & t & t\\
w e^{ika} & 0 & 0 & 0 & 0 & 0 & t & \epsilon & 0\\
w e^{ika} & 0 & 0 & 0 & 0 & 0 & t & 0 & \epsilon  \\
\end{array}
\right ] 
\label{ham-2b}
\end{equation}

\vspace{0.5cm}
\begin{equation}
\hat{\mathcal{H}}_{6}(k) = \left[ \begin{array}{cccccccccccccccc}

\epsilon & t & t & 0 & 0 & 0 & 0 & 0 & 0 & 0 & w e^{-ika} & w e^{-ika} \\
t & \epsilon & 0 & t & 0 & 0 & 0 & 0 & 0 & 0 & 0 & 0\\
t & 0 & \epsilon & t & 0 & 0 & 0 & 0 & 0 & 0 & 0 & 0\\
0 & t & t & \epsilon & t & t & 0 & 0 & 0 & 0 & 0 & 0\\
0 & 0 & 0 & t & \epsilon & 0 & v & 0 & 0 & 0 & 0 & 0\\
0 & 0 & 0 & t & 0 & \epsilon & v & 0 & 0 & 0 & 0 & 0\\
0 & 0 & 0 & 0 & v & v & \epsilon & t & t & 0 & 0 & 0\\
0 & 0 & 0 & 0 & 0 & 0 & t & \epsilon & 0 & t & 0 & 0 \\ 
0 & 0 & 0 & 0 & 0 & 0 & t & 0 & \epsilon & t & 0 & 0 \\ 
0 & 0 & 0 & 0 & 0 & 0 & 0 & t & t & \epsilon & t & t \\
w e^{ika} & 0 & 0 & 0 & 0 & 0 & 0 & 0 & 0 & t & \epsilon & 0 \\ 
w e^{ika} & 0 & 0 & 0 & 0 & 0 & 0 & 0 & 0 & t & 0 & \epsilon 

\end{array}
\right ] 
\label{ham-2c}
\end{equation}
When Four and Six GSSHC-I  are cross-linked (as shown in Fig.~\ref{fig3} ) then kernels of the Hamiltonian ($ \hat{\mathcal{H}}_{7}(k)$, $ \hat{\mathcal{H}}_{8}(k)$)  become,
\vspace{0.5cm}
\begin{equation}
\hat{\mathcal{H}}_{7}(k) = \left[ \begin{array}{cccccccccccccccc}

\epsilon & 0 & 0 & t & t & 0 & 0 & 0 & w e^{-ika} & w e^{-ika} \\ 
0 & \epsilon & t & t & 0 & 0 & 0 & w e^{-ika} & e^{-ika} & 0 \\ 
0 & t & \epsilon & 0 & 0 & v & 0 & 0 & 0 & 0 \\  
t & t & 0 & \epsilon & 0 & v & v & 0 & 0 & 0 \\ 
t & 0 & 0 & 0 & \epsilon & 0 & v & 0 & 0 & 0 \\ 
0 & 0 & v & v & 0 & \epsilon & 0 & t & t & 0 \\ 
0 & 0 & 0 & v & v & 0 & \epsilon & 0 & t & t \\  
0 & w e^{ika} & 0 & 0 & 0 & t & 0 & \epsilon & 0 & 0 \\ 
w e^{ika} &  w e^{ika} & 0 & 0 & 0 & t & t & 0 & \epsilon & 0 \\  
w e^{ika} & 0 & 0 & 0 & 0 & 0 & t & 0 & 0 & \epsilon \\
\end{array}
\right ] 
\label{ham-3a}
\end{equation}

\vspace{0.5cm}
\begin{equation}
\hat{\mathcal{H}}_{8}(k) = \left[ \begin{array}{cccccccccccccccc}

\epsilon & 0 & 0 & 0 & 0 & t & t & 0 & 0 & 0 & 0 & 0 & w e^{-ika} & w e^{-ika} \\ 
0 & \epsilon & 0 & 0 & t & t & 0 & 0 & 0 & 0 & 0 & w e^{-ika} & w e^{-ika} & 0 \\ 
0 & 0 &\epsilon & t & t & 0 & 0 & 0 & 0 & 0 & w e^{-ika} & w e^{-ika} & 0 & 0 \\ 
0 & 0 & t & \epsilon & 0 & 0 & 0 & v & 0 & 0 & 0 & 0 & 0 & 0 \\ 
0 & t & t & 0 & \epsilon & 0 & 0 & v & v & 0 & 0 & 0 & 0 & 0 \\ 
t & t & 0 & 0 & 0 & \epsilon & 0 & 0 & v & v & 0 & 0 & 0 & 0 \\ 
t & 0 & 0 & 0 & 0 & 0 & \epsilon & 0 & 0 & v & 0 & 0 & 0 & 0 \\  
0 & 0 & 0 & v & v & 0 & 0 & \epsilon & 0 & 0 & t & t & 0 & 0  \\  
0 & 0 & 0 & 0 & v & v & 0 & 0 & \epsilon & 0 & 0 & t & t & 0  \\ 
0 & 0 & 0 & 0 & 0 & v & v & 0 & 0 & \epsilon & 0 & 0 & t & t  \\ 
0 & 0 & w e^{ika} &  0 & 0 & 0 & 0 & t & 0 & 0 & \epsilon& 0 & 0 & 0 \\ 
0 & w e^{ika} &  w e^{ika} & 0 & 0 & 0 & 0 & t & t & 0 & 0 & \epsilon & 0 & 0 \\ 
w e^{ika} & w e^{ika} & 0 & 0 & 0 & 0 & 0 & 0 & t & t & 0 & 0 &\epsilon & 0 \\ 
w e^{ika} & 0 & 0 & 0 & 0 & 0 & 0 & 0 & 0 & t & 0 & 0 & 0 &\epsilon \\

\end{array}
\right ] 
\label{ham-3b}
\end{equation}

\label{appendixB}
\clearpage
\section{Magnified dispersion curves}

\begin{figure}[ht]
\centering

(a)\includegraphics[width=0.22\columnwidth]{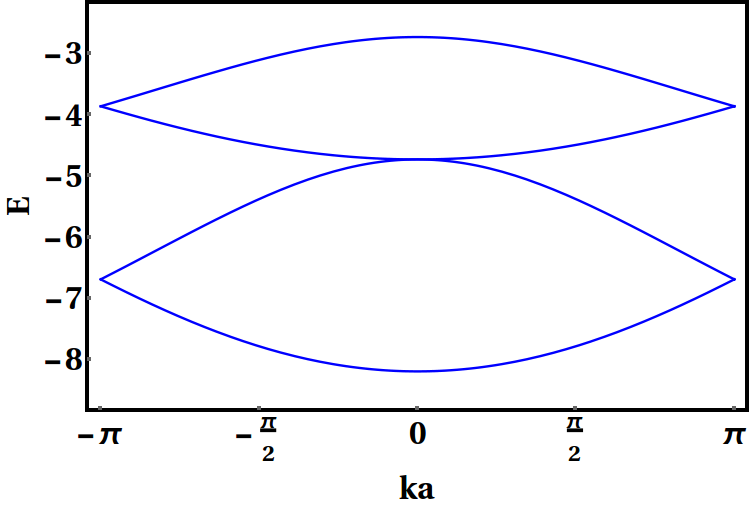}
(b)\includegraphics[width=0.22\columnwidth]{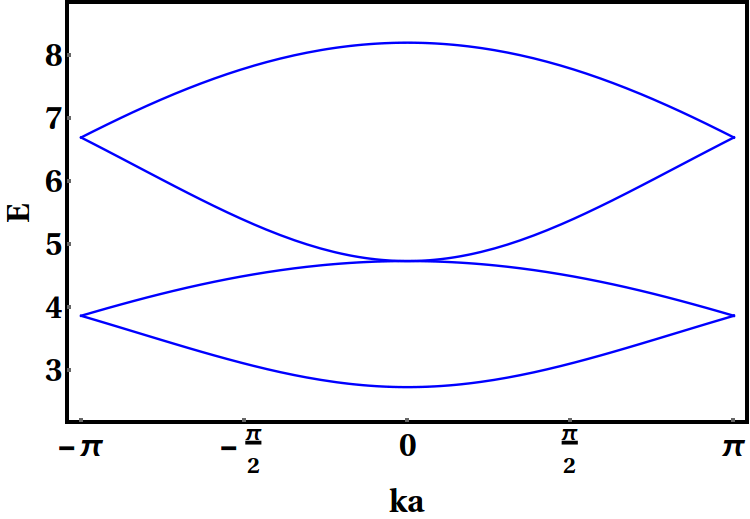}
(c)\includegraphics[width=0.22\columnwidth]{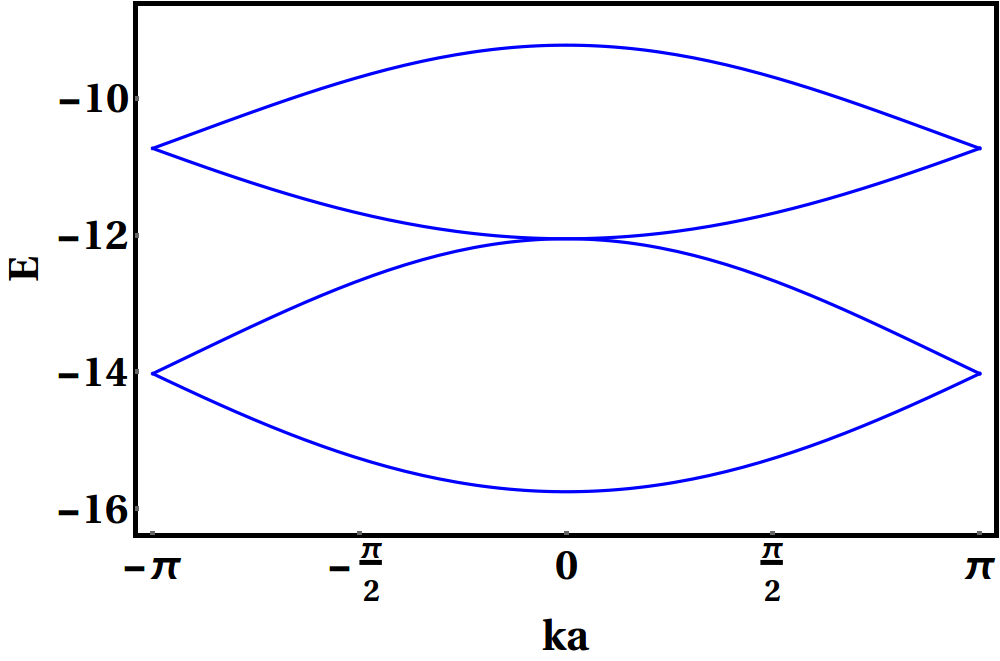}
(d)\includegraphics[width=0.22\columnwidth]{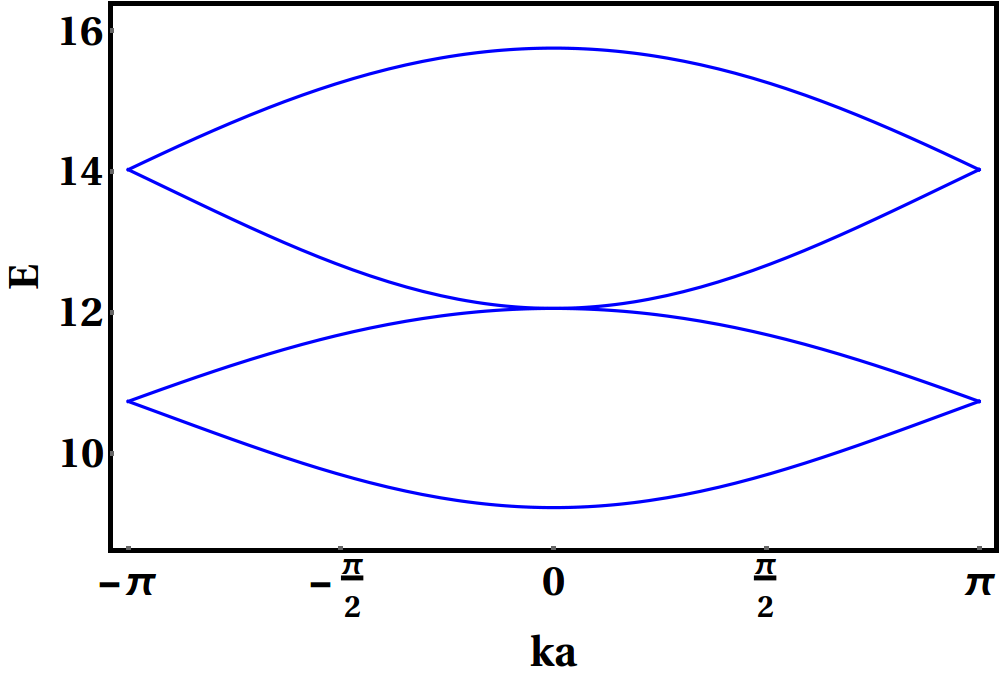}
(e)\includegraphics[width=0.22\columnwidth]{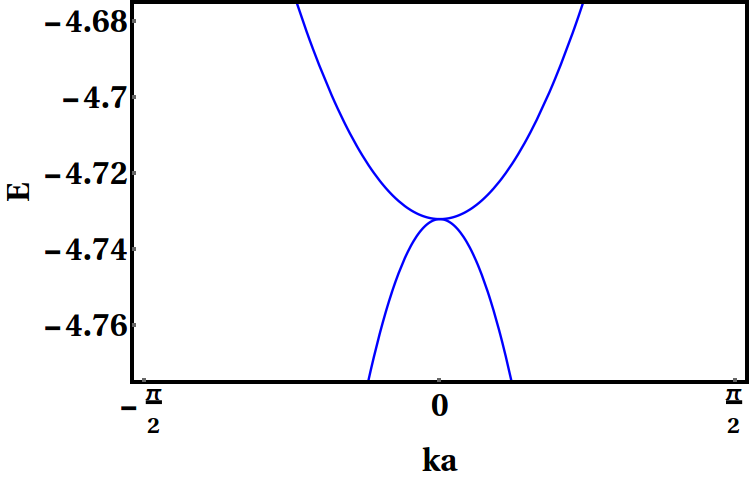}
(f)\includegraphics[width=0.22\columnwidth]{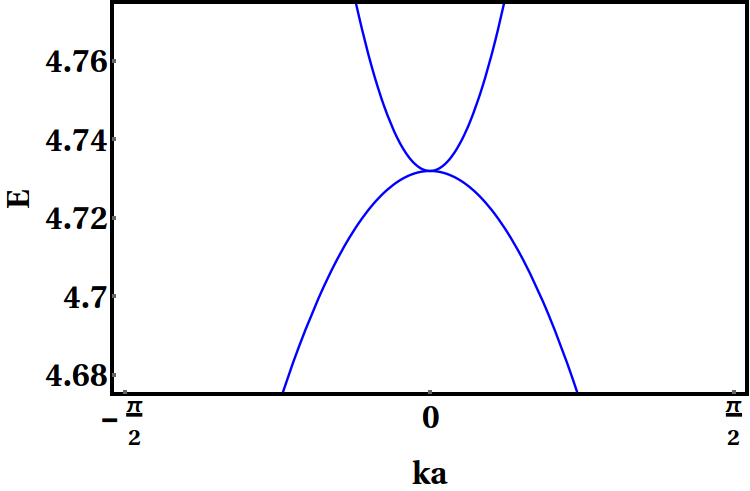}
(g)\includegraphics[width=0.22\columnwidth]{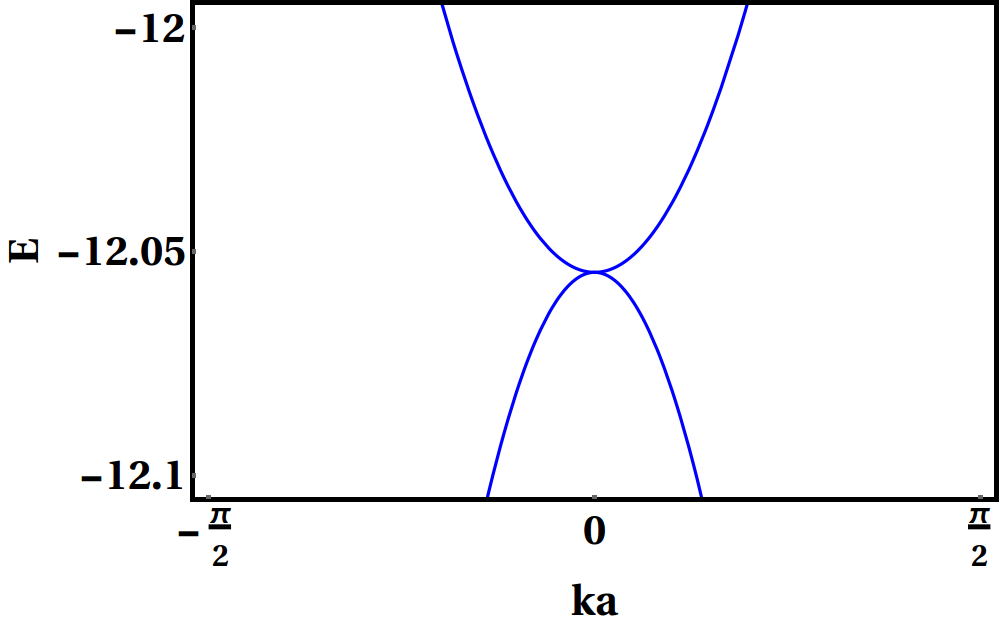}
(h)\includegraphics[width=0.22\columnwidth]{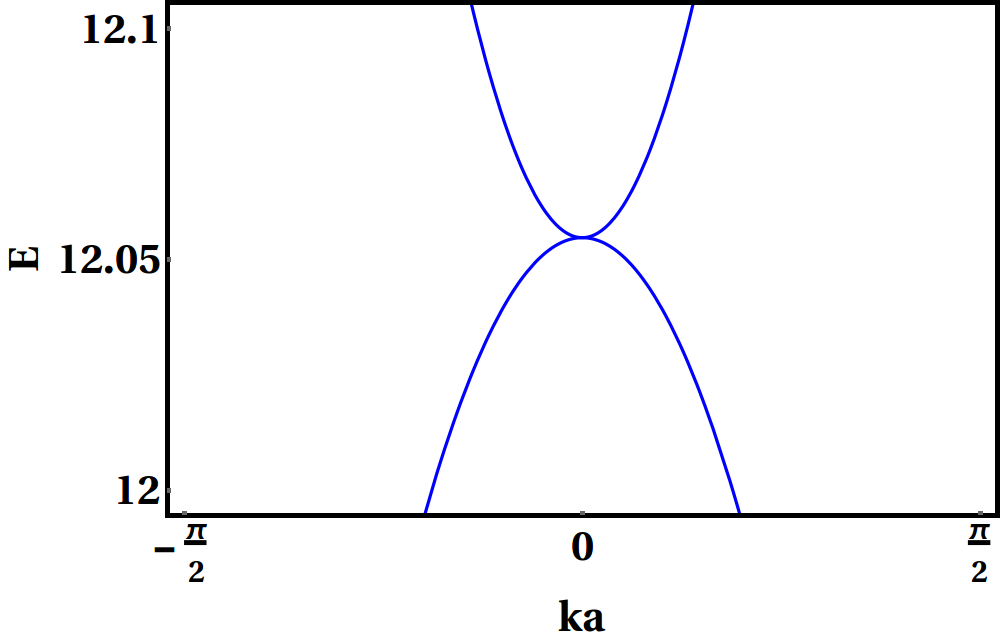}

\caption{(Color online)  (a)-(b) and (c)-(d)  are the magnified versions of Fig.~\ref{e-k3}(b) and Fig.~\ref{e-k3}(d) respectively. (e), (f), (g), and (h) are more zoomed structures of (a), (b), (c), and (d) respectively. }  
\label{mag-e-k3}
\end{figure}



\begin{figure}[ht]
\centering

(a)\includegraphics[width=0.22\columnwidth]{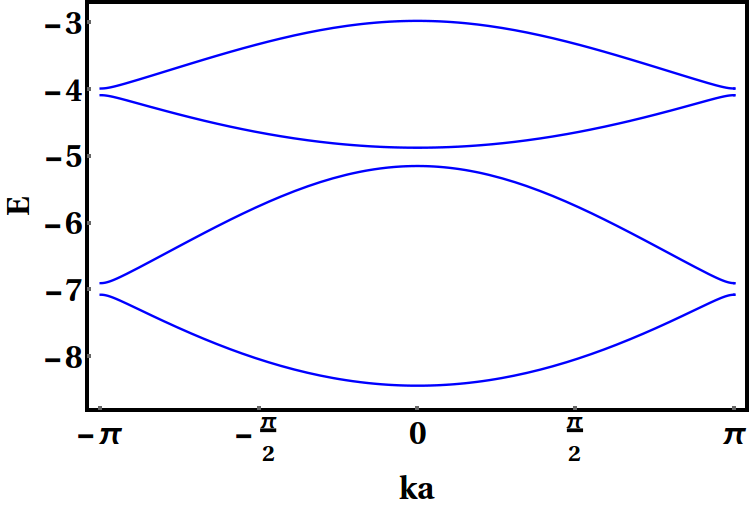}
(b)\includegraphics[width=0.22\columnwidth]{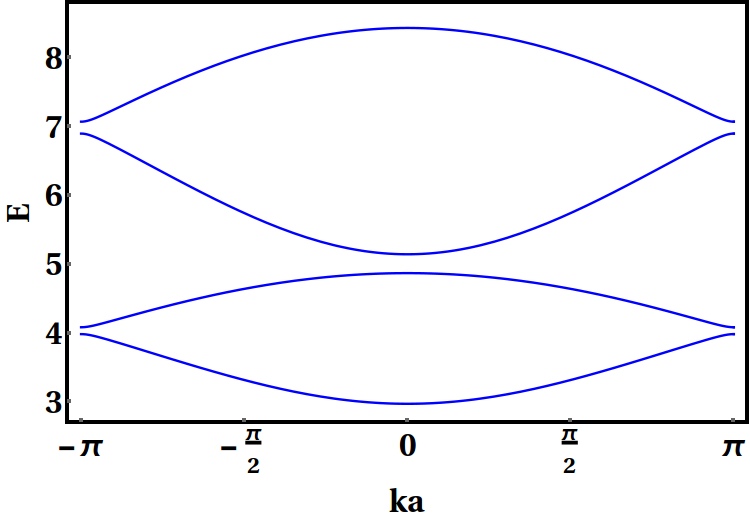}
(c)\includegraphics[width=0.22\columnwidth]{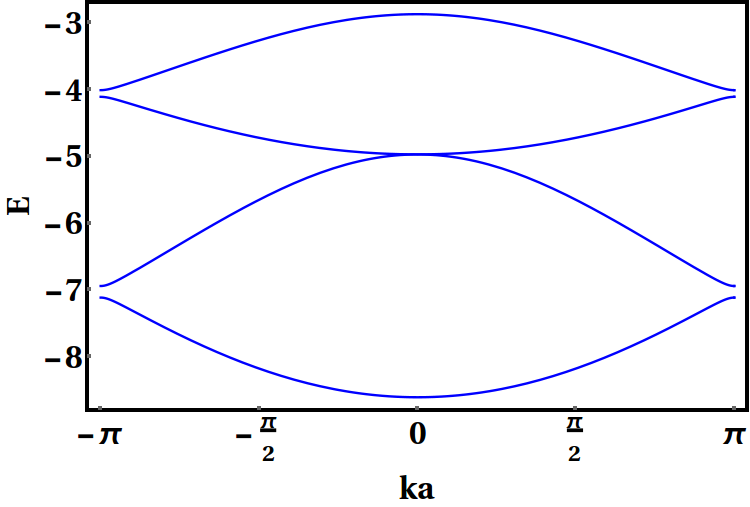}
(d)\includegraphics[width=0.22\columnwidth]{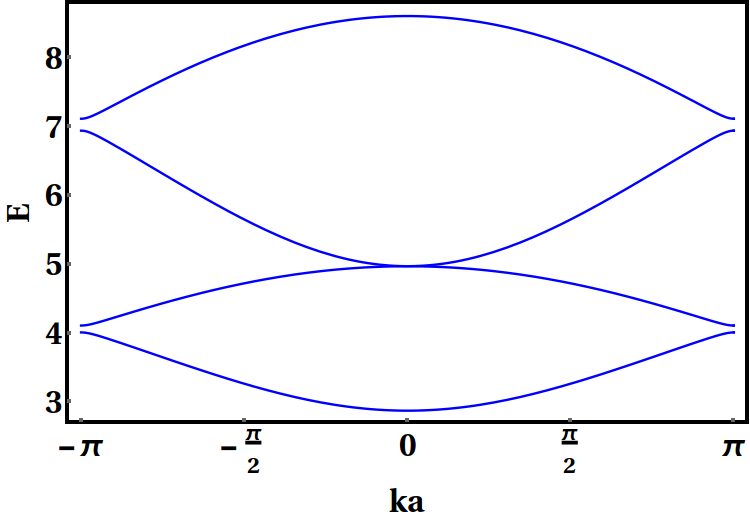}
(e)\includegraphics[width=0.25\columnwidth]{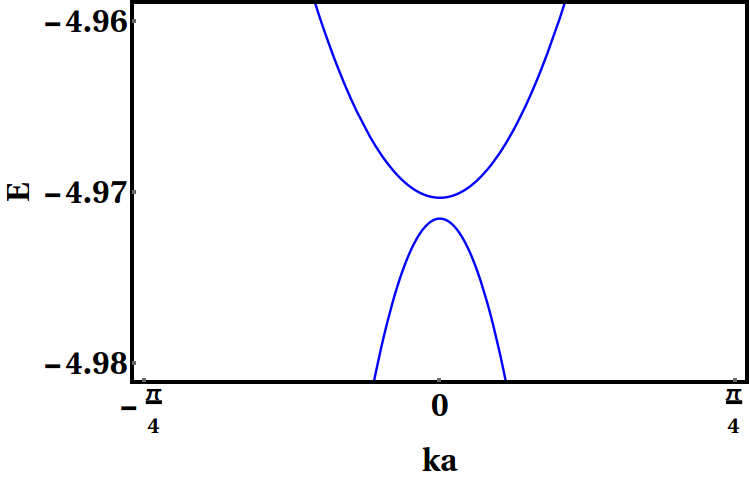}
(f)\includegraphics[width=0.25\columnwidth]{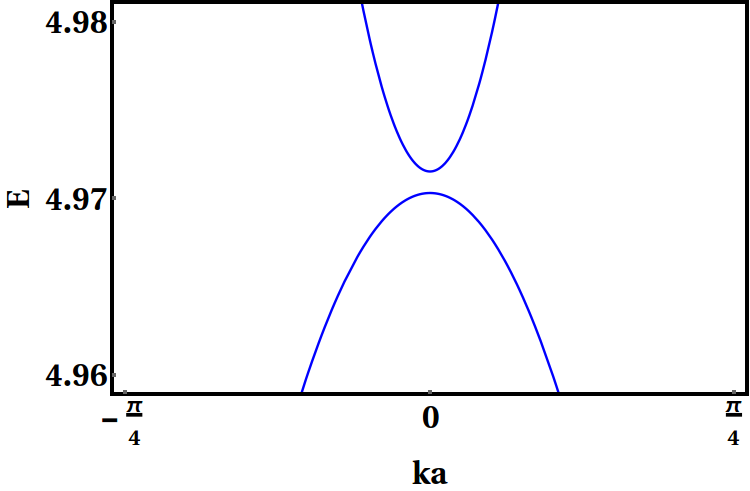}\\
(g)\includegraphics[width=0.22\columnwidth]{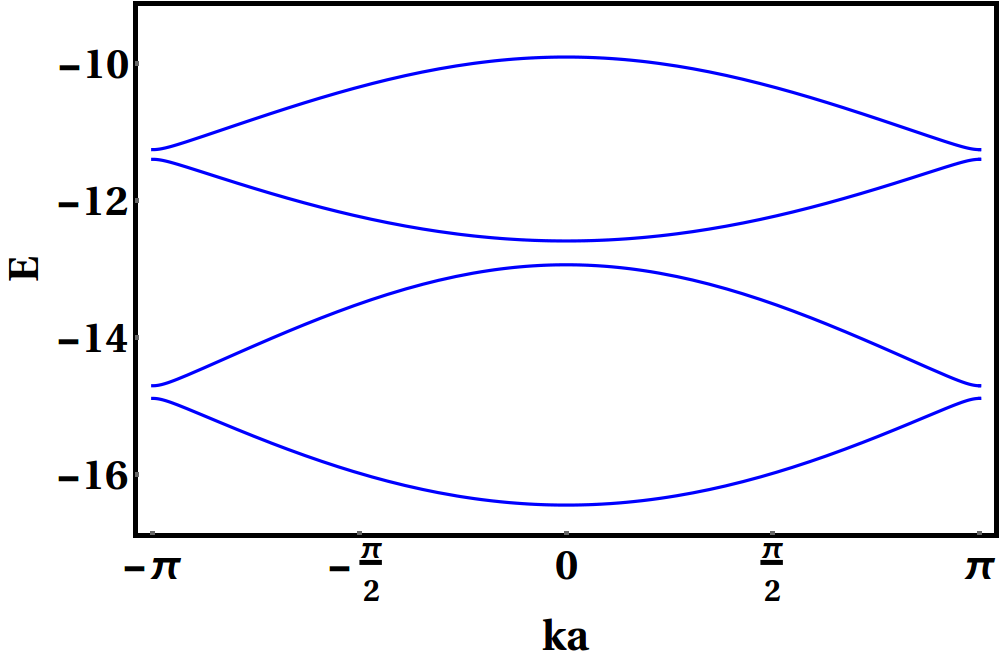}
(h)\includegraphics[width=0.22\columnwidth]{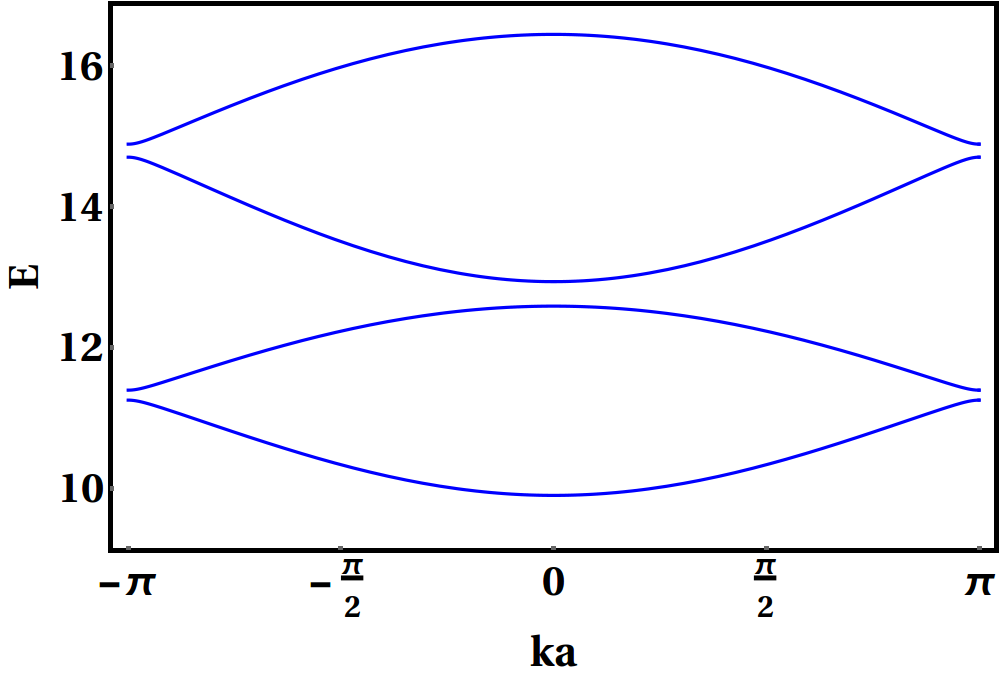}
(i)\includegraphics[width=0.22\columnwidth]{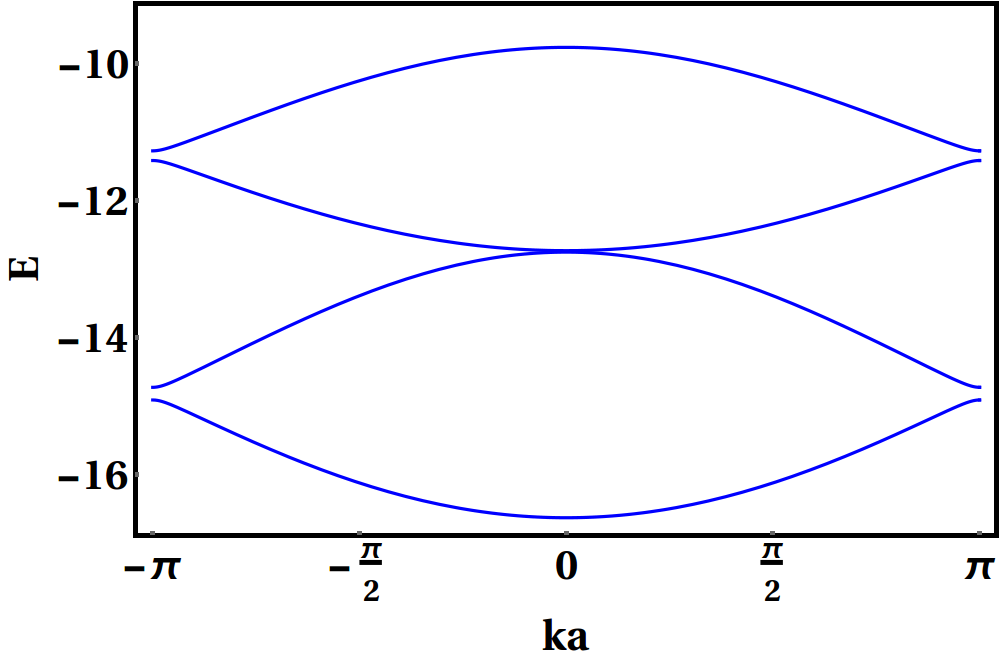}
(j)\includegraphics[width=0.22\columnwidth]{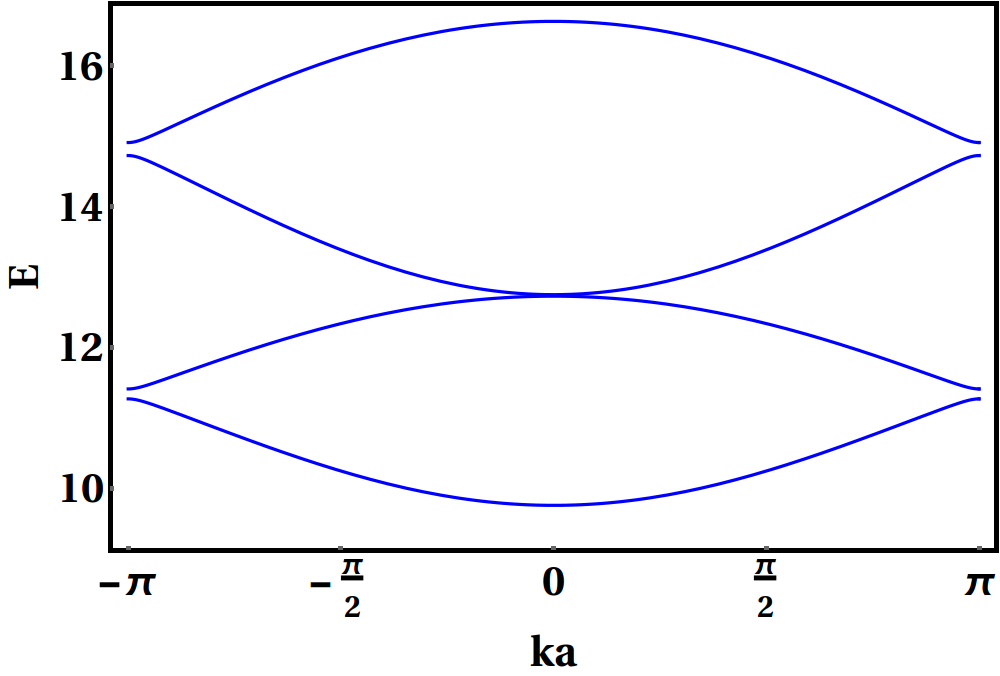}
(k)\includegraphics[width=0.25\columnwidth]{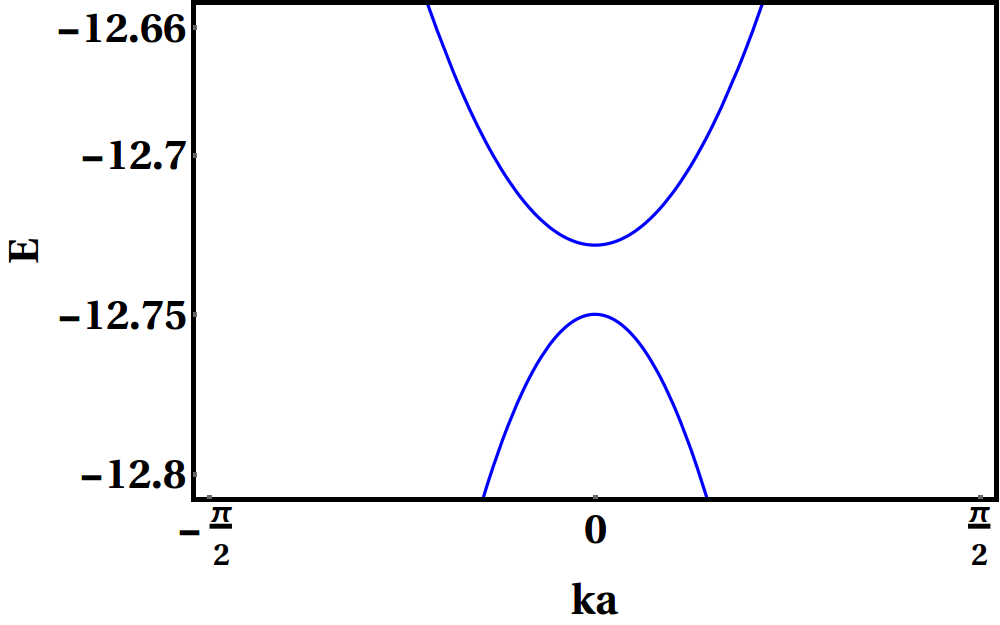}
(l)\includegraphics[width=0.25\columnwidth]{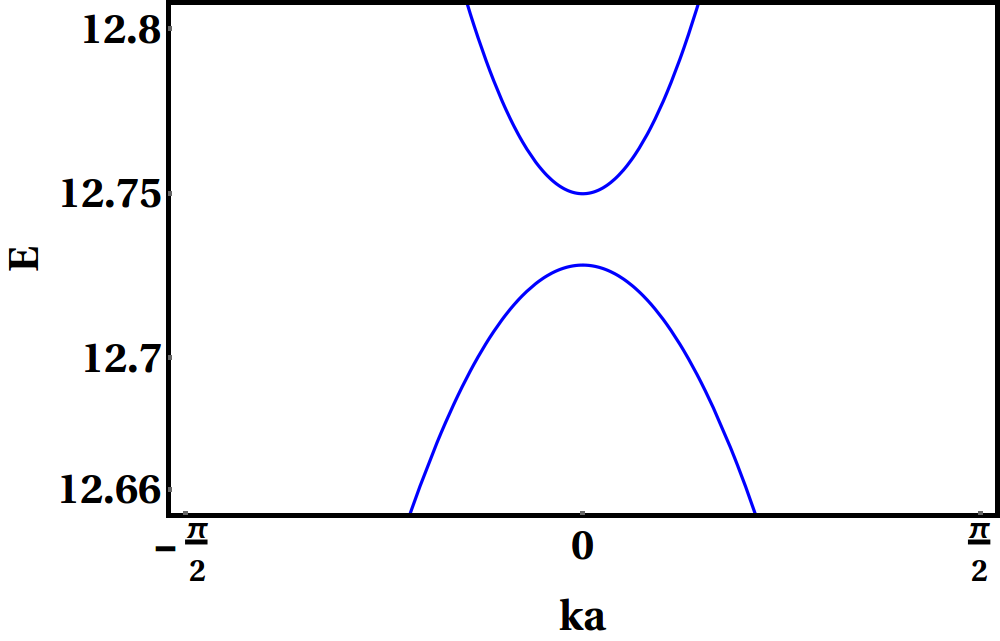}

\caption{(Color online) (a)-(b) and (c)-(d) are the magnified versions of Fig.~\ref{e-k4} (a) and (c) respectively. (e) and (f) are more zoomed structures of (c) and (d). Similarly, (g)-(h) and (i)-(j) are the magnified versions of Fig.~\ref{e-k4} (d) and (f) respectively. (k) and (l) are more zoomed structures of (i) and (j).}  
\label{zoom}
\end{figure}
\label{appendixC}

\clearpage
\section{The Zak Phases}

\begin{table}[ht]
\centering
\caption{Zak phase calculation (only for dispersive bands)}
\vspace{0.5cm}
\begin{tabular}{|p{3cm}|p{1.1cm}|p{1.1cm}|p{1.1cm}|p{1.1cm}|p{1.1cm}|p{1.1cm}|p{1.1cm}|p{1.1cm}|p{1.1cm}|p{1.3cm}|p{1.3cm}|p{1.3cm}|}
 \hline
 \multicolumn{13}{|c|}{\textbf{GSSHC-I (as shown in Fig.~\ref{fig1}(a)), Band diagram(Fig.~\ref{e-k1}(a)-(c))  } }\\
 \hline
 Choice of parameters & band(1) & band(2) & band(3) & band(4) & band(5) & band(6) & band(7) & band(8) & band(9) & band(10) & band(11) & band(12)\\
 \hline
t=1.5, v=1, w=0.9 & 0 & 0 & 0 & 0 & - & - & - & - & - & - & - & - \\
 \hline
t=1.5, v=1, w=1.1 & $\pi$ & $\pi$ & $\pi$ & $\pi$ & - & - & - & -& - & - & - & -  \\
\hline
\multicolumn{13}{|c|}{\textbf{ GSSHC-II (as shown in Fig.~\ref{fig1}(b)), Band diagram(Fig.~\ref{e-k1}(d)-(f)) } }\\
\hline
t=1.5, v=1, w=0.9 & 0 & $\pi$  & $\pi$ & $\pi$ & $\pi$ & 0 & - & - & - & - & - & -  \\
\hline
t=1.5, v=1, w=1.1 & $\pi$ & $\pi$  & 0 & 0 & $\pi$ & $\pi$ & - & - & - & - & - & -  \\
\hline
\multicolumn{13}{|c|}{\textbf{ GSSHC-III (as shown in Fig.~\ref{fig1}(c)), Band diagram(Fig.~\ref{e-k1}(g)-(i)) } }\\
\hline
t=1.5, v=1, w=0.9 & 0 & 0  & $\pi$ & $\pi$ & $\pi$ & $\pi$ & 0 & 0 & - & - & - & - \\
\hline
t=1.5, v=1, w=1.1 & $\pi$ & $\pi$  & 0 & 0 & 0 & 0 & $\pi$ & $\pi$ & - & - & - & - \\
\hline
\multicolumn{13}{|c|}{\textbf{ Two cross-linked GSSHC-I (as shown in Fig.~\ref{fig2}(a)), Band diagram(Fig.~\ref{e-k2}(a)-(c)) } }\\
\hline
t=1.5, v=1, w=0.9 & 0 & 0 & 0 & 0 & - & - & - & - & - & - & - & -  \\
 \hline
t=1.5, v=1, w=1.1 & $\pi$ & $\pi$ & $\pi$ & $\pi$ & - & - & - & - & - & - & - & -  \\
\hline
\multicolumn{13}{|c|}{\textbf{ Two cross-linked GSSHC-II (as shown in Fig.~\ref{fig2}(b)), Band diagram(Fig.~\ref{e-k2}(d)-(f)) } }\\
\hline
t=1.5, v=1, w=0.9 & 0 & 0 & 0 & 0 & 0 & 0 & - & - & - & - & - & -  \\
 \hline
t=1.5, v=1, w=1.1 & $\pi$ & $\pi$ & $\pi$ & $\pi$ & $\pi$ & $\pi$ & - & - & - & - & - & - \\
\hline
\multicolumn{13}{|c|}{\textbf{Two cross-linked  GSSHC-III (as shown in Fig.~\ref{fig2}(c)), Band diagram(Fig.~\ref{e-k2}(g)-(i)) } }\\
\hline
t=1.5, v=1, w=0.9 & 0 & $\pi$  & 0 & 0 & 0 & 0 & $\pi$ & 0 & - & - & - & - \\
\hline
t=1.5, v=1, w=1.1 & 0 & $\pi$  & $\pi$ & $\pi$ & $\pi$ & $\pi$ & $\pi$ & 0 & - & - & - & - \\
\hline
\multicolumn{13}{|c|}{\textbf{ Four cross-linked GSSHC-I (as shown in Fig.~\ref{fig3}(a)), Band diagram(Fig.~\ref{e-k4}(a)-(c)) } }\\
\hline
t=3.92, v=1, w=0.9 & 0 & 0  & 0 & 0 & 0 & 0 & 0 & 0 & - & - & - & - \\
\hline
t=3.92, v=1, w=1.1 & $\pi$ & $\pi$  & $\pi$ & $\pi$ & $\pi$ & $\pi$ & $\pi$ & $\pi$ & - & - & - & - \\
\hline
\multicolumn{13}{|c|}{\textbf{ Six cross-linked GSSHC-I (as shown in Fig.~\ref{fig3}(b)), Band diagram(Fig.~\ref{e-k4}(d)-(f))} }\\
\hline
t=7.95, v=1, w=0.9 & 0 & 0  & 0 & 0 & 0 & 0 & 0 & 0 & 0 & 0 & 0 & 0 \\
\hline
t=7.95, v=1, w=1.1 & $\pi$ & $\pi$  & $\pi$ & $\pi$ & $\pi$ & $\pi$ & $\pi$ & $\pi$ &  $\pi$ &  $\pi$ &  $\pi$ &  $\pi$ \\
\hline
\end{tabular}
\label{z}
\end{table}
 \begin{itemize}
     \item 
 band(1), band(2),...denote only the dispersive energy bands ordering from higher to lower energy as shown in the energy-band diagram for all networks.
 \end{itemize}
\end{widetext}

\label{appendixD}

\end{document}